\documentclass[aps,prd,twocolumn,superscriptaddress,nofootinbib,floatfix]{revtex4-2}
\usepackage{graphicx}
\usepackage[colorlinks=true,linkcolor=blue,citecolor=blue,urlcolor=blue]{hyperref}
\usepackage{amsmath,amssymb,bm}
\usepackage{xcolor}
\usepackage{xcolor}
\definecolor{osu}{HTML}{ba0c2f}
\usepackage{xurl}

\newif\iffinalbuild\finalbuildtrue
\newif\ifjoint\jointfalse
\newcommand{\figorbox}[2]{%
  \IfFileExists{#1}{\includegraphics[width=#2]{#1}}{%
    \iffinalbuild\errmessage{Missing figure file #1}\fi
    \fbox{\parbox[c][0.5#2][c]{0.94#2}{\centering\small
      Figure placeholder\\[2pt]\texttt{\detokenize{#1}}}}}}
\newcommand{\orcid}[1]{\texorpdfstring{\begingroup
  \hypersetup{hidelinks}\href{https://orcid.org/#1}{\includegraphics[width=10pt]{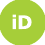}\,}%
\endgroup}{}}
\newcommand{\rst}{r_{\rm stack}}
\newcommand{\rS}{r^{s}_{\rm stack}}
\newcommand{\rP}{r^{p}_{\rm stack}}
\newcommand{\kzo}{k_{z,0}}
\newcommand{\shat}{\hat{\bm s}}
\newcommand{\pd}{\hat{\bm p}_{\rm d}}
\newcommand{\pu}{\hat{\bm p}_{\rm u}}
\newcommand{\kd}{\hat{\bm k}_{\rm d}}
\newcommand{\ku}{\hat{\bm k}_{\rm u}}
\newcommand{\xhat}{\hat{\bm x}}
\newcommand{\zhat}{\hat{\bm z}}
\newcommand{\Amp}{E_0}
\newcommand{\ph}{e^{i\kappa(x\cos\beta+y\sin\beta)}}

\begin{document}

\title{Lunar regolith effects on low-frequency antennas for Dark Ages 21-cm measurements: a self-consistent layered-medium model tested against Apollo~17 data}

\author{Paramita Dasgupta\,\orcid{0000-0002-9559-4803}}
\email{dasgupta.80@osu.edu}
% \affiliation{Center for Cosmology and AstroParticle Physics (CCAPP), The Ohio State University, Columbus, OH 43210, USA}
% \affiliation{Department of Physics, The Ohio State University, Columbus, OH 43210, USA}
% \affiliation{Department of Astronomy, The Ohio State University, Columbus, OH 43210, USA}
\affiliation{Center for Cosmology and AstroParticle Physics (CCAPP), \href{https://ror.org/00rs6vg23}{\color{osu}The Ohio State University}, Columbus, OH 43210, USA}
\affiliation{Department of Physics, \href{https://ror.org/00rs6vg23}{\color{osu}The Ohio State University}, Columbus, OH 43210, USA}
\affiliation{Department of Astronomy, \href{https://ror.org/00rs6vg23}{\color{osu}The Ohio State University}, Columbus, OH 43210, USA}

\begin{abstract}
The lunar far side is a promising site for measuring the global 21-cm absorption signal from the Dark Ages, expected to reach about 40~mK near 16~MHz. At these frequencies, the regolith is part of the antenna system. Its layered structure affects the antenna current, impedance, beam, coupling to the ground, and thermal emission. We calculate the response of a wire antenna above an arbitrary stack of planar lossy layers, retaining both propagating and evanescent fields and solving self-consistently for its current. We validate the calculation against exact limits, energy conservation, and independent numerical codes for layered media and antennas above a half-space, and test it against the Apollo~17 Surface Electrical Properties measurements along the Taurus-Littrow traverse. A graded profile fitted to the low-frequency measurements predicts the higher-frequency data held out from the fit. The best-fitting two-layer profile fits the low-frequency data better but predicts a higher-frequency interference ripple with an rms up to 2.6 times that of the measured fluctuations. For the tested profile families, the measurements place the rise of permittivity to about 5 at 12--18~m depth and disfavor the two-layer parameter range adopted in a LuSEE-Night subsurface study, but do not determine the sharpness of the transition or exclude all two-layer profiles. For a stand-alone LuSEE-Night-like dipole above a far-side profile informed by Chang'E-4 radar, the impedance-boundary approximation used in LuSEE-Night simulations differs from the layered-medium calculation by 0.13--0.17 in ground fraction over the tested antenna heights. An illustrative recovery exercise, not a forecast for the flight instrument, shows that keeping the bias below the signal requires a subsurface model broad enough to include the underlying profile, a model of cooling in the upper centimeters, and tight priors on receiver-gain slope and curvature. Even under these conditions, the statistical error for a twelve-night observation remains larger than the signal. These results show that antenna response, subsurface layering, and thermal emission must be treated together when designing and interpreting lunar low-frequency experiments.
\end{abstract}

\maketitle

% =====================================================================
\section{Introduction}
\label{sec:intro}
% =====================================================================

Between recombination and the formation of the first stars, neutral hydrogen in the Universe cooled below the temperature of the cosmic microwave background (CMB) and absorbed part of it at 21~cm, the wavelength of its hyperfine transition \citep{Furlanetto2006,PritchardLoeb2012}. During these Dark Ages, at redshifts of about 30--200, stars and galaxies had not yet formed. The signal therefore depends on the thermal history of the gas and on processes that alter it, including interactions between dark matter and ordinary matter and energy injection from exotic sources \citep{LoebZaldarriaga2004,Barkana2018,MondalBarkana2023}. The expected sky-averaged, or global, signal is an absorption of about 40~mK near 16~MHz, extending over 7--46~MHz, beneath Galactic radio emission more than $10^5$ times brighter \citep{PritchardLoeb2012,MondalBarkana2023,NovacoBrown1978}.

Observing this band from Earth is difficult. Below about 30~MHz, the ionosphere refracts and absorbs radio waves and introduces time-dependent spectral structure \citep{Datta2016}. Terrestrial radio transmitters also produce interference far brighter than the signal. The far side of the Moon is shielded from both \citep{JesterFalcke2009,Koopmans2021}. The plasma immediately above the lunar surface is too tenuous to pose a comparable obstacle: at the electron density measured by the Chandrayaan-3 lander, it blocks radio waves only below about 0.2~MHz \citep{Manju2025}. The Moon offers a quiet observing environment, but the absence of an ionosphere does not remove the electromagnetic influence of the ground.

Experiments on Earth illustrate how strongly the ground can affect a low-frequency measurement. EDGES, the Experiment to Detect the Global Epoch of Reionization Signature, reported an absorption feature at 78~MHz \citep{Bowman2018}; a metal ground plane over moist soil can form a resonant cavity that produces a feature of this kind \citep{Bradley2019}. SARAS~3, a Shaped Antenna measurement of the background RAdio Spectrum, floated its antenna on a lake to provide a uniform ground and did not find the reported absorption \citep{Raghunathan2021,Singh2022,Bevins2022}. On the Moon, the ground is the regolith, the broken rock and fine soil produced by meteoroid impacts. Its bulk density, permittivity and dielectric loss vary with depth, and these properties determine how radio waves interact with the surface and the material beneath it.

Lunar radio experiments use several antenna configurations. Some place antennas directly on the surface, as in the Apollo~17 Surface Electrical Properties (SEP) experiment \citep{Simmons1972} and the proposed FARSIDE \citep{Burns2019,Burns2021FARSIDE}, FarView \citep{Polidan2024} and DEX \citep{Ghosh2026DEX} arrays. Others mount antennas a few meters above the ground on landers, as on Chang'E-4 \citep{Zhu2021} and the Odysseus lander \citep{Gopalswamy2026,Hibbard2025}, and as planned for LuSEE-Night, the Lunar Surface Electromagnetics Experiment at Night \citep{Bale2023,BaleSlosar2024}. A third group observes from lunar orbit. These include the flown RAE-2 mission \citep{Alexander1975} and the planned or proposed DSL, PRATUSH, CosmoCube and DARE experiments \citep{Chen2021DSL,Shi2022,PRATUSH2023,Artuc2025,Burns2017}, for which the Moon occupies a substantial part of the antenna's field of view.

At 7--46~MHz, the wavelength is 6.5--43~m. An antenna on the surface or a few meters above it is therefore less than one wavelength from the regolith, which becomes part of the antenna system. The regolith affects the input impedance and hence the coupling to the receiver, the beam, which sets the response to radiation from different directions, and the ground fraction, the fraction of the power accepted by the antenna that the ground absorbs. By reciprocity, the ground fraction is also the weight of the ground's thermal emission in the antenna temperature. These effects arise from both propagating and evanescent waves. The antenna field can be decomposed into plane waves (Sec.~\ref{sec:formalism}): propagating waves reach the surface at angles ranging from normal to grazing incidence, while evanescent waves form the near field and can couple to the ground even when the antenna is several meters above it. Reflection from the regolith changes the field at the antenna and, in turn, its current and response.

The size of the ground contribution makes this a central concern for a Dark Ages measurement. For a stand-alone LuSEE-Night-like dipole 1.5--3.5~m above a far-side reference profile, the ground fraction across this band is 0.36--0.88 (Sec.~\ref{sec:inputs}). At temperatures of 90--250~K \citep{Hayne2017,Williams2017}, the ground contributes tens to hundreds of kelvin to the antenna temperature, compared with a signal of about 40~mK. Even a small error in the frequency dependence of this contribution can introduce a spectral residual comparable to or larger than the Dark Ages signal. Recovering the global signal requires a model of the ground whose range of allowed depth profiles includes the actual, as yet unknown, subsurface at the landing site. Its contribution must then be fitted together with the sky foreground and the receiver response.

Current LuSEE-Night simulations represent the regolith using the layered impedance boundary implemented in the HFSS solver \citep{HFSSintro2025}. This provides an efficient approximation and reproduces the exact reflection coefficient at normal incidence. However, an antenna close to the surface responds to both oblique propagating waves and evanescent waves, whose contributions are not adequately captured by this boundary condition \citep{SeniorVolakis1995}. For a stand-alone LuSEE-Night-like dipole above the far-side reference profile, we find that the approximation produces an absolute error of 0.13--0.17 in the ground fraction at all antenna heights tested, from 0.3 to 3.5~m (Sec.~\ref{sec:ibcantenna}). This discrepancy also persists when the antenna is mounted on an illustrative lander model. The lander introduces an additional change in the antenna response above 20~MHz (Sec.~\ref{sec:lander}).

\begin{table*}[t]
  \caption{Subsurface models and electromagnetic methods used in previous low-frequency radio studies of the Moon and in this work. Here $\varepsilon$ is the complex relative permittivity and $\varepsilon'$ its real part.\label{tab:prior}}
  \begin{ruledtabular}
  \begin{tabular}{p{2.3cm}p{3.4cm}p{3.6cm}p{3.5cm}p{2.9cm}}
  Study & Subsurface model & Electromagnetic method & Quantities calculated & Comparison with lunar data \\
  \hline
  Grimm \citep{Grimm2018}
  & Layered porosity profile
  & Layered-earth inversion
  & Subsurface structure
  & Apollo~17 SEP field measurements (inversion) \\

  Yousuf {\it et al.} \citep{Yousuf2026}
  & Two uniform layers: a 0.5--3~m top layer with $\varepsilon'=3.2$--4.2 over a lower layer with $\varepsilon'=3.8$--4.8
  & HFSS with a layered impedance boundary
  & LuSEE-Night beam; joint inference of subsurface properties and foreground
  & -- \\

  Camacho {\it et al.} \citep{Camacho2026}
  & Eight layers, with $\varepsilon'$ ranging from 3.7 at the surface to 6.1 at 100~m
  & HFSS with a layered impedance boundary
  & Antenna beams for sky-map reconstruction
  & -- \\

  Rotermund {\it et al.} \citep{Rotermund2026}
  & Uniform, lossless half-space
  & Approximate analytic field; HFSS integral-equation method
  & Input impedance as a function of antenna height
  & -- \\

  Burns {\it et al.} \citep{Burns2019}
  & Uniform medium, $\varepsilon=3+0.005i$
  & NEC-4.2
  & FARSIDE antenna-element beam
  & -- \\

  Smith {\it et al.} \citep{DAPowerSpec2026}
  & One- and four-layer dielectric models
  & FarView dipole simulations
  & Foreground leakage into the 21-cm power spectrum
  & -- \\

  This work
  & Planar stacks of any number of lossy layers, including graded and transition profiles
  & Layered-medium Green's function, including evanescent waves; antenna currents solved using the method of moments, with and without a lander
  & SEP fields, input impedance, antenna beam, ground fraction, thermal emission, calibrator response, reflectivity and signal bias
  & Apollo~17 SEP measurements: fits at 2.1 and 4.0~MHz; predictions tested at 8.1--32.1~MHz \\
  \end{tabular}
  \end{ruledtabular}
\end{table*}

The subsurface models adopted in lunar radio design studies vary considerably (Table~\ref{tab:prior}). The two LuSEE-Night studies use either two uniform layers \citep{Yousuf2026} or eight layers \citep{Camacho2026}; the former identifies the unknown subsurface as the main limitation on instrument calibration. FARSIDE and FarView simulations have considered uniform or few-layer grounds \citep{NEC42,Burns2019,FARSIDEpol2025,DAPowerSpec2026}. Sensitivity studies for orbiting experiments, in turn, treat the Moon as a body at 180~K that reflects a constant 7\% of the sky \citep{Liu2024VZOP,Shi2022,McKinley2018,Evans1969,Vedantham2015}. To our knowledge, none of the subsurface models adopted in these studies has been tested by forward prediction of radio propagation measured on the Moon.

Such measurements are available from Apollo~17. The SEP experiment measured radio propagation along the lunar surface at 1--32~MHz \citep{Simmons1972,Simmons1973}, directly within the band of interest to current lunar radio experiments. Its archived data \citep{SEPdata2021} have also been used to infer the depth profile of the regolith \citep{Grimm2018}. The experiment's final report had already found that no plane-layer model reproduced the measurements accurately at all frequencies \citep{SEPFinal1975}. A forward model that reproduces the SEP measurements provides a quantitative test of how a layered regolith reflects and absorbs radio waves in this band. Such a test is particularly valuable because these same electromagnetic properties affect the antenna response and the calibration of a lunar radiometer. No equivalent surface-propagation measurement exists for the far side. There, the Chang'E-4 rover and Chang'E-6 lander used penetrating radars to sound the upper 40~m and 3~m of the regolith, respectively \citep{Li2020,Zhang2025CE6}. We use these radar results as site-specific inputs to construct far-side profiles and examine how the antenna response varies with the assumed subsurface (Sec.~\ref{sec:farside}).

Previous studies have addressed individual parts of this problem, but have not combined them in a calculation tested against lunar measurements and applied to the response of a radiometer. Studies of antennas over a uniform half-space include near-field effects but do not account for layering \citep{Rotermund2026,Burns2019}. LuSEE-Night studies include multiple layers but represent them through a surface impedance, in beam calculations \citep{Camacho2026} and in a joint inference of the subsurface and the Galactic foreground \citep{Yousuf2026}. FarView simulations of dipoles over dielectric layers examine how the antenna beam couples foreground emission into the 21-cm power spectrum \citep{DAPowerSpec2026}. The inversion by Grimm \citep{Grimm2018} infers a depth profile from the SEP measurements but does not calculate the response of a radio antenna or radiometer. Here, we use the measurements to test a forward calculation. A regolith profile fitted over one frequency range must predict the measurements at frequencies excluded from the fit, and the same calculation must determine the observables of a radiometer antenna.

We bring these elements together in a single framework that models the response of an antenna above a layered lunar regolith, tests the subsurface model against Apollo~17 measurements, and examines the implications for recovering the Dark Ages 21-cm signal. Building on established methods for electromagnetic fields above layered media \citep{Sommerfeld1909,Weyl1919,Chew1995,Michalski1998} and the method of moments for wire antennas \citep{Harrington1968}, we calculate the fields and antenna observables for a current above any number of planar, laterally uniform lossy layers, including the evanescent waves, and solve for the current itself over the same layers. The calculation gives both the fields measured by SEP and the input impedance, beam, ground fraction, thermal emission and calibrator response of an antenna above the regolith (Sec.~\ref{sec:formalism}). It extends the plane-wave formalism of Ref.~\cite{Dasgupta2026}, developed for NASA's balloon-borne ANITA (Antarctic Impulsive Transient Antenna) and tested against the HiCal calibration pulses recorded after reflection from Antarctic ice \citep{DasguptaJain2021,Prohira2018,Gorham2017}. We verify the calculation against exact limits and energy conservation, against an independent code for layered media, and against an independent antenna code over a homogeneous half-space (Sec.~\ref{sec:verification}). We then test it against Apollo~17 by fitting subsurface profiles at 2.1 and 4.0~MHz and predicting the measurements at higher frequencies (Sec.~\ref{sec:sep}). The comparison accounts explicitly for correlated residuals, the receiver noise floor and uncertainty in the transmitter current. 

The paper has a methodological core, an empirical test and their applications. Section~\ref{sec:regolith} describes the regolith models, Sec.~\ref{sec:formalism} develops the electromagnetic and antenna calculation, and Sec.~\ref{sec:verification} verifies this calculation. Section~\ref{sec:sep} tests the calculation and the models against Apollo~17 measurements. This is the central empirical result. Sections~\ref{sec:radiometer}--\ref{sec:bias} apply the calculation to a far-side radiometer: the antenna and its lander, the orbiting calibrator as a probe of the local regolith, the thermal emission of the regolith, and an illustrative study of signal recovery. Section~\ref{sec:discussion} discusses the results and the response of orbiting and surface-based experiments and summarizes the approximations and tests. Appendix~\ref{app:numerics} describes the numerical evaluation of the integrals. Appendices~\ref{app:decomp}--\ref{app:mom} derive the fields and antenna quantities. Appendices~\ref{app:sepdata}--\ref{app:stats} give the layout of the archived Apollo~17 data, the robustness tests and the statistical treatment of the comparison and of the Fisher-matrix estimates, including the correction of the calibrator passes for the response of the lander. Appendices~\ref{app:signal} and \ref{app:secondary} give the sky and signal models and further results for orbiting and surface experiments. \ifjoint A companion Letter, submitted jointly, presents the Apollo~17 test in a focused form \citep{DasguptaLetter}.\fi

% =====================================================================
\section{Regolith models}
\label{sec:regolith}
% =====================================================================

Two depth-dependent quantities describe the regolith in our calculations: the real relative permittivity $\varepsilon'$ and the loss tangent $\tan\delta$. The vertical coordinate $z$, defined in Sec.~\ref{sec:formalism}, points upward, with the surface at $z=0$. The depth below the surface is $|z|$. This section describes the dielectric properties and depth profiles used in our calculations and identifies the source of each input, whether it comes from published measurements, a fit to the Apollo~17 SEP data, or an assumption made in this work. The regolith affects the fields above the surface through the reflection coefficients of the layered ground (Sec.~\ref{sec:stack}). To calculate the power absorbed in each layer and its contribution to thermal emission, we also need the fields within the regolith (Sec.~\ref{sec:absorption}).

\subsection{Density, permittivity and loss}
\label{sec:density}

The bulk density $\rho$ is the mass of solid grains per unit volume
of regolith, including pore space. It increases rapidly with depth
as the material becomes more compact. From measurements of
nighttime surface cooling by the Diviner radiometer aboard the
Lunar Reconnaissance Orbiter, \citet{Hayne2017} derived the profile
\begin{equation}
  \rho(z) = \rho_{\rm deep}
  -(\rho_{\rm deep}-\rho_{\rm surf})\,e^{-|z|/z_\rho},
  \label{eq:density}
\end{equation}
where $\rho_{\rm surf}=1.10$~g\,cm$^{-3}$ is the surface density
and $\rho_{\rm deep}=1.80$~g\,cm$^{-3}$ is the asymptotic density
at greater depths. At a depth $|z|=z_\rho$, the density has
completed $1-e^{-1}\simeq63\%$ of its increase from the surface
to the deep value. The Diviner data do not constrain all three
parameters independently, so \citet{Hayne2017} fixed the two
densities and fitted $z_\rho$. We adopt $z_\rho=0.07$~m, the
mean of their values of $6.8\pm0.7$~cm at latitudes within
$\pm60^\circ$. The density is close to its deep value below
about 0.3~m. Measurements from Apollo drive tubes and drill
cores are consistent with this profile, giving densities of
about 1.1~g\,cm$^{-3}$ near the surface and
1.8--1.9~g\,cm$^{-3}$ at depths of 30--60~cm
\citep{Carrier1991,Hayne2017}.

For the time dependence $e^{-i\omega t}$ used throughout this
work, where $\omega=2\pi\nu$ is the angular frequency and $\nu$
the frequency, the complex relative permittivity of a nonmagnetic
medium is
\begin{equation}
  \varepsilon=\varepsilon'+i\varepsilon'',
\end{equation}
where $\varepsilon'$ and $\varepsilon''$ are its real and
imaginary parts, respectively
\citep{BornWolf1999,BalanisAEE2012}. The real part determines
the phase velocity, while both parts enter the reflection
coefficient at a boundary. For low-loss media, reflection is
governed mainly by the real part. The complex refractive index is
$n=\sqrt{\varepsilon}$, with the root chosen so that
$\mathrm{Im}\,n\geq0$. We treat the regolith as nonmagnetic, with
relative permeability $\mu_r=1$ (Sec.~\ref{sec:limitations}).
The loss tangent is defined as
\begin{equation}
  \tan\delta=\frac{\varepsilon''}{\varepsilon'},
  \label{eq:tandelta}
\end{equation}
where $\delta$ is the loss angle. For $\tan\delta\ll1$, the
wave power decreases by a factor of $e$ over the attenuation
length
\[
  L_{\rm att} =
  \frac{c}{2\pi\nu\sqrt{\varepsilon'}\tan\delta},
\]
where $c$ is the speed of light in vacuum. For example, at
10~MHz, $\varepsilon'=3$ and
$\tan\delta=0.01$ give $L_{\rm att}\approx280$~m.

Laboratory measurements of lunar samples relate the dielectric
properties to bulk density and composition
\citep{Olhoeft1975,Carrier1991}:
\begin{equation}
  \varepsilon'=1.919^{\,\rho},\qquad
  \tan\delta=10^{\,0.038\,f_{\rm FeTi}+0.312\rho-3.260},
  \label{eq:epsdensity}
\end{equation}
where $\rho$ is in g\,cm$^{-3}$ and $f_{\rm FeTi}$ is the
combined abundance of TiO$_2$ and FeO, expressed as a
percentage by weight. The first relation gives
$\varepsilon'=2.05$ at the adopted surface density and
$\varepsilon'=3.23$ at the deep density.

The near-surface density also varies between landing sites.
The ChaSTE probe (Chandra's Surface Thermophysical Experiment)
on Chandrayaan-3, at $69^\circ$S, inferred a mean density of
1.94~g\,cm$^{-3}$ over the upper 10~cm
\citep{Mathew2025}. Over the same depth, Eq.~(\ref{eq:density})
gives a mean density of 1.43~g\,cm$^{-3}$. Through
Eq.~(\ref{eq:epsdensity}), these values correspond to
permittivities of about 3.5 and 2.5, respectively. This
difference illustrates the variation in near-surface density
between sites and cautions against assuming a single profile
for the entire Moon. We therefore leave the surface permittivity
free when fitting the SEP measurements and vary it in our
sensitivity analyses. The orbiting calibrator planned for LuSEE-Night
could provide an additional constraint at its landing site.
A transmitter on an orbiter will send a signal that the antenna on
the lander records during each pass overhead
(Sec.~\ref{sec:calibrator}).

The relation for $\tan\delta$ also accounts for compositional
differences between lunar soils. Basaltic mare soils are
generally richer in FeO and TiO$_2$ than highland soils. At
$\rho=1.5$~g\,cm$^{-3}$, the relation gives
$\tan\delta=0.004$ for $f_{\rm FeTi}=10$ and 0.014 for
$f_{\rm FeTi}=25$. Remote sensing of the central floor of the
Taurus-Littrow valley, where the SEP experiment was conducted,
gives 16--20~wt\% FeO and 6--9~wt\% TiO$_2$, with mean
abundances of 17.6 and 8.3~wt\%, respectively
\citep{RobinsonJolliff2002}. Thus,
$f_{\rm FeTi}=22$--29, with a mean of 26. For this mean
composition, the relation gives $\tan\delta=0.016$ at
$\rho=1.5$~g\,cm$^{-3}$ and 0.026 at
$\rho=2.2$~g\,cm$^{-3}$. Over the composition range
$f_{\rm FeTi}=22$--29 and density range
$\rho=1.5$--2.2~g\,cm$^{-3}$, it gives values of
0.011--0.034.

Where site-specific constraints are available, we use them
instead of relying solely on this composition-based estimate.
At Taurus-Littrow, we constrain the loss tangent by fitting
the Apollo~17 SEP measurements (Sec.~\ref{sec:sep}). For
the far-side reference profile, we adopt the loss tangent
inferred from the Chang'E-4 radar observations and test the
Chang'E-6 value as an alternative
(Sec.~\ref{sec:farside}).

We represent each regolith profile by $N$ laterally uniform
layers, indexed by $j=1,\ldots,N$, over a semi-infinite
substrate with complex relative permittivity
$\varepsilon_{\rm sub}$. Layer $j$ has thickness $d_j$ and
complex relative permittivity
\begin{equation}
  \varepsilon_j=\varepsilon'_j(1+i\tan\delta_j).
\end{equation}
Its refractive index is $n_j=\sqrt{\varepsilon_j}$, with
the root chosen so that $\mathrm{Im}\,n_j\geq0$. The
substrate refractive index is defined similarly as
$n_{\rm sub}=\sqrt{\varepsilon_{\rm sub}}$. Continuous
profiles are approximated by successively finer layers until
further refinement no longer changes the calculated results
(Appendix~\ref{app:numerics}).

The model assumes planar, laterally uniform layers and does not resolve individual buried rocks, lateral variations, or surface roughness. We examine the effects of these simplifications in several ways. At Taurus-Littrow, the residuals of the SEP fits constrain the combined effects of unmodeled variations in the site and the idealized transmitter and receiver (Sec.~\ref{sec:limitations}). We also estimate how surface roughness reduces the coherent reflection (Sec.~\ref{sec:limitations}). For the far side, we consider 72 alternative layer profiles and variations in the top unit based on the Chang'E-4 and Chang'E-6 radar values. These tests show how sensitive our results are to the assumed subsurface structure (Sec.~\ref{sec:ensemble}).

\subsection{The Apollo~17 site}
\label{sec:tldescriptions}

The SEP experiment measured radio propagation along the regolith surface in the Taurus-Littrow valley on the near side of the Moon. The valley is floored by basalt at the southeastern edge of Mare Serenitatis. We compare the measurements with five regolith models, ranging from a homogeneous half-space and a two-layer profile to smoothly varying profiles. This comparison tests whether the data require a change in permittivity with depth and, if so, whether they can distinguish a sharp interface from a gradual transition. We fit the parameters of models (a), (b), (d), and (e) to the measurements (Sec.~\ref{sec:sep}). Model (c), the Grimm-like profile, is held fixed. It is a smooth surrogate for the subsurface structure inferred by
\citet{Grimm2018} from an inversion of the SEP measurements, rather than a model family with free parameters to be fitted.
\begin{enumerate}
\item[(a)] A homogeneous half-space, with one permittivity and one loss tangent at all depths. It is the ground of the FARSIDE and impedance studies of Table~\ref{tab:prior} \citep{Burns2019,Rotermund2026}. It tests whether the data require any change with depth.
\item[(b)] A two-layer model: a uniform layer of thickness $d_1$ and permittivity $\varepsilon'_1$ over a substrate of permittivity $\varepsilon'_2$, with one loss tangent. It is the simplest description with a buried interface and the form used in the LuSEE-Night inference of \citet{Yousuf2026}.
\item[(c)] A Grimm-like profile. \citet{Grimm2018} inverted the SEP measurements for the depth profile of the regolith. The inversion gives a porosity that falls steeply over the top 20--30~m. The LuSEE-Night simulations of \citet{Camacho2026} represent this result by eight layers. Their permittivity rises from about 3.7 at the surface to 6.1 at 100~m, and their loss tangent is 0.01. The layer values are not published, so we construct a smooth surrogate,
\[
\varepsilon'(z)=6.1-2.4\,e^{-|z|/(12~{\rm m})},
\]
with $\tan\delta=0.012$, the mean loss tangent of the inversion \citep{Grimm2018}. We chose the 12~m scale so that the surrogate completes 81\% of its rise by 20~m and 92\% by 30~m, the depths over which the porosity falls. We also test fifteen eight-layer step profiles between the same end values, all with $\tan\delta=0.01$. They combine three layer spacings with five shapes: exponential rises on scales of 6, 12 and 25~m, and rises linear in depth or in its logarithm (Sec.~\ref{sec:sepfits}).
\item[(d)] A graded model,
\begin{equation}
  \varepsilon'(z)=\varepsilon'_\infty-(\varepsilon'_\infty-\varepsilon'_0)\,e^{-|z|/z_g},
  \label{eq:gradedfamily}
\end{equation}
with surface permittivity $\varepsilon'_0$, deep permittivity $\varepsilon'_\infty$, grading depth $z_g$ and one loss tangent. At $|z|=z_g$, the permittivity has completed 63\% of its rise. The model has the form of Eq.~(\ref{eq:density}) on a larger scale. It describes a regolith whose porosity falls gradually with depth, as \citet{Grimm2018} found, without a sharp interface.
\item[(e)] A transition model,
\begin{equation}
  \varepsilon'(z)=\varepsilon'_{\rm top}+\frac{\varepsilon'_{\rm deep}-\varepsilon'_{\rm top}}{1+e^{-(|z|-z_t)/\zeta}},
  \label{eq:transitionfamily}
\end{equation}
with permittivity $\varepsilon'_{\rm top}$ above and $\varepsilon'_{\rm deep}$ below a transition, and one loss tangent. At the depth $z_t$, the permittivity is midway between the two. The scale $\zeta$ sets how sharp the transition is: the permittivity covers 10--90\% of the step over $2\zeta\ln9\approx4.4\,\zeta$. The model contains a single interface ($\zeta\to0$) as well as gradual rises (larger $\zeta$), so fitting $\zeta$ measures how sharply the permittivity rises.
\end{enumerate}
The grading depth $z_g$ and the 12~m scale of the surrogate describe a rise over meters to tens of meters, the depths that the SEP measurements sample. The compaction depth $z_\rho$ of Eq.~(\ref{eq:density}) is a few centimeters and describes only the top decimeters.

\subsection{The far side}
\label{sec:farside}

LuSEE-Night and other proposed Dark Ages experiments will observe from the far side, where the subsurface is known only from radar soundings. The Lunar Penetrating Radar on the Yutu-2 rover of Chang'E-4 sounded the upper 40~m of Von K\'arm\'an crater \citep{Li2020}, while the Lunar Regolith Penetrating Radar on the Chang'E-6 lander sounded the upper 3~m at its landing site in the Apollo basin \citep{Zhang2025CE6}. Orbital measurements include normal-incidence echoes recorded by the Apollo~17 Lunar Sounder at 5, 15, and 150~MHz \citep{Phillips1973} and by the Kaguya Lunar Radar Sounder at 5~MHz \citep{Ono2010LRS}, whose polar orbit covered the far side.

The published far-side permittivities and loss tangents are inferred from radar echoes using methods specific to each
instrument. Fitting the echoes directly would require a forward model of each radar, including its antennas, bandwidth, viewing geometry, and data processing. These radars also operate above the frequency range considered here: the Chang'E-4 analysis we adopt uses data at 500~MHz \citep{CE4SPA}, while the Chang'E-6 radar operates at 1--3~GHz \citep{Zhang2025CE6}. How the loss tangent varies between these frequencies and the 5--45~MHz band considered here is not
established. Laboratory measurements of returned samples show that the loss tangent depends on frequency. In Chang'E-5 soils measured at 0.2--12.2~GHz the real permittivity varies little across the range, while the loss tangent decreases as the frequency rises \citep{Liu2026CE5}. Apollo samples measured at 0.3--5~GHz show magnetic as well as dielectric loss \citep{Linton2026}, which we discuss in Sec.~\ref{sec:limitations}. Neither set of measurements extends below 0.2~GHz, so neither fixes the dielectric or magnetic properties at 5--45~MHz.
 We therefore use the published far-side values as site-informed effective inputs, not as measurements at the
frequencies of interest, and vary them in our sensitivity analyses (Sec.~\ref{sec:ensemble}). Holding each loss tangent constant across 5--45~MHz is also a
modeling choice. Our comparison with measured
surface-propagation data in Sec.~\ref{sec:sep} is limited to Taurus-Littrow,
so it constrains the loss there, not on the far side. The orbital radar echoes provide a possible
independent test of far-side profiles: they depend on the normal-incidence surface reflectivity, which we calculate using
the same framework (Sec.~\ref{sec:orbit}).

For our far-side calculations, we adopt the Chang'E-4 landing site as a reference. The Yutu-2 radar identified fine regolith
down to 12~m, coarser material containing embedded rock fragments from 12 to 24~m, and alternating coarse and fine layers from
24 to 40~m \citep{Li2020}. An analysis of the same radar data gives $\varepsilon'=2.64$--3.85 and $\tan\delta=0.0032$--0.0044 for the regolith \citep{CE4SPA}. For the upper layer, extending from the surface to 12~m, we calculate $\varepsilon'$ using
Eqs.~(\ref{eq:density}) and (\ref{eq:epsdensity}). The density profile is based on Diviner measurements that include the far
side. We adopt $\tan\delta=0.004$ from the Chang'E-4 radar analysis \citep{CE4SPA}. The resulting permittivity increases
from 2.05 at the surface to 3.21 at 0.3~m and approaches 3.23 at greater depths.

The radar-derived permittivities for the upper 30~cm are
averages over that depth. Our density-based profile gives an
average of 2.93 over the same interval. Two analyses of the
Yutu-2 data give $2.35\pm0.20$ \citep{Dong2021CE4} and
$2.78\pm0.01$ \citep{Chen2022CE4}. Using
Eq.~(\ref{eq:epsdensity}), these values correspond to mean
densities of 1.31 and 1.57~g\,cm$^{-3}$, respectively, compared
with 1.64~g\,cm$^{-3}$ for our profile. This comparison suggests
that the upper decimeters at the Chang'E-4 site may be less
compact than assumed by Eq.~(\ref{eq:density}), consistent
with the site-to-site variations discussed in
Sec.~\ref{sec:density}. Below 0.3~m, the permittivity of our
profile lies within the range inferred from the radar data.
More recently, \citet{Dossi2026} estimated average permittivities
to a depth of about 13~m along a 1.43~km Yutu-2 traverse,
using the velocities of diffraction hyperbolas.

We keep the density-based permittivity in the reference profile for two reasons. First, the two radar estimates disagree: they differ by more than twice the larger of their errors. Second, the antenna response hardly depends on the difference. With the top 30~cm at 2.35 or 2.78, the impedance changes by at most 1.3\% and the ground fraction by at most 0.012 (Sec.~\ref{sec:ensemble}).

\begin{figure}[t]
  \centering
  \figorbox{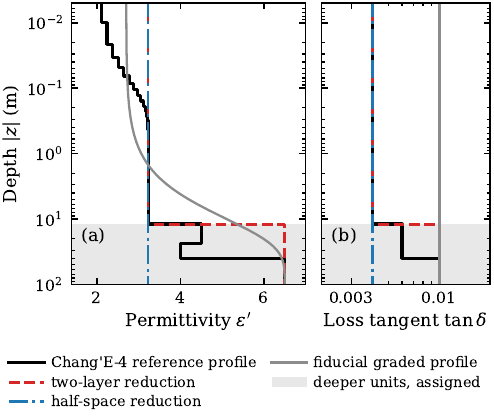}{\columnwidth}
  \caption{Far-side regolith profiles used in this work.
  (a) Real relative permittivity $\varepsilon'$ and (b) loss
  tangent $\tan\delta$ as functions of depth for the
  Chang'E-4 reference profile, its two-layer and half-space
  reductions, and the fiducial graded profile used in the
  Fisher-matrix analyses. In the upper 12~m the permittivity
  of the reference profile follows from the density,
  Eqs.~(\ref{eq:density}) and (\ref{eq:epsdensity}). The
  shading below 12~m marks the deeper units, whose
  permittivities and loss tangents are assigned in this work
  (Sec.~\ref{sec:farside}).}
  \label{fig:profiles}
\end{figure}

The radar located the deeper units but did not measure their permittivity. We assign the values that Eq.~(\ref{eq:epsdensity}) gives for plausible densities. The coarse unit at 12--24~m has $\varepsilon'=4.5$ (2.3~g\,cm$^{-3}$), and the layered unit at 24--40~m has $\varepsilon'=4.0$ (2.1~g\,cm$^{-3}$), both with $\tan\delta=0.006$. The substrate below 40~m has $\varepsilon'=6.5$ (2.9~g\,cm$^{-3}$), close to that of lunar rock \citep{Carrier1991}, and $\tan\delta=0.01$. At Taurus-Littrow the SEP fits require a similar rise, from 2--4.3 near the surface to about 5 at 11.5--18.3~m and 6 at depth (Sec.~\ref{sec:sep}), which makes these values plausible but does not measure them on the far side. We call the result the Chang'E-4 reference profile. It is a radar-informed reference profile, not a measured one: the radar sets its units, and densities set its permittivities. The radar shows the structure changing along the path of the rover \citep{Li2020,Chen2022CE4}, so Sec.~\ref{sec:ensemble} tests 72 variations of it. For comparison, we also use two reductions of the reference profile. The two-layer reduction is a 12~m layer with the depth-averaged values of the top unit, $\varepsilon'=3.22$ and $\tan\delta=0.004$, over the same substrate. The half-space reduction has $\varepsilon'=3.22$ and $\tan\delta=0.004$ at all depths (Fig.~\ref{fig:profiles}).

The Chang'E-6 site gives an independent far-side constraint. Its radar found a fine-grained layer about 1.7~m thick over coarser ejecta that extends to 3~m, the sounded depth \citep{Zhang2025CE6}. The permittivity, $2.22\pm0.72$, comes from the reflection at the air--regolith interface, calibrated against a metal plate. It describes the surface, not every depth to 3~m. The loss tangent, $0.010\pm0.002$, comes from the attenuation of the echoes and the decrease of their central frequency. It is an effective value over the sounded depth, 2.5 times that of the Chang'E-4 reference. The two sites lie in different terrains and were sounded by different radars, so we do not average their values. Instead, Sec.~\ref{sec:ensemble} uses the Chang'E-6 values as an effective homogeneous description of the shallow material, not a reconstruction of its layers, and propagates their uncertainties.

The Fisher-matrix analyses of Secs.~\ref{sec:calibrator} and \ref{sec:bias} differentiate the observables with respect to the regolith parameters. They need a profile with a few smooth parameters. For them, we use the graded model of Eq.~(\ref{eq:gradedfamily}) with $\varepsilon'_0=2.7$, $\varepsilon'_\infty=6.5$, $z_g=10$~m and $\tan\delta=0.01$, which we call the fiducial graded profile (Fig.~\ref{fig:profiles}). Its surface permittivity lies within the near-surface range of Sec.~\ref{sec:density} and within the radar range for the Chang'E-4 regolith, and its deep permittivity is that of the reference substrate. The grading depth is close to the 11--12~m fitted at Taurus-Littrow (Sec.~\ref{sec:sep}). The loss tangent equals the Chang'E-6 value and lies within the SEP values of \citet{Grimm2018}. All other far-side calculations use the Chang'E-4 reference profile or one of its variations.

% =====================================================================
\section{Formalism}
\label{sec:formalism}
% =====================================================================

The calculations in this paper are based on a common result: the
electromagnetic field produced by a current on a wire above the
layered regolith. For the SEP experiment, the current is that of
the transmitting wire, and the rover measures the resulting field.
For a radiometer, we use reciprocity to calculate the response of
the receiving antenna. Its response to an incoming wave from a
given direction and polarization is determined by the far-field
pattern that the same antenna would radiate in that direction
with the corresponding polarization \citep{Balanis2016}. This
allows us to calculate the response to both the 21-cm signal and
the Galactic foreground, which arrive from all directions on the
sky. The orbiting calibrator is also treated using reciprocity.
Its transmitter is 100--1000~km away, so although the emitted
field is spherical, the curvature of its wavefront is negligible
across the antenna and the surrounding ground relevant to the
measurement. We can therefore treat the incident field as a
plane wave. Thermal emission originates within the regolith.
By reciprocity, the contribution from each layer is related to
the power dissipated in that layer when the antenna is driven.

Because Maxwell's equations are linear, the field of a wire
current can be obtained by adding the fields of its individual
current elements. We therefore start with the field of an
elementary horizontal electric dipole above the layered ground,
which defines the Green's function of the problem. Integrating
this Green's function along the wire, weighted by the current
distribution, gives the field of the complete antenna
(Sec.~\ref{sec:sepfields}). The method of moments uses the same
Green's function to solve for the antenna current
(Sec.~\ref{sec:mom}).

The direct field of an elementary dipole has an outgoing
spherical-wave form. To calculate the field above the layered
ground, we express this spherical-wave Green's function as a
superposition of plane-wave components using the
Sommerfeld--Weyl identity \citep{Sommerfeld1909,Weyl1919}.
This is an exact representation, not a plane-wave or
far-field approximation, and includes both propagating and
evanescent components. We then resolve each component into its
$s$- and $p$-polarized parts, calculate its propagation through
the layers, and account for all multiple reflections through
the corresponding reflection coefficients. Finally, we sum
the direct field and the components reflected back into free
space to obtain the complete Green's function. For a planar,
laterally uniform, isotropic stack, this treatment is exact
for any number of layers, following the formulation of
Ref.~\cite{Dasgupta2026}. Throughout this paper, ``exact'' refers to this layered-medium calculation for a planar, laterally uniform, isotropic and nonmagnetic stack, in contrast to the approximate descriptions of the ground considered here. Effects beyond these assumptions are discussed in Sec.~\ref{sec:limitations}. The remaining approximations concern the antenna model. We
treat the antenna as a thin wire, with an idealized feed for
the radiometer and a prescribed current for the SEP transmitter
(Secs.~\ref{sec:mom} and \ref{sec:sepfields}, respectively).
We assess the numerical errors in the field integrals in
Sec.~\ref{sec:verification}.
Appendices~\ref{app:decomp}--\ref{app:mom} give the
derivations of the quantities used in our calculations.

\subsection{Geometry}
\label{sec:geometry}

\begin{figure*}[t]
  \centering
  \figorbox{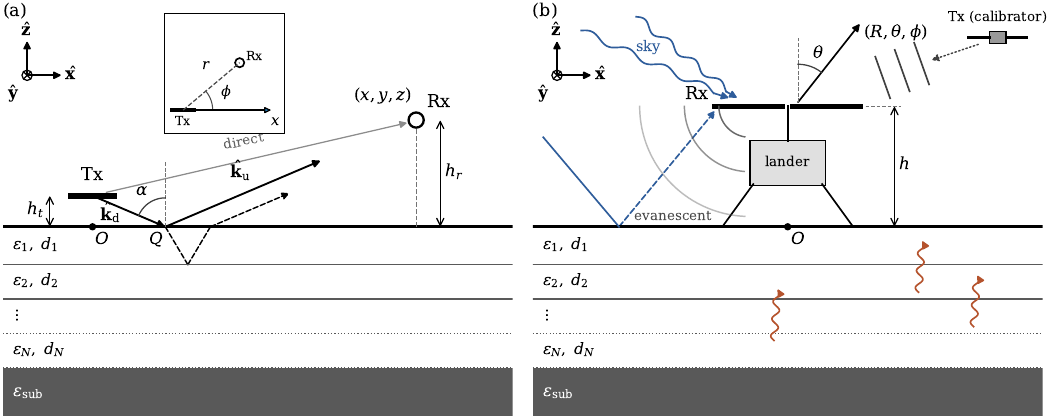}{\textwidth}
\caption{Geometry of the field calculation (not to scale). In both panels, the lunar surface is at $z=0$, above a stack of $N$ planar regolith layers. Layer $j=1,\dots,N$, counted downward from the surface, has thickness $d_j$ and complex relative permittivity $\varepsilon_j$; the stack rests on a semi-infinite substrate with complex relative permittivity $\varepsilon_{\rm sub}$. The origin $O$ lies on the surface below the center of the antenna, the transmitting wire in (a) and the receiving dipole in (b). The unit vectors $\zhat$, $\xhat$, and $\hat{\bm y}$ point upward, along the wire, and into the page ($\otimes$), respectively.
(a) SEP measurement. The transmitting wire (Tx), at height $h_t$, illuminates the surface, and the rover receiver (Rx) measures the field at $(x,y,z)$, at height $h_r$. The inset shows the horizontal range $r$ and azimuth $\phi$ of the receiver. The measured field is the sum of the direct field and the field returned by the regolith. The spherical field of the transmitter is decomposed into plane waves; one component, traveling downward along $\kd$, strikes the surface at $Q$ with incidence angle $\alpha$ and azimuth $\beta$ (not shown). Part is reflected at the surface, while the remainder enters the regolith and undergoes reflection and transmission at the layer interfaces, with part returning to free space (dashed path).
(b) Radiometer on a lander. The receiving dipole (Rx), at height $h$ on a mast, receives sky radiation directly and after reflection from the ground (blue paths), the signal from an orbiting calibrator (Tx), and thermal emission from the regolith (orange arrows). The calibrator, at a distance of 100--1000~km, produces wavefronts that are effectively planar across the antenna and surrounding ground. By reciprocity, the antenna response to radiation from direction $(\theta,\phi)$ is obtained from the field radiated toward the far-field point $(R,\theta,\phi)$, where $\theta$ is measured from the vertical and $\phi$ from the $x$ axis. The thermal-emission contribution of each layer is determined by the power dissipated in it by the driven antenna. The calculation also includes the antenna's evanescent field (gray arcs), which decays with distance.}

  \label{fig:geometry}
\end{figure*}

We place the origin at the surface directly below the center of
the antenna. The $z$ axis points upward, and the $x$ axis is
parallel to the antenna, the SEP transmitter in
Fig.~\ref{fig:geometry}(a) and the radiometer antenna in
Fig.~\ref{fig:geometry}(b). Free space
occupies $z>0$, while the layered regolith occupies $z<0$.
The source is a horizontal electric dipole at height $h$,
oriented along $\xhat$, with current moment $Il$, where $I$
is the current and $l$ is the length of the short wire element.
An observation point above the surface has coordinates
$(x,y,z)$, horizontal distance
$r=(x^2+y^2)^{1/2}$ from the origin, and azimuth $\phi$
measured from the $x$ axis. A far-field direction is specified
by the spherical angles $(\theta,\phi)$, where $\theta$ is
measured from the positive $z$ axis. We also describe each
downward-propagating plane-wave component by its incidence
angle $\alpha$, measured from the downward surface normal,
and its azimuth $\beta$, measured from the $x$ axis.

Over the spatial scales relevant to an antenna on a lander,
the lunar surface can be treated as planar. The effect of
surface curvature over the kilometer-scale SEP traverses is
examined in Sec.~\ref{sec:sep}. Following
Ref.~\cite{Dasgupta2026}, we obtain the dipole field from its
Hertz vector \citep{Chew1995},
\begin{equation}
  \bm\Pi=\Pi_0\,\frac{e^{ikR}}{R}\,\xhat,\qquad
  \Pi_0=\frac{i\,Il\,\eta_0}{4\pi k},
  \label{eq:hertz}
\end{equation}
where $R$ is the distance from the dipole,
$k=2\pi\nu/c$ is the free-space wavenumber at frequency
$\nu$, $c$ is the speed of light, and
$\eta_0=376.73~\Omega$ is the impedance of free space.
The electric and magnetic fields follow from
\begin{equation}
  \bm E=\nabla(\nabla\cdot\bm\Pi)+k^2\bm\Pi,\qquad
  \bm H=-\frac{ik}{\eta_0}\,\nabla\times\bm\Pi .
  \label{eq:EHfromPi}
\end{equation}

\subsection{Plane-wave decomposition}

The spherical wave of Eq.~(\ref{eq:hertz}) is a sum of plane waves by the Weyl identity \citep{Weyl1919,Chew1995},
\begin{multline}
  \frac{e^{ikR}}{R}=\frac{ik}{2\pi}\int_0^{2\pi}\!d\beta\int_C\!d\alpha\,\sin\alpha\\
  \times\exp\!\big[ik\big(x\sin\alpha\cos\beta+y\sin\alpha\sin\beta+|z-h|\cos\alpha\big)\big].
  \label{eq:weyl}
\end{multline}
The contour $C$ runs along real $\alpha$ from $0$ to $\pi/2$, the propagating plane waves, and then from $\pi/2$ to $\pi/2-i\infty$, the evanescent ones, for which $\sin\alpha>1$ and $\cos\alpha$ is imaginary. Each plane wave has the transverse wavenumber $\kappa=k\sin\alpha$, the same in every layer, the vertical wavenumber $\kzo=k\cos\alpha$ above the surface, and
\begin{equation}
  k_{z,j}=\big(k^2\varepsilon_j-\kappa^2\big)^{1/2},\qquad \mathrm{Im}\,k_{z,j}\ge0 ,
  \label{eq:kzj}
\end{equation}
in layer $j$. An evanescent plane wave, $\kappa>k$, decays as $e^{-(\kappa^2-k^2)^{1/2}|z-h|}$ away from the dipole. For a receiver far above the surface these waves never arrive. For an antenna at a height $h\lesssim1/k$ they reach the surface with little decay and carry most of the coupling to the ground.

\subsection{Polarization basis and incident amplitudes}
\label{sec:basis_and_amp}
We resolve each plane-wave component into $s$ and $p$
polarizations, defined with respect to its plane of incidence.
This plane contains the wave vector and the surface normal
(the $z$ axis). The $s$-polarized electric field is
perpendicular to this plane, while the $p$-polarized field
lies within it. For a downward-propagating wave with unit
wave vector
$\kd=(\sin\alpha\cos\beta,\sin\alpha\sin\beta,-\cos\alpha)$,
the corresponding unit polarization vectors are
\begin{equation}
  \begin{aligned}
  \shat&=(-\sin\beta,\ \cos\beta,\ 0),\\
  \pd&=\shat\times\kd
  =-(\cos\alpha\cos\beta,\ \cos\alpha\sin\beta,\ \sin\alpha).
  \end{aligned}
  \label{eq:basis}
\end{equation}
For the upward-propagating wave,
$\ku=(\sin\alpha\cos\beta,\sin\alpha\sin\beta,\cos\alpha)$,
the $s$-polarization vector is unchanged, while the
$p$-polarization vector is
$\pu=\shat\times\ku
=(\cos\alpha\cos\beta,\cos\alpha\sin\beta,-\sin\alpha)$.

The electric field of each plane-wave component is the
projection of $k^2\bm\Pi$ perpendicular to its propagation
direction (Appendix~\ref{app:decomp}). At the surface, its
components along $\shat$ and $\pd$ define the incident
amplitudes:
\begin{equation}
  E^{\rm inc}_{s}=-\Amp\,\sin\beta,\qquad
  E^{\rm inc}_{p}=-\Amp\,\cos\alpha\cos\beta,
  \label{eq:incident}
\end{equation}
where
\begin{equation}
  \Amp\,d\alpha\,d\beta
  =k^2\,\Pi_0\,\frac{ik}{2\pi}\,
  \sin\alpha\,e^{i\kzo h}\,d\alpha\,d\beta
  \label{eq:amplitude}
\end{equation}
is the amplitude associated with the angular element
$d\alpha\,d\beta$ at the surface directly below the dipole.
Each component also has a horizontal phase factor $\ph$,
which is the same in free space and in the layered medium
and is suppressed in the expressions above. For $\beta=0$,
the plane of incidence coincides with the vertical plane
containing the dipole, and the field is purely $p$-polarized.
For $\beta=90^\circ$, the field is purely $s$-polarized.

\subsection{Reflection by the layered regolith}
\label{sec:stack}

Each downward-propagating plane-wave component reaches the
surface at an incidence angle $\alpha$. Part of the field is
reflected into free space, while the rest enters the upper
regolith layer. Within the stack, the transmitted field
undergoes further reflections and transmissions at the layer
interfaces. Because the layers are planar and isotropic, the
$s$- and $p$-polarized components do not mix and can be treated
separately. The transverse wavenumber $\kappa$ is conserved
across all interfaces, while the vertical wavenumber
$k_{z,j}$, given by Eq.~(\ref{eq:kzj}), depends on the layer.
All upward-propagating contributions from the stack have the
same transverse wavenumber and combine into a single reflected
plane wave. Its electric-field amplitudes are
\begin{equation}
  E^{\rm ref}_{s}=\rS E^{\rm inc}_{s},\qquad
  E^{\rm ref}_{p}=\rP E^{\rm inc}_{p},
\end{equation}
along $\shat$ and $\pu$, respectively. The reflection
coefficients $\rS$ and $\rP$ account for the complete layered
structure.

We calculate these coefficients following
Ref.~\cite{Dasgupta2026}. For each polarization, the
dimensionless admittance parameter of layer $j$ is
\begin{equation}
  q_j=
  \begin{cases}
    k_{z,j}/k, & s\text{ polarization},\\
    k_{z,j}/(k\varepsilon_j), & p\text{ polarization},
  \end{cases}
  \label{eq:admittance}
\end{equation}
where $j=0$ denotes free space and $j={\rm sub}$ denotes
the substrate. A wave crossing layer $j$ once accumulates
a phase $\xi_j=k_{z,j}d_j$, where $d_j$ is the layer
thickness. The reflection coefficient at the interface
between media $j-1$ and $j$ is
\begin{equation}
  r_{j-1,j}=\frac{q_{j-1}-q_j}{q_{j-1}+q_j}.
\end{equation}

Let $\mathcal R_j$ denote the effective reflection coefficient
at the lower boundary of layer $j$, including all the layers
and the substrate below it. At the deepest interface,
\begin{equation}
  \mathcal R_N=r_{N,{\rm sub}}.
\end{equation}
Starting there, we calculate the reflection coefficient at
each successive upper interface using
\begin{equation}
  \mathcal R_{j-1}=
  \frac{r_{j-1,j}+\mathcal R_j e^{2i\xi_j}}
       {1+r_{j-1,j}\mathcal R_j e^{2i\xi_j}}.
  \label{eq:parratt}
\end{equation}
The recursion is applied from the deepest layer up to the
surface, where $\mathcal R_0$ gives the reflection coefficient
of the entire stack, $\rst$ \citep{Parratt1954}. Each step
accounts for the reflection at the upper interface and all
subsequent round trips within that layer
(Appendix~\ref{app:stack}). The recursion also remains stable
for evanescent waves, since
$|e^{2i\xi_j}|\leq1$ for the chosen branch of $k_{z,j}$.

Our convention for the $p$-polarized reflection coefficient
uses the ratio of tangential magnetic-field amplitudes. With
the polarization basis defined in Sec.~\ref{sec:basis_and_amp},
a perfect conductor therefore has $\rS=-1$ and $\rP=+1$.
For a planar, laterally uniform, isotropic stack, this
recursion gives the exact reflection coefficients, including
all multiple reflections within the layers.

\subsection{The reflected field}

The field returned to free space is the sum of the plane-wave
components reflected by the layered regolith. Each component
propagates upward along $\ku$, giving
\begin{multline}
  \bm E^{\rm ref}=\int_0^{2\pi}\!\!d\beta\int_C\!d\alpha\,
  \big[\rS E^{\rm inc}_{s}\,\shat
  +\rP E^{\rm inc}_{p}\,\pu\big]\\
  \times\ph\,e^{i\kzo z}.
  \label{eq:Ereflected}
\end{multline}
For an upward-propagating plane wave in free space,
$\bm H=\eta_0^{-1}\,\ku\times\bm E$. Applying this relation
to each reflected component and using the incident amplitudes
from Eq.~(\ref{eq:incident}), we obtain
\begin{multline}
  \bm H^{\rm ref}=\int_0^{2\pi}\!\!d\beta\int_C\!d\alpha\,\frac{\Amp}{\eta_0}\big[\rS\sin\beta\,\pu\\
  -\rP\cos\alpha\cos\beta\,\shat\big]\,\ph\,e^{i\kzo z}.
  \label{eq:Href}
\end{multline}

These expressions show that the effect of the layered regolith on the
fields above the surface is contained in the reflection coefficients
$\rS$ and $\rP$. We use these fields to calculate the observables
considered in this work. The SEP fields are obtained by integrating
over the plane-wave components. The change in antenna impedance is
calculated from the reaction of the reflected field with the antenna
current, while the beam pattern and ground fraction follow from the
far-field radiation pattern. We also calculate the fields inside
each regolith layer to determine the absorbed power and thermal
emission (Appendices~\ref{app:emission}--\ref{app:mom}).
In this way, the same layered-medium model is used throughout.
The SEP measurements provide a direct test of its predicted field
response, although they do not uniquely determine the subsurface
profile.

\subsection{The SEP fields}
\label{sec:sepfields}

The SEP transmitter was a wire laid on the lunar surface, which
we place at height $h_t=5$~cm. The receiver, consisting of three orthogonal
loops mounted on the rover, measured the magnetic field at height
$h_r\simeq 2.5$~m (Sec.~\ref{sec:sepstats}). The measurement
position is specified by its horizontal range $r$ and azimuth
$\phi$ relative to the transmitting dipole.

To calculate the reflected field at the receiver, we write
Eq.~(\ref{eq:Href}) in Cartesian components and change the
integration variable from $\alpha$ to $\kappa$, using
$\sin\alpha\,d\alpha=\kappa\,d\kappa/(k\kzo)$.
The integration over the azimuth $\beta$ can then be performed
analytically using
\begin{equation}
  \int_0^{2\pi}\!e^{im\beta}\,
  e^{i\kappa r\cos(\beta-\phi)}\,d\beta
  =2\pi\,i^{m}J_m(\kappa r)\,e^{im\phi},
  \label{eq:besselid}
\end{equation}
where $J_m$ is the Bessel function of the first kind of order
$m$. This identity reduces the double integrals to single
integrals over $\kappa$ (Appendix~\ref{app:sep}):
\begin{align}
  H^{\rm ref}_x &= \frac{Il}{8\pi}\sin2\phi\int_0^\infty\!
  \kappa\,(\rS+\rP)\,J_2(\kappa r)\,\mathcal E\,d\kappa,\nonumber\\
  H^{\rm ref}_y &= -\frac{Il}{8\pi}\int_0^\infty\!
  \kappa\,\big[(\rS-\rP)J_0(\kappa r)\nonumber\\
        &\qquad+(\rS+\rP)\cos2\phi\,J_2(\kappa r)\big]\,
        \mathcal E\,d\kappa,\nonumber\\
  H^{\rm ref}_z &= \frac{iIl}{4\pi}\sin\phi\int_0^\infty\!
  \frac{\kappa^2}{\kzo}\,\rS\,J_1(\kappa r)\,
  \mathcal E\,d\kappa ,
  \label{eq:sep}
\end{align}
where $\mathcal E=e^{i\kzo(h_t+h_r)}$.
The total field measured by the rover is the sum of this reflected
field and the direct field radiated by the dipole through free space
(Appendix~\ref{app:sep}). For a perfectly conducting surface,
$\rS+\rP=0$ and $\rS-\rP=-2$, and the reflected field reduces to
that of the image dipole. Only $\rS$ contributes to the vertical
magnetic field, since a $p$-polarized wave has no vertical magnetic
field component.

The SEP transmitter used a 70~m wire. Traps and suppressors made
it a half-wave resonant dipole at 2.1, 4, 8.1, 16 and 32.1~MHz,
with the resonance tuned for a surface permittivity
$\varepsilon_{\rm tune}=3.3$. At 1~MHz, the entire wire was loaded
to operate as a short dipole \citep{Simmons1972}. The length of
the resonant element at frequency $\nu$ is approximately
\begin{equation}
  L_{\rm el} =
  \frac{c}{2\nu[(1+\varepsilon_{\rm tune})/2]^{1/2}},
\end{equation}
ranging from 48.7~m at 2.1~MHz to 3.2~m at 32.1~MHz.

We calculate the field of the center-fed wire of length $L$ by
summing the contributions from its individual elements:
\begin{equation}
  \bm H(x,y,z)=\int_{-L/2}^{L/2}
  I(x')\,\bm H_{\rm HED}(x-x',y,z)\,dx',
  \label{eq:wire}
\end{equation}
where $\bm H_{\rm HED}$ is the field of a Hertzian dipole of unit
moment. For the resonant elements, we use the half-wave current
distribution $I(x')=\cos(\pi x'/L_{\rm el})$ over
$|x'|\leq L_{\rm el}/2$. At 1~MHz, we instead use a linearly
tapered current over the full 70~m wire (Appendix~\ref{app:wire}).

\subsection{Impedance, beam and ground fraction}
\label{sec:antennaobs}

We use the calculated antenna current to determine three quantities that govern the response of a surface antenna: its input impedance, its far-field beam pattern, and the fraction of accepted power absorbed by the ground.

The ground changes the input impedance through the field reflected back onto the antenna by the regolith. For a dipole of length $L$, oriented along $\xhat$ at height $h$ and carrying current $I(x)$, the reflected-field contribution to the input impedance is
\[
Z_{\rm ref}=-\frac{1}{I_{\rm feed}^2}
\int E^{\rm ref}_x(x)\,I(x)\,dx,
\]
where $I_{\rm feed}$ is the current at the feed. In the plane-wave decomposition, the current distribution enters through its Fourier transform,
\[
\tilde I(\kappa_x)=\int I(x)\,e^{-i\kappa_x x}\,dx,
\]
where $\kappa_x=\kappa\cos\beta$. The field reflected back onto the wire therefore couples to $\tilde I(-\kappa_x)$. Projecting Eq.~(\ref{eq:Ereflected}) onto $\xhat$ and changing variables to $\kappa$ gives (Appendix~\ref{app:impedance})
\begin{multline}
  Z_{\rm ref} = \frac{k\eta_0}{8\pi^2 I_{\rm feed}^2}\int_0^\infty\!\frac{\kappa\,d\kappa}{\kzo}\int_0^{2\pi}\!d\beta\,\tilde I(\kappa_x)\,\tilde I(-\kappa_x)\\
  \times\Big[\rS\sin^2\beta-\rP\,\frac{\kzo^2}{k^2}\cos^2\beta\Big]\,e^{2i\kzo h}.
  \label{eq:Zr}
\end{multline}
The two current factors appear as a product, not as an absolute square, because this is a field-current reaction rather than a power integral. They reduce to $|\tilde I(\kappa_x)|^2$ only for a real current distribution, as in the sinusoidal-current approximation, and not for the complex current induced by the ground. The integral also includes evanescent components with $\kappa>k$. For these components, the height-dependent factor becomes
$e^{-2h\sqrt{\kappa^2-k^2}}$, describing their decay away from the antenna. They contribute to near-field coupling between the antenna and the ground, and are retained in the impedance calculation. The sinusoidal-current approximation does not account for the change in current caused by the ground. For the 6~m dipole, it also gives an inaccurate free-space impedance above 25~MHz. We therefore solve for the current distribution (Sec.~\ref{sec:mom}) and use the sinusoidal current only for comparison.

We next calculate the far-field beam pattern. At large distances, the angular-spectrum integral is dominated by the plane wave traveling toward the observer. For an observation direction $(\theta,\phi)$, this corresponds to $\alpha=\theta$ and $\beta=\phi$. Evaluating the integral by stationary phase (Appendix~\ref{app:spa}) gives the far-field components $E_\theta$ and $E_\phi$ along the spherical unit vectors $\hat{\bm\theta}$ and $\hat{\bm\phi}$, each the sum of the direct field and the field reflected by the ground:
\begin{equation}
  \begin{aligned}
  E_\theta&\propto\cos\theta\cos\phi\,\tilde I\,(1-\rP e^{2ikh\cos\theta}),\\
  E_\phi&\propto-\sin\phi\,\tilde I\,(1+\rS e^{2ikh\cos\theta}),
  \end{aligned}
  \label{eq:pattern}
\end{equation}
where $\tilde I=\tilde I(k\sin\theta\cos\phi)$ and the reflection coefficients are evaluated at $\alpha=\theta$. The phase factor $e^{2ikh\cos\theta}$ accounts for the extra path traveled by the reflected wave. Only propagating plane waves contribute to the far field. The power radiated into the upper hemisphere is obtained by integrating the power per unit solid angle, $dP/d\Omega\propto|E_\theta|^2+|E_\phi|^2$.

Let $P_{\rm sky}$ denote this total radiated power. The fraction of power absorbed by the ground is then
\begin{equation}
  f_g=1-\frac{P_{\rm sky}}{\tfrac12|I_{\rm feed}|^2\,\mathrm{Re}\,Z_{\rm in}},
  \label{eq:fg}
\end{equation}
where $Z_{\rm in}$ is the input impedance. The denominator is the total power accepted at the antenna feed. Thus, $f_g$ is the fraction of accepted power that is absorbed below the surface.

The same beam pattern determines how the antenna receives radiation from the sky. By reciprocity, its sensitivity to radiation from direction $(\theta,\phi)$ is proportional to $dP/d\Omega$. The ground also emits thermal radiation. By Kirchhoff's law, its contribution is weighted by the power the antenna deposits in the regolith. We can therefore write the antenna temperature as
\begin{equation}
  T_A=(1-f_g)\,\frac{\int (dP/d\Omega)\,T_{\rm sky}\,d\Omega}{\int (dP/d\Omega)\,d\Omega}+f_g\,T_{\rm eff},
  \label{eq:TA}
\end{equation}
where both integrals are over the upper hemisphere. Here, $T_{\rm sky}(\theta,\phi)$ is the sky brightness temperature, and $T_{\rm eff}$ is the effective regolith temperature, weighted by the power absorbed at different depths (Sec.~\ref{sec:absorption}).

The global 21-cm signal is isotropic, so its beam-weighted average is independent of the beam shape. Its contribution to $T_A$ is therefore reduced by the factor $1-f_g$. The Galactic foreground, in contrast, varies across the sky. Its contribution depends on the shape of the beam, which changes with frequency as a result of reflection from the regolith.

\subsection{Calibrator response}
\label{sec:calresponse}

The orbiting calibrator is in the far field of the antenna. By
reciprocity, the voltage induced by the calibrator is proportional
to the projection of the antenna's far field onto the calibrator's
polarization. For a polarization unit vector $\hat{\bm e}$ and
calibrator direction $(\theta_c,\phi_c)$, the received power is
therefore
\begin{equation}
  P_{\rm cal}(\theta_c,\phi_c,\nu)\propto
  \big|\hat{\bm e}^{*}\cdot
  \big[E_\theta\hat{\bm\theta}+E_\phi\hat{\bm\phi}\big]\big|^2 .
  \label{eq:calib}
\end{equation}
Here, $E_\theta$ and $E_\phi$ include both the direct and
ground-reflected fields, with $\rS$ and $\rP$ evaluated at the
incidence angle $\alpha=\theta_c$. The received power thus
depends on the regolith reflection coefficients. As the calibrator
moves across the sky, its response samples this dependence at
different incidence angles and azimuths above the landing site.

\subsection{Antenna current}
\label{sec:mom}

The antenna current changes in response to the field reflected
by the regolith. We therefore solve for the current with the
ground included, rather than assume a fixed current
distribution. For a perfectly conducting wire, the tangential
component of the total electric field must vanish along the
wire, except across the feed gap where the voltage is applied.
The field produced by the wire current is obtained by integrating
the contributions from all points along the wire. The Green's
function gives the field produced by an elementary dipole above
the layered regolith. Enforcing the conducting-wire boundary
condition therefore gives an integral equation for the unknown
current, known as the electric-field integral equation.

We divide a wire of length $L$ into $N_{\rm seg}$ equal
segments of length $\Delta x=L/N_{\rm seg}$. The current
vanishes at the two ends of the open wire, leaving
$N_{\rm seg}-1$ interior division points where the current
is unknown. At each interior point $x_n$, we define a
triangular basis function $\Lambda_n(x)$ that is unity at
$x_n$ and decreases linearly to zero at the two neighboring
division points:
\begin{equation}
  \Lambda_n(x)=
  \begin{cases}
    1-|x-x_n|/\Delta x, & |x-x_n|\leq\Delta x,\\
    0, & |x-x_n|>\Delta x.
  \end{cases}
\end{equation}
We approximate the current by a sum of these overlapping
triangular functions,
\begin{equation}
  I(x)=\sum_{n=1}^{N_{\rm seg}-1}I_n\Lambda_n(x),
\end{equation}
where $I_n$ is the current at the $n$th interior division
point.

To solve the integral equation, we use the method of moments
\citep{Harrington1968}. We substitute the current expansion
into the electric-field integral equation and test the resulting
equation with a set of functions: we multiply it by each
testing function and integrate along the wire. In the Galerkin
form used here, the testing functions are the same triangular
functions used to represent the current. This gives one linear
equation for each unknown current coefficient:
\begin{equation}
  \sum_{n=1}^{N_{\rm seg}-1}
  \big(Z^{\rm fs}_{mn}+Z^{\rm ref}_{mn}\big)\,I_n=v_m,
  \label{eq:mom}
\end{equation}
where $m=1,\ldots,N_{\rm seg}-1$ labels the testing functions.
The applied voltage is $v_m=1$~V for the function centered at
the feed and zero for all the others. The matrix has two
contributions. $Z^{\rm fs}_{mn}$ describes the interaction
between basis functions $m$ and $n$ in free space, while
$Z^{\rm ref}_{mn}$ accounts for the field that travels from
one basis function to the ground and is reflected back onto
the other (Appendix~\ref{app:mom}).

The reflected contribution is
\begin{equation}
  Z^{\rm ref}_{mn}=
  \int\!\!\int\Lambda_m(x)\,\Lambda_n(x')\,
  K_{\rm ref}(|x-x'|)\,dx\,dx',
\end{equation}
where $-K_{\rm ref}(u)$ is the $x$ component of the reflected
field on the wire at a distance $u$ from an $x$-directed
Hertzian dipole of unit moment. Applying the plane-wave
decomposition used to obtain Eq.~(\ref{eq:Zr}) and performing
the azimuthal integral with Eq.~(\ref{eq:besselid}) gives
\begin{multline}
  K_{\rm ref}(u)=\frac{k\eta_0}{8\pi}\int_0^\infty\!
  \frac{\kappa\,d\kappa}{\kzo}\,e^{2i\kzo h}\Big\{
  \rS\big[J_0(\kappa u)\\
  +J_2(\kappa u)\big]-\rP\,\frac{\kzo^2}{k^2}
  \big[J_0(\kappa u)-J_2(\kappa u)\big]\Big\}.
  \label{eq:Kr}
\end{multline}
Thus, the ground enters the current calculation through the
same reflection coefficients and integration contour used
for the other field calculations.

Solving Eq.~(\ref{eq:mom}) gives the current distribution and
the feed current $I_{\rm feed}$. For the 1~V feed excitation,
the input impedance is
$Z_{\rm in}=(1~{\rm V})/I_{\rm feed}$.
We also need the Fourier transform of the solved current for
the far-field and power calculations. For the triangular
basis expansion, it is
\begin{equation}
  \tilde I(\kappa_x)=
  \Delta x\left[\frac{\sin(\kappa_x\Delta x/2)}{\kappa_x\Delta x/2}\right]^2
  \sum_{n=1}^{N_{\rm seg}-1}I_n\,e^{-i\kappa_x x_n},
  \label{eq:momft}
\end{equation}
where $x_n$ is the center
of the $n$th triangular basis function. The factor
before the sum is the Fourier transform of the triangular
basis function. We use this current transform to calculate
the beam pattern, ground fraction, power absorbed at different
depths, and calibrator response.

We solve for the current separately for each ground model,
including the far-side profiles of Fig.~\ref{fig:profiles} and
the impedance-boundary model of Sec.~\ref{sec:ibcformal}, so
each ground model is compared using the current distribution
obtained for that particular model.

For the 6~m dipole considered in Sec.~\ref{sec:radiometer},
we use $N_{\rm seg}=80$ ($\Delta x=7.5$~cm) and a wire
radius $a=1$~cm. The method also applies to structures made
of several wires with different orientations, such as the
four monopoles and the illustrative wire-grid lander of
Sec.~\ref{sec:lander}. In that case,
the reflected coupling between each pair of basis functions
is calculated by summing the plane-wave contributions,
using the Fourier transforms of the corresponding currents,
projected onto the $s$ and $p$ polarizations, and the stack
reflection coefficients of Sec.~\ref{sec:stack}
(Appendix~\ref{app:mom}).

\subsection{Ground absorption and thermal emission}
\label{sec:absorption}

We calculate how the antenna's power is absorbed at different
depths in the regolith. For each plane-wave component, we
propagate the tangential electric and magnetic fields through
the layered medium and calculate the normal Poynting flux at
each interface. The difference between the downward power
fluxes at the top and bottom of a layer gives the power
absorbed within it. Dividing this by the power entering the
ground gives the absorption fraction $f^{s,p}_j(\alpha)$
for layer $j$ and polarization $s$ or $p$
(Appendix~\ref{app:emission}).

We combine these fractions over the plane-wave spectrum of
the antenna to obtain $f_j$, the fraction of its total
ground-absorbed power deposited in layer $j$. By Kirchhoff's
law, the thermal emission from the ground contributes
$f_g T_{\rm eff}$ to the antenna temperature, where the
effective ground temperature is
\begin{equation}
  T_{\rm eff}(\nu)=
  \sum_{j=1}^{N} f_j(\nu)\,T(z_j)
  +f_{\rm sub}(\nu)\,T_{\rm sub},
  \label{eq:Teff}
\end{equation}
Here, $T(z_j)$ is the physical temperature at the center
$z_j$ of layer $j$, $T_{\rm sub}$ is the substrate
temperature, and $f_{\rm sub}$ is the fraction of the
ground power absorbed in the substrate. The fractions
satisfy $\sum_j f_j+f_{\rm sub}=1$.

The same absorption calculation gives the thermal brightness
of the regolith as seen from orbit. For a plane wave emitted
at an angle $\theta$ from the surface normal, the brightness
temperature in polarization $s$ or $p$ is
\begin{multline}
  T_B^{s,p}(\theta)=
  \big[1-|r_{\rm stack}^{s,p}(\theta)|^2\big]
  \Bigg[
  \sum_{j=1}^{N}f_j^{s,p}(\theta)\,T(z_j)\\
  +f_{\rm sub}^{s,p}(\theta)\,T_{\rm sub}
  \Bigg]
  +|r_{\rm stack}^{s,p}(\theta)|^2\,T_{\rm sky},
  \label{eq:TB}
\end{multline}
where $r_{\rm stack}^{s,p}$ is the reflection coefficient
of the full layered stack, and $T_{\rm sky}$ is the sky
brightness temperature incident from the specular direction.
The factor $1-|r_{\rm stack}^{s,p}|^2$ is the fraction
of incident power absorbed by the ground, while the
$ f_j^{s,p}$ and $f_{\rm sub}^{s,p}$ give the fractions
of that absorbed power deposited in the individual layers
and substrate, respectively.

A distant observer measures the brightness averaged over the
visible lunar disc, with the projected-area weight
$2\sin\theta\cos\theta\,d\theta$. For an orbiter at altitude
$h_{\rm orb}$, a line of sight at nadir angle $\psi$
corresponds to an emission angle $\theta$ given by
\[
  \sin\theta=
  \left(1+\frac{h_{\rm orb}}{R_{\rm M}}\right)\sin\psi,
\]
where $R_{\rm M}=1737.4$~km is the lunar radius. The
observable extends to the limb, where $\theta=\pi/2$.

\subsection{The impedance-boundary approximation}
\label{sec:ibcformal}

An impedance boundary replaces the layered ground with a single
effective surface impedance. The standard Leontovich impedance
boundary condition (IBC) \citep{SeniorVolakis1995} obtains this
impedance from the reflection coefficient at normal incidence
and applies it at all angles of incidence. In units of the
free-space impedance $\eta_0$, the surface impedance is
\begin{equation*}
  \eta_{\rm IBC}=
  \frac{1+\rS(0)}{1-\rS(0)},
\end{equation*}
where $\rS(0)$ is the $s$-polarized reflection coefficient
at normal incidence. The corresponding reflection coefficients
are
\begin{equation}
  r^{s}_{\rm IBC}=
  \frac{\eta_{\rm IBC}\cos\alpha-1}
       {\eta_{\rm IBC}\cos\alpha+1},
  \qquad
  r^{p}_{\rm IBC}=
  \frac{\cos\alpha-\eta_{\rm IBC}}
       {\cos\alpha+\eta_{\rm IBC}},
  \label{eq:ibc}
\end{equation}
where $\cos\alpha=\kzo/k$, continued onto the evanescent
part of the integration contour $C$.

The approximation is appropriate when the wave transmitted
into the ground propagates nearly normal to the surface,
even for oblique incidence from free space. This requires
$|n_j|\gg1$ in the ground, as is approximately the case
for good conductors such as seawater. Lunar regolith has a
much smaller refractive index, $n\approx1.5$--$1.8$ in
the upper layers. The transmitted angle therefore varies
with the angle of incidence, and an impedance determined
at normal incidence cannot fully describe the reflection
at oblique angles. The approximation also does not reproduce
the full response of the layered medium to evanescent waves.

Related layered-impedance boundary conditions are used in
commercial finite-element solvers. HFSS describes its
layered-impedance model as representing multiple layers
by a single equivalent impedance surface
\citep{HFSSintro2025}. The COMSOL documentation likewise
states that its layered impedance condition assumes that
the wave propagates essentially normal to the surface
within the layers \citep{COMSOL2025}.

Our calculation does not reduce the layered ground to a surface
impedance. We retain the individual regolith layers and
calculate the $s$- and $p$-polarized reflection coefficients
of the full stack for each plane-wave component. Both
propagating and evanescent waves are included, the latter
being important for the near-field coupling between the
antenna and the ground. These reflection coefficients enter
the calculation of the fields, the antenna current and
input impedance, the far-field beam, and the power absorbed
and emitted by the regolith. We use the impedance-boundary
approximation as a comparison to assess the effects of
replacing the layered ground with a single surface
impedance. We call it the impedance boundary in the text and
the figures, and IBC in the tables.

% =====================================================================
\section{Verification}
\label{sec:verification}
% =====================================================================

Before comparing the calculation with measurements, we test each stage of it against exact limits,
conservation of power and independent codes.
Table~\ref{tab:verification} lists the checks and the largest
difference found in each. Differences below $10^{-12}$ are
rounding errors. The others come from numerical integration
(Appendix~\ref{app:numerics}) or discretization, or from
differences between independent codes.

The layered-medium response is exact to rounding error. The
Parratt recursion, Eq.~(\ref{eq:parratt}), agrees with a direct
solution of the boundary conditions at every interface, each
plane wave conserves power, and a half-space gives the Fresnel
coefficients. The fields agree with the independent
layered-medium code empymod \citep{Werthmuller2017} to
$1.4\times10^{-8}$ for the SEP fields and to $2.8\times10^{-7}$
for the field reflected along the antenna.

For the antenna, we compute the power accepted at the feed from
the input impedance, the power radiated into the sky from the
far field, and the power entering the ground from the
plane-wave spectrum. The first equals the sum of the other two
to $2\times10^{-5}$, which tests the solved current, the
reflected field and the far field together. Over a perfect
conductor the calculation reproduces the field of the image
dipole to rounding error, and a second implementation of the
method of moments (Appendix~\ref{app:mom}) agrees with the first
to $5.5\times10^{-6}$.
\begin{table}[t]
  \caption{Numerical checks: relative differences, except
  absolute ones for the ground fraction $f_g$ and the
  normalized beam; for 40 vs 160 layers, the larger of the
  relative difference in $Z_{\rm in}$ and the absolute one
  in $f_g$. At 35--45~MHz, near the antiresonance,
  impedances are compared through the input conductance
  $\mathrm{Re}(1/Z_{\rm in})$, which does not depend on
  the model of the feed gap (Sec.~\ref{sec:verification}).}
  \label{tab:verification}
  \begin{ruledtabular}
    \begin{tabular}{lc}
      Check & Difference \\
      \hline
      \multicolumn{2}{l}{Layered medium} \\
      Recursion vs direct solution & $8\times10^{-15}$ \\
      Power in vs absorbed plus transmitted & $2\times10^{-14}$ \\
      Half-space vs Fresnel & $3\times10^{-14}$ \\
      40 vs 160 layers, $Z_{\rm in}$ and $f_g$ & $1.4\times10^{-5}$ \\

      \multicolumn{2}{l}{Antenna} \\
      $Z_{\rm ref}$ over a conductor vs image dipole & $3\times10^{-13}$ \\
      $K_{\rm ref}$ over a conductor vs image dipole & $2\times10^{-15}$ \\
      $K_{\rm ref}$, layered, vs path below axis & $1.4\times10^{-8}$ \\
      $K_{\rm ref}$, layered, vs empymod & $2.8\times10^{-7}$ \\
      Input vs sky plus ground power, sinusoid & $6\times10^{-6}$ \\
      Input vs sky plus ground power, solved current & $2\times10^{-5}$ \\
      Eq.~(\ref{eq:Zr}), solved current, vs Galerkin reaction & $4\times10^{-13}$ \\
      Quasi-static image over a dielectric & $3\times10^{-6}$ \\
      Second implementation vs first & $5.5\times10^{-6}$ \\
      Lander, symmetry subspace vs full & $1.6\times10^{-14}$ \\
      Hall\'en solver, free space, $Z_{\rm in}$, 5--30~MHz & 1--4\% \\
      Hall\'en solver, $\mathrm{Re}(1/Z_{\rm in})$, 35--45~MHz & 0.7--1.4\% \\
      40, 160 vs 80 segments, $Z_{\rm in}$, 5--30~MHz & 0.7--1.9\% \\
      40, 160 vs 80 segments, $\mathrm{Re}(1/Z_{\rm in})$, 35--45~MHz & 0.1--0.4\% \\
      40, 160 vs 80 segments, $f_g$ and beam & $8\times10^{-4}$ \\

      \multicolumn{2}{l}{NEC-2, homogeneous half-space} \\
      $f_g$, dipole, 5--45~MHz & $9\times10^{-4}$ \\
      Normalized beam, dipole & $2\times10^{-3}$ \\
      $Z_{\rm in}$, dipole, 5--30~MHz & 0.7--2.4\% \\
      $\mathrm{Re}(1/Z_{\rm in})$, dipole, 35--45~MHz & 0.3--0.7\% \\
      Change of $1/Z_{\rm in}$ by the ground, dipole & 3.8\% \\
      $f_g$, dipole above the lander body & $3\times10^{-3}$ \\
      $Z_{\rm in}$, dipole above the lander body & 1.3--2.7\% \\

      \multicolumn{2}{l}{Sky, SEP and thermal emission} \\
      Sky power, map pixels vs quadrature & $10^{-5}$ \\
      SEP fields vs image dipole & $1.2\times10^{-7}$ \\
      SEP fields vs empymod, three profiles & $1.4\times10^{-8}$ \\
      SEP integrals, layered, vs path below axis & $2.3\times10^{-9}$ \\
      Short wire vs point dipole & $3\times10^{-5}$ \\
      Interpolated vs direct integrals & $2.2\times10^{-3}$ \\
      Isothermal $T_B$, Kirchhoff form & $3\times10^{-16}$ \\
    \end{tabular}
  \end{ruledtabular}
\end{table}

The input impedance itself is less certain, because thin-wire
solvers represent the feed gap differently. The gap acts as a
capacitance of a few tenths of a picofarad in parallel with the
antenna terminals (Appendix~\ref{app:impedance}). It matters
most near the antiresonance at about 43~MHz, where the 6~m
dipole is a little shorter than a wavelength, its input
resistance peaks and its reactance changes sign
\citep{Balanis2016}. At 5--30~MHz a solver of Hall\'en's
integral equation \citep{Hallen1938} gives free-space impedances
within 1--4\% of ours. The Numerical Electromagnetics Code NEC-2 \citep{NEC2,necpp}, which treats a
homogeneous lossy half-space but not layered ground, gives
impedances within 0.7--2.4\% of ours for the 6~m dipole 2.5~m
above the half-space reduction of the far-side model
($\varepsilon'=3.22$, $\tan\delta=0.004$). At 35--45~MHz the
input conductance, $\mathrm{Re}(1/Z_{\rm in})$, which the
capacitance does not change, agrees with NEC-2 to 0.7\%.

The quantities that carry the effect of the regolith agree more
closely. With NEC-2 the ground fraction agrees to
$9\times10^{-4}$ and the normalized beam to $2\times10^{-3}$,
and above the lander model (Sec.~\ref{sec:lander}) the ground
fraction agrees to 0.003. The change of the admittance
$1/Z_{\rm in}$ by the ground agrees to 3.8\%. The ground changes
the admittance of this dipole by at most 9\%, so this is less
than 0.4\% of the admittance. With 40 or 160 segments instead of
80 the impedance changes by 0.7--1.9\%, but the ground fraction
and the beam by less than $8\times10^{-4}$.

The layered-medium calculation is therefore limited only by
numerical precision, and an independent solver confirms the
ground fraction and the beam to about $10^{-3}$. The absolute
impedance is uncertain by a few percent through the model of the
feed. This uncertainty does not depend on the ground and largely
cancels between descriptions of the regolith computed with the
same solver and feed (Secs.~\ref{sec:radiometer}--\ref{sec:bias}).
Near the antiresonance the impedance depends on the actual feed,
so the values quoted there are not predictions for the flight
antenna (Sec.~\ref{sec:limitations}).

% =====================================================================
\section{Test against Apollo~17}
\label{sec:sep}
% =====================================================================

The SEP experiment placed two orthogonal 70~m wire antennas on the floor of the Taurus-Littrow valley. They transmitted at 1.0, 2.1, 4.0, 8.1, 16.0 and 32.1~MHz, while loop antennas on the lunar rover recorded the fields as the astronauts drove away from the transmitter, over distances of several kilometers \citep{Simmons1972,Simmons1973,SEPFinal1975}. These measurements probe radio propagation along the lunar surface, directly in the frequency band of Dark Ages experiments, and provide a test of the calculation in Sec.~\ref{sec:formalism}. Equations~(\ref{eq:sep}) and (\ref{eq:wire}), together with the direct field, give the magnetic field at the rover above a planar stack of layers. The calculation retains the full response of the layered ground, including both propagating and evanescent plane waves and reflections from all interfaces. We calculate the field along the traverse for each regolith description and compare it with the measurements.

The measured field forms an interference pattern \citep{Kong1972,Tsang1973,Grimm2018}. One part of the transmitted signal travels through the space above the surface, while another travels through the regolith near the surface, where its speed is reduced by the refractive index $n$. Over uniform ground, with $n=\sqrt{\varepsilon'}$, these waves interfere and produce a pattern of alternating maxima and minima with spatial period $c/[\nu(n-1)]$, where $c$ is the speed of light and $\nu$ is the frequency. The period therefore depends on the permittivity. At 4~MHz it is 155~m for $\varepsilon'=2.2$ and 70~m for $\varepsilon'=4.3$, the range of surface permittivities found in Sec.~\ref{sec:sepfits}. Reflections from buried layers add further structure to the pattern. If the permittivity increases with depth, the wave in the regolith bends downward and the interference pattern weakens. The positions and spacing of the minima therefore probe how the permittivity changes over the upper tens of meters, at frequencies relevant to Dark Ages radiometers.

We first decode the archived measurements (Sec.~\ref{sec:sepdata}) and define the score used to compare them with the models (Sec.~\ref{sec:sepstats}). We then fit the near-side regolith descriptions of Sec.~\ref{sec:tldescriptions} at 2.1 and 4.0~MHz (Sec.~\ref{sec:sepfits}) and use the fitted parameters to predict the measurements at higher frequencies (Sec.~\ref{sec:sepholdout}). Section~\ref{sec:sharpness} examines how sharply the permittivity rises with depth. Finally, we test two elements of the far-side models used for LuSEE-Night, the two-layer parameter range of its subsurface study \citep{Yousuf2026} and the impedance-boundary approximation (Sec.~\ref{sec:sepluseeibc}). Here, a description means either a model from Sec.~\ref{sec:tldescriptions} with fitted parameters or the fixed Grimm-like profile. Its profile is the corresponding depth dependence of the permittivity.

\subsection{The archived data}
\label{sec:sepdata}

The calibrated archive \citep{SEPdata2021} contains a westward traverse extending 4.3~km from the transmitter. We decode it using the bundle description, which follows the format of the instrument team's final report \citep{SEPFinal1975}. Appendix~\ref{app:sepdata} describes the record layout and decoding procedure, which we checked against both the final report and the data themselves. At each of the six frequencies, a record contains the measurement ranges and six received levels in dBm: the X, Y and Z components of the receiver for each of the two transmitting wires. A mode word identifies the receiver mode for each measurement cycle and the frequency pair used for its calibration and transmitter-off records.

The traverse ran about $11^\circ$ from the axis of one wire and nearly perpendicular to the other (Appendix~\ref{app:sepdata}). Following the archive and the final report \citep{SEPFinal1975}, we call the first the endfire transmitter, since the rover was beyond the end of the dipole, and the second the broadside transmitter, since the rover was to its side.

The receiver operated normally out to 1.7~km. Beyond this range, it entered synchronization-acquisition mode, in which only partial records were collected. A false resynchronization at 4.3~km ended the field-strength record \citep{SEPFinal1975}. In these partial records, the arrays from the three loops contain a single signal, but the archive does not identify which loop recorded it. We therefore use only records from normal operation and restrict every fit to ranges below 1.5~km. For each receiver component and frequency, we estimate the noise floor from the median level recorded with the transmitter off. We exclude samples above $-33$~dBm, near the upper end of the dynamic range at about $-30.5$~dBm, where the receiver saturates. Changing the noise-floor estimate changes the comparison at 2.1--4.0~MHz by no more than 1.4 in the score defined below and does not change any conclusion (Appendix~\ref{app:robust}).

\begin{figure*}[t]
  \centering
  \figorbox{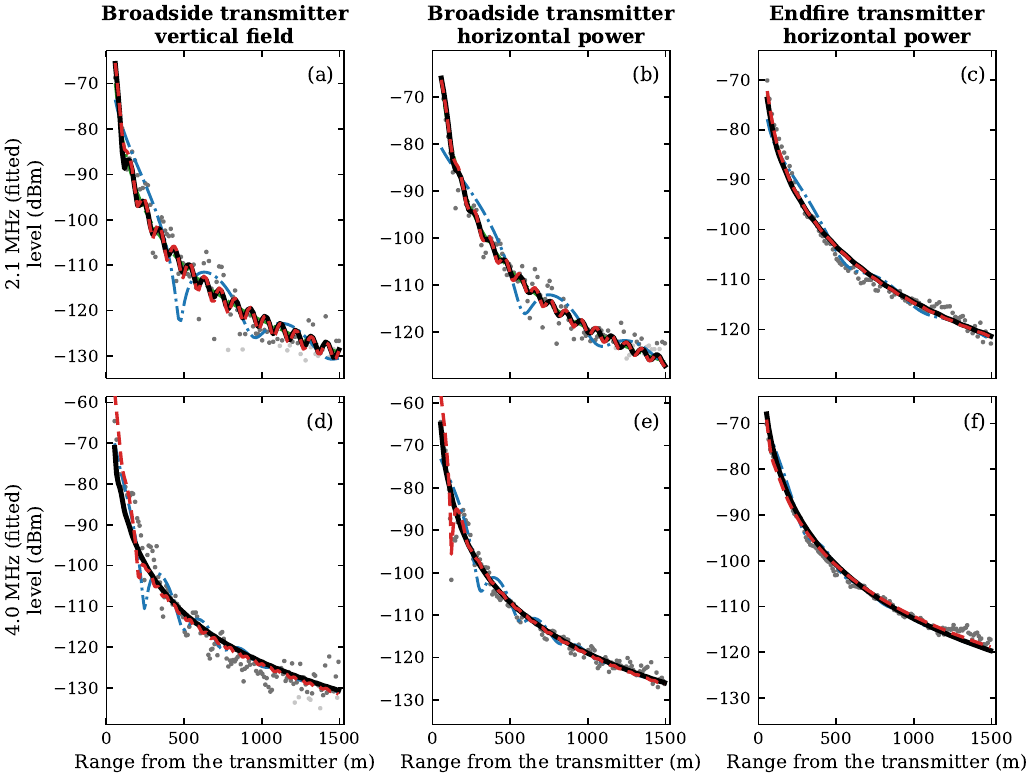}{\textwidth}
  \caption{Fits to the Apollo~17 SEP measurements at 2.1~MHz, (a)--(c), and 4.0~MHz, (d)--(f). The three columns show the measured quantities defined in Sec.~\ref{sec:sepdata}. These are the broadside vertical field in (a) and (d), the broadside horizontal power in (b) and (e), and the endfire horizontal power in (c) and (f). Gray points show the measurements in 10~m bins, with lighter gray indicating points within 3~dB of the noise floor. The curves show the graded (black, solid), two-layer (red, dashed) and half-space (blue, dash-dotted) descriptions fitted at each frequency, together with the fixed Grimm-like profile evaluated using the impedance-boundary approximation (green, dash-dot-dotted). The latter lies mostly below the other curves. Each curve includes its fitted channel offset (Sec.~\ref{sec:sepstats}).}
  \label{fig:sep}
\end{figure*}

The Z loop measures the vertical magnetic field, while the X and Y loops measure the horizontal field along axes fixed to the rover. We use the vertical field and the total horizontal power, defined as the sum of the X and Y powers. This sum is independent of the rover's heading. A horizontal dipole produces no vertical or radial magnetic field, $H_z$ or $H_r$, along its axis, and no azimuthal field, $H_\phi$, in the broadside direction. The endfire vertical field is therefore weak and is not used. We retain three quantities at each frequency: the broadside vertical field, the broadside horizontal power and the endfire horizontal power. These are shown, from left to right, in the three columns of Figs.~\ref{fig:sep} and \ref{fig:holdout}. The quantities probe the ground differently: the vertical field depends only on the reflection of $s$-polarized waves (Sec.~\ref{sec:sepfields}), while the horizontal power depends on both polarizations. The broadside vertical field shows clear interference minima, including one near 500~m at 4~MHz [Fig.~\ref{fig:sep}(d)], whereas the endfire horizontal power, in the right columns, is smooth at every frequency shown, as anticipated by \citet{Simmons1972}. Figure~\ref{fig:sep} shows the two frequencies at which we fit the descriptions, 2.1 and 4.0~MHz (Sec.~\ref{sec:sepstats}). At 1.0~MHz the whole wire was loaded as a short dipole (Sec.~\ref{sec:sepfields}) whose loading is not documented in detail, so we hold 1.0~MHz out of the fit, as we do 8.1--32.1~MHz, and show it with 8.1 and 16.0~MHz in Fig.~\ref{fig:holdout}.

Beyond its far-field distance, which is at most 33~m at 2.1~MHz, the transmitter can be treated as a point dipole. Replacing the actual transmitter by a point dipole changes the field by 0.22~dB rms at 2.1~MHz and by at most 0.02~dB at 8.1--32.1~MHz. This difference is confined to ranges within the far-field distance (Appendix~\ref{app:wire}). We prescribe the transmitter current rather than solve for it, because the traps and suppressors that determine its distribution are not documented in enough detail to model. The free offsets in the fits (Sec.~\ref{sec:sepstats}) absorb the overall current amplitude, but not its distribution along the wire. Appendix~\ref{app:robust} therefore repeats the comparison for nine current distributions, including one solved by the method of moments over each profile.

Our calculation treats the lunar surface as planar. At a horizontal range $r$ from the transmitter, the curved surface lies $r^2/(2R_{\rm M})$ below the plane tangent at the transmitter, 0.83~m at $r=1.7$~km, where $R_{\rm M}=1737.4$~km is the lunar radius. The waves that reach the rover travel close to the surface and follow its curvature. For such propagation the departure from the planar result is set by the diffraction parameter $(kR_{\rm M}/2)^{1/3}\,r/R_{\rm M}$, where $k$ is the free-space wavenumber, and the planar result holds when this parameter is much smaller than 1 and the antenna heights are well below $(kR_{\rm M}/2)^{1/3}/k$ \citep{Fock1965}. Over the traverse the parameter is at most 0.08, and at most 0.12 for the wave in the regolith, whose wavenumber is larger by the refractive index, below 3.2. The height scale is at least 120~m, against the 2.5~m of the receiver, so the planar calculation holds along the whole traverse. At 32.1~MHz we nevertheless keep to ranges below 1~km, where even the phase $kr^2/(2R_{\rm M})$ that a vertically traveling wave would gain over this drop stays below 0.2~rad.

\subsection{Statistical comparison}
\label{sec:sepstats}

The SEP measurements give the field along the traverse at several frequencies and for different transmitter and receiver orientations. We use them to test whether a regolith description reproduces the measured variation of the field along the rover traverse, including the positions and amplitudes of the interference features. Since the archived levels contain an instrumental constant that is not part of the propagation model, we allow the overall normalization to vary independently in each measurement channel. This leaves the spatial dependence of the field as the main quantity being tested.

A channel $c$ is the sequence of measurements along the traverse at one frequency, for one transmitter orientation, broadside or endfire, and one receiver quantity, the vertical field or the total horizontal power. We denote the measured levels, in dB, by $D_{ci}$ and the corresponding model predictions from Eqs.~(\ref{eq:sep}) and (\ref{eq:wire}) by $D^{\rm mod}_{ci}$, where $i$ labels the measurement range $r_i$. We allow one additive offset $o_c$ for each channel and no other nuisance parameter. Measurements within 3~dB of the noise floor are treated as upper limits (censored samples), $D_c^{\rm up}$, set 3~dB above the floor. Assuming Gaussian scatter with standard deviation $\sigma_d$ per sample, the channel log-likelihood is
\begin{multline}
  \ln\mathcal L_c=\sum_{i\in{\rm det}}\ln\mathcal N\big(D_{ci}\,\big|\,D^{\rm mod}_{ci}+o_c,\,\sigma_d\big)\\
  +\sum_{i\in{\rm cens}}\ln\Phi\Big(\frac{D^{\rm up}_c-D^{\rm mod}_{ci}-o_c}{\sigma_d}\Big),
  \label{eq:likelihood}
\end{multline}
where the first sum is over detections and the second over upper limits, $\mathcal N$ is the normal probability density, and $\Phi$ is its cumulative distribution. The offset $o_c$ is determined by maximizing this likelihood, including the upper-limit contributions. The maximum is unique because the log-likelihood is concave in $o_c$.

The measurements within a channel are not statistically independent. Residuals from all the regolith descriptions vary smoothly along the traverse, typically over distances of tens of meters. Such correlations can arise from effects absent from the model, including scattering by rocks, lateral variations in the regolith and details of the transmitter. Treating every sample as independent would therefore give excessive weight to extended, correlated features in the residuals. To reduce this effect, we estimate the effective number of independent samples in each channel,
\[
  N_{{\rm eff},c}=\frac{N_c}{\tau_c},
\]
where $N_c$ is the number of detected samples and
\begin{equation}
  \tau_c=1+2\sum_{m\ge1}\gamma_{c,m}
  \label{eq:tauint}
\end{equation}
is the integrated autocorrelation time of the residuals in bins. Here $\gamma_{c,m}$ is the autocorrelation at lag $m$. We truncate the sum using the initial-positive-sequence rule of \citet{Geyer1992}. The resulting $N_{{\rm eff},c}$ estimates how many independent samples would provide approximately the same information as the correlated measurements in that channel.

We use these estimates to construct an effective log-likelihood,
\begin{equation}
  \ell_{\rm eff}=\sum_c\frac{N_{{\rm eff},c}}{N_c}\,\ln\mathcal L_c ,
  \label{eq:elleff}
\end{equation}
which gives less weight to channels with stronger correlations. We calculate $N_{{\rm eff},c}$ once for each set of measurements, taking the median over six reference descriptions, and use the resulting weights for every description being compared (Appendix~\ref{app:stats}). This keeps the weighting fixed across models. The estimated correlation times are substantial: $\tau_c=2$--27 bins at 2.1 and 4.0~MHz, and 6--85 bins at 8.1--32.1~MHz. Thus, particularly at the higher frequencies, the number of recorded samples substantially overstates the amount of independent information.

The score $\ell_{\rm eff}$ is a practical means of ranking the descriptions while accounting approximately for the reduction in information caused by correlations. It is not the likelihood of the correlated data: the channel likelihoods in Eq.~(\ref{eq:likelihood}) are constructed from independent-sample contributions and are then reweighted by the effective sample fractions. We therefore do not interpret differences in $\ell_{\rm eff}$ as odds, statistical significance or information criteria. Instead, we test whether the resulting ranking is supported by predictions at frequencies not used in the fit and by two further treatments of the residual correlations.

First, we fit the regolith parameters at 2.1 and 4.0~MHz and use them to predict the measurements at the held-out frequencies, 8.1--32.1~MHz (Sec.~\ref{sec:sepholdout}). These higher-frequency measurements were recorded along the same traverse but are not used to determine the fitted parameters. This tests whether a description that reproduces the lower-frequency data also predicts the interference patterns at higher frequencies.

Second, we evaluate the comparisons using an explicit covariance model. We describe the residuals in each channel as an Ornstein--Uhlenbeck (OU) process along the traverse, with its variance and correlation range fitted separately for each channel and regolith description. This gives a log-likelihood, $\ln\mathcal L_{\rm OU}$, that accounts explicitly for correlations among the detected samples within each channel. It is exact for the detected samples under the assumed OU covariance model. The channel likelihoods are nevertheless evaluated separately and summed. They do not constitute a joint likelihood for channels that share the same ground (Appendix~\ref{app:stats}).

Third, we use block bootstraps to assess how sensitive the ranking is to the particular stretches of the traverse included in the data. Resampling individual points would destroy the correlations between neighboring measurements, so we resample contiguous blocks at least as long as the estimated autocorrelation time. We use five schemes: circular blocks \citep{Politis1992}; common blocks of two different lengths, which select the same stretches of range for every channel; stationary blocks \citep{PolitisRomano1994}; and blocks stratified in range (Appendix~\ref{app:stats}). For each scheme, we draw 500 replicates and refit the models using a linearization about the fit to the full data. We test this approximation against full refits (Appendix~\ref{app:robust}). The fraction of replicates in which one description leads indicates the stability of the ranking under resampling, not the probability that the description is correct.

We also repeat the fits with the full calculation after removing all samples within 100, 200 or 300~m of the transmitter. This tests whether the conclusions depend on the measurements nearest the source, where details of the transmitter and the limitations of its idealized representation may have a greater effect. Unless stated otherwise, ``leads'' and ``trails'' below refer to the ranking by $\ell_{\rm eff}$; comparisons based on $\ln\mathcal L_{\rm OU}$ are identified explicitly.

We fit the measurements at 2.1 and 4.0~MHz, where the transmitter is a documented half-wave element and the wavelengths, 75--143~m, are long compared with the rocks that scatter at the higher frequencies \citep{Simmons1972}. We leave 1.0, 8.1, 16.0 and 32.1~MHz out of the fit. At each fitted frequency, we use the three quantities defined in Sec.~\ref{sec:sepdata}, binned in 10~m intervals from 50 to 1500~m, and take $\sigma_d=2$~dB.

The receiving antenna consists of three orthogonal loops, each about a third of a meter across \citep{SEPStudy1973}, mounted on the rover with their upper edges about 2.5~m above the surface. We use 2.5~m for the loop height and test the sensitivity of the results over the range 1.5--3.5~m (Appendix~\ref{app:robust}). We maximize $\ell_{\rm eff}$ within the physical parameter bounds using a trust-region least-squares method. Since the score has multiple local maxima, we start the optimization from points distributed across the allowed parameter ranges (Appendix~\ref{app:robust}). Finally, we evaluate the Grimm-like profile of Sec.~\ref{sec:tldescriptions} using both the full layered-medium calculation and the impedance boundary of Eq.~(\ref{eq:ibc}). This provides a direct test of the impedance-boundary representation used in LuSEE-Night simulations against the SEP measurements.

\subsection{Fits at 2.1 and 4.0 MHz}
\label{sec:sepfits}

\begin{table*}[t]
  \caption{Descriptive scores \(\ell_{\rm eff}\) for the Taurus-Littrow regolith models. The first five columns give the scores at the fitted frequencies, 2.1 and 4.0~MHz, including their sum and the differences from the best score in the sum and in the OU log-likelihood, \(\ln\mathcal L_{\rm OU}\). The last two columns give the score at the held-out frequency of 1.0~MHz and its difference from the best score. IBC denotes evaluation with the impedance-boundary approximation. For the graded IBC case, the profile parameters are fixed to those obtained with the full layered-medium calculation. The receiver-loop height is 2.5~m. Higher scores indicate better agreement. The score differences are descriptive and should not be interpreted as odds or statistical significance (Sec.~\ref{sec:sepstats}).}
  \label{tab:sep}
  \begin{ruledtabular}
  \begin{tabular}{lccccccc}
  & \multicolumn{5}{c}{Fitted: 2.1 and 4.0~MHz} & \multicolumn{2}{c}{Held out: 1.0~MHz} \\
  Description & 2.1~MHz & 4.0~MHz & Sum & Behind best & Behind best, OU & $\ell_{\rm eff}$ & Behind best \\
  \hline
  half-space & $-380.3$ & $-165.9$ & $-546.2$ & 169.0 & 111.2 & $-233.7$ & 34.9 \\
  two-layer & $-254.0$ & $-123.2$ & $-377.2$ & 0 & 0 & $-234.0$ & 35.2 \\
  graded & $-257.8$ & $-137.7$ & $-395.4$ & 18.2 & 60.1 & $-251.7$ & 52.9 \\
  graded, IBC & $-272.2$ & $-138.1$ & $-410.3$ & 33.1 & 69.5 & $-198.8$ & 0 \\
  Grimm-like & $-272.7$ & $-140.9$ & $-413.7$ & 36.5 & 84.6 & $-244.6$ & 45.8 \\
  Grimm-like, IBC & $-272.1$ & $-137.9$ & $-409.9$ & 32.7 & 68.0 & $-201.8$ & 3.0 \\
  \end{tabular}
  \end{ruledtabular}
\end{table*}

\begin{figure}[t]
  \centering
  \figorbox{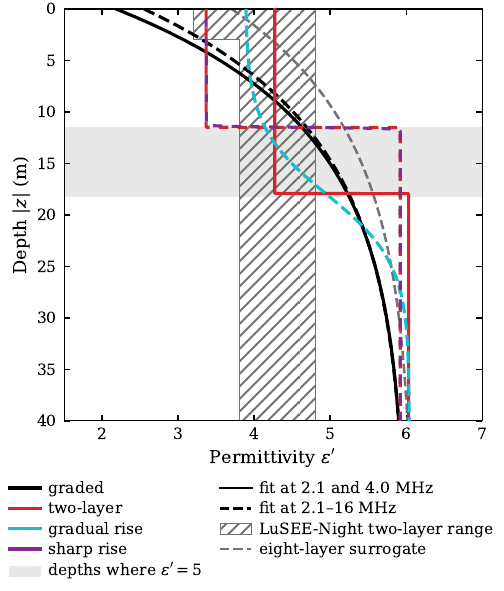}{\columnwidth}
  \caption{Taurus-Littrow profiles fitted to the Apollo~17 SEP measurements at 2.1 and 4.0~MHz (solid curves) or over 2.1--16~MHz (dashed curves). The profiles shown are the graded (black), two-layer (red), gradual-rise (cyan) and sharp-rise (purple) models. The two-layer and sharp-rise profiles have transitions in permittivity that occur within a meter, whereas the graded and gradual-rise profiles have transitions extending over meters to tens of meters. The two-layer and sharp-rise profiles fitted over 2.1--16~MHz coincide and are shown with alternating dashes. Also shown in gray are the profiles considered in the LuSEE-Night studies, the range spanned by the two-layer profiles of \citet{Yousuf2026} (hatched), and our surrogate for the eight-layer profile of \citet{Camacho2026} (dashed). The shaded band marks the depths at which the fitted profiles reach \(\varepsilon'=5\).}
  \label{fig:tlfits}
\end{figure}

Figure~\ref{fig:sep} compares the three measured quantities along the traverse with the predictions of the different regolith descriptions. The fitted parameters indicate a lower-permittivity surface region over material with $\varepsilon'\approx6$ (Fig.~\ref{fig:tlfits}). The two-layer fit gives $\varepsilon'=4.27$ in a 17.9~m upper layer and $\varepsilon'=6.03$ below it, with $\tan\delta=0.008$. The best-fitting graded profile has surface and deep permittivities $\varepsilon'_0=2.18$ and $\varepsilon'_\infty=6.01$, respectively, and a grading depth $z_g=11.3$~m. Thus, both descriptions require a substantial increase in permittivity with depth, although they represent the transition differently.

At the two fitted frequencies, the two-layer description gives the highest score, $\ell_{\rm eff}=-377.2$, ahead of the graded profile by 18.2 (Table~\ref{tab:sep}). The corresponding difference in the OU log-likelihood is 60.1. The two-layer score has a second maximum, 24.1 lower, at a 14.9~m layer with $\varepsilon'=2.89$ over material with $\varepsilon'=5.06$ (Appendix~\ref{app:robust}). At these frequencies, the wavelength in the regolith is 30--100~m, longer than the depth over which the permittivity rises. The measurements therefore constrain the depth and magnitude of the increase more strongly than its detailed shape. The preference for the two-layer description applies to the fit at 2.1 and 4.0~MHz; its ability to predict the higher-frequency measurements is tested separately in Sec.~\ref{sec:sepholdout}.

The half-space gives a substantially poorer fit. It falls behind the graded profile by 151 in $\ell_{\rm eff}$ and by 51 in $\ln\mathcal L_{\rm OU}$, even though its fitted loss tangent reaches the largest allowed value, 0.05. The fixed Grimm-like profile, with surface permittivity 3.7, falls behind the graded profile by 18.3 in $\ell_{\rm eff}$ and 24.6 in the OU comparison. When both profiles are evaluated using the impedance boundary, their scores are nearly equal: the Grimm-like profile is ahead of the graded profile by 0.4 in $\ell_{\rm eff}$ and 1.5 in the OU comparison. The measurements therefore favor a subsurface increase in permittivity over a uniform half-space and, when evaluated with the full layered-medium calculation, the fitted increase over the fixed Grimm-like profile.

The layer values used in the simulated eight-layer LuSEE-Night structure have not been published. We therefore compare the measurements with 15 eight-layer step profiles that follow its general description. These use three choices of layer spacing over 0--100~m, five profile shapes spanning permittivities from 3.7 to 6.1, and $\tan\delta=0.01$. With the full layered-medium calculation, their scores fall behind the graded profile by 3.4--193 in $\ell_{\rm eff}$. Under the impedance boundary used in the simulations, they fall behind by 11.8--15.6, comparable to the reduction in score when the same boundary is applied to the graded profile. Their predictions at 8.1--32.1~MHz are within 5 score units of the corresponding graded-profile predictions. The measurements therefore do not exclude the eight-layer description itself. How well it fits depends on the specific layer values, which are not available for a direct comparison.

\begin{table*}[t]
  \caption{Bootstrap and range-cut tests of the SEP model comparisons. The upper part summarizes the bootstrap results. For each resampling scheme, the columns give the fraction of replicates in which the graded model outperforms the two-layer model (2L), the 2.5th-percentile score lead of the graded model over the half-space (HS), Grimm-like and best tested LuSEE-Night profiles, the fraction of replicates in which the graded model predicts the held-out 8.1--16~MHz measurements better than the two-layer model, and the fraction in which a sharp rise is preferred over a gradual rise in the 2.1--16~MHz fits (Sec.~\ref{sec:sharpness}). The parameters are held fixed when evaluating the predictions at the held-out frequencies. Each bootstrap scheme uses 500 linearized refits, except for the final common-block test, which uses a separate set of 80 replicates, refitted in full where the linearization fails. Parentheses give 95\% Wilson intervals. Entries marked with \(\approx\) are approximate because the linearized common-block refits fail when the fitted surface permittivity reaches its upper bound (Appendix~\ref{app:robust}). The lower part shows differences in \(\ell_{\rm eff}\) after all models are refitted with samples within \(r_{\rm min}\) of the transmitter removed. The columns give the graded-model lead over the two-layer, half-space, Grimm-like and best tested LuSEE-Night profiles, followed by its lead over the two-layer model at 8.1--16~MHz and the lead of the sharp-rise model over the gradual-rise model. Negative values in the first column favor the two-layer model, while negative values in the last column favor the gradual rise.}
  \label{tab:seprobust}
  \begin{ruledtabular}
  \begin{tabular}{lcccccc}
  & Ahead of & \multicolumn{3}{c}{Lead, 2.5th percentile} & Ahead of 2L, & Sharp \\
  Scheme & 2L (\%) & HS & Grimm-like & LuSEE range & 8.1--16~MHz (\%) & preferred (\%) \\
  \hline
  circular blocks & 0.8 (0.3--2.0) & 78 & 5.2 & 236 & 100 (99.2--100) & 35.4 (31.3--39.7) \\
  common blocks & $\approx7.2$ (5.2--9.8) & $\approx71$ & $\approx5.9$ & $\approx273$ & 95.6 (93.4--97.1) & 39.6 (35.4--44.0) \\
  common blocks, half length & $\approx12.2$ (9.6--15.4) & $\approx73$ & $\approx7.2$ & $\approx277$ & 98.4 (96.9--99.2) & 35.8 (31.7--40.1) \\
  common blocks, 80, full refits & 5.0 (2.0--12.2) & -- & -- & -- & -- & -- \\
  stationary & 1.6 (0.8--3.1) & 80 & 6.6 & 227 & 100 (99.2--100) & 35.8 (31.7--40.1) \\
  stratified by range & 0.6 (0.2--1.7) & 85 & 7.2 & 225 & 100 (99.2--100) & 36.2 (32.1--40.5) \\
  \hline
  & \multicolumn{6}{c}{Range cuts: lead of the graded model in $\ell_{\rm eff}$} \\
  all samples, from 50~m & $-18.2$ & $151$ & $18.3$ & $328$ & $67.4$ & $-3.8$ \\
  $r_{\rm min}=100$~m & $-21.1$ & $112$ & $11.3$ & $322$ & $90$ & $-18.2$ \\
  $r_{\rm min}=200$~m & $-2.2$ & $117$ & $9.1$ & $281$ & $192$ & $-10.4$ \\
  $r_{\rm min}=300$~m & $-1.1$ & $87$ & $6.4$ & $273$ & $201$ & $-6.5$ \\
  \end{tabular}
  \end{ruledtabular}
\end{table*}

The preference for the two-layer model at the fitted frequencies is stable under changes to the analysis (Table~\ref{tab:seprobust}). It leads by 8.4--31.8 in every analysis variant of Appendix~\ref{app:robust}, by 17.7--18.6 for every tested SEP transmitter current, and in 88--99\% of the bootstrap replicates under every scheme. In a separate set of 80 common-block replicates, full refits where the linearization fails give a two-layer preference in 95\% of cases (95\% Wilson interval: 88--98\%). With all measurements within 100, 200 or 300~m of the transmitter removed and every model refitted with the full calculation, the two-layer model remains ahead by 1.1--21.1. The preference therefore depends neither on particular stretches of the traverse nor on the linearized refits.

The other model comparisons are also stable. The 2.5th-percentile bootstrap margins of the graded profile over the half-space, the Grimm-like profile and the best tested LuSEE-Night profile remain positive for every resampling scheme. When the measurements nearest the transmitter are removed, the graded profile remains ahead of these three descriptions by 87--117, 6.4--11.3 and at least 273, respectively. Thus, although the two-layer model is preferred over the graded profile at the fitted frequencies, the graded profile consistently outperforms the half-space, the fixed Grimm-like profile and the tested realizations of the LuSEE-Night structure.

The residuals of the graded profile have an rms of 1.3--1.7~dB in the endfire horizontal power and 2.6--3.9~dB in the broadside channels. The broadside measurements contain the interference pattern and are more sensitive to scattering and lateral variations in the regolith. The residuals reflect the combined effects of scattering, lateral structure and the idealized transmitter and receiver; the present data do not allow these contributions to be separated.

\begin{figure*}[t]
  \centering
  \figorbox{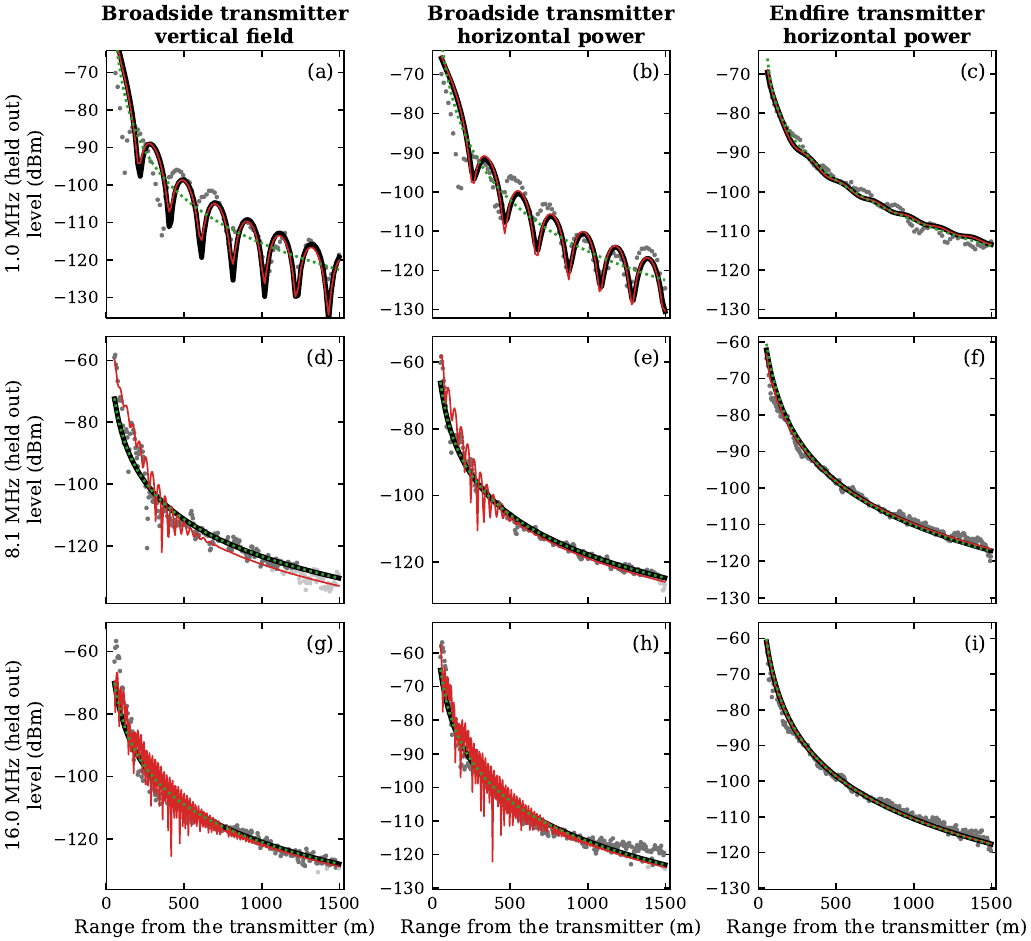}{\textwidth}
  \caption{Predictions for the Apollo~17 SEP measurements at the frequencies held out of the fit, 1.0~MHz in (a)--(c), 8.1~MHz in (d)--(f), and 16.0~MHz in (g)--(i). The columns show the same three measured quantities as in Fig.~\ref{fig:sep}. Gray points show the measurements in 10~m bins at 1.0~MHz and 5~m bins at 8.1 and 16.0~MHz. Points within 3~dB of the noise floor are shown in light gray. The curves show the graded profile calculated with the exact layered-medium model (thick black) and the impedance-boundary approximation (dotted green), and the two-layer description (thin red). All profile parameters are fixed to the values obtained from the fits at 2.1 and 4.0~MHz. Only the channel offset is fitted at each held-out frequency.}
  \label{fig:holdout}
\end{figure*}

The bootstrap refits also provide estimates of the uncertainty in the graded-profile parameters (Appendix~\ref{app:stats}). With circular blocks, the 16th and 84th percentiles are 1.7 and 2.9 for $\varepsilon'_0$, 5.98 and 6.06 for $\varepsilon'_\infty$, and 10.1 and 13.0~m for $z_g$. With common blocks, $\varepsilon'_\infty$ and $z_g$ remain well constrained, with percentile ranges of 5.96--6.08 and 9.5--14.3~m. The surface permittivity, however, reaches the upper search bound of 5.0 in 16--19\% of the replicates. The linearization fails in these cases, and full refits give $\varepsilon'_0=3.4$--5.0. Away from the upper bound, the full and linearized refits agree (Appendix~\ref{app:robust}). Predictions at the held-out frequencies use the fitted parameters without further adjustment.

The deep permittivity is more robustly constrained than the surface permittivity. The latter governs only the upper few meters of the regolith, a small fraction of a wavelength, and is therefore less well determined by these measurements. Correlations between channels further limit the information available on this parameter. The samples nearest the transmitter provide most of the constraint: removing measurements within 100, 200 and 300~m raises the fitted surface permittivity from 2.18 to 2.59, 3.67 and 4.20, respectively, and increases $z_g$ from 11.3 to 12.0, 14.9 and 17.5~m. The deep permittivity remains nearly unchanged, between 6.02 and 6.10.

The loss tangent is constrained mainly from above. The best-fitting graded profile reaches the lower search bound, $\tan\delta=0.001$. After refitting the other three parameters, increasing the loss tangent lowers $\ell_{\rm eff}$ by 0.7 at $\tan\delta=0.004$, the Chang'E-4 radar value; by 1.7 at 0.008; and by 2.2--2.8 at 0.010--0.012, the in situ SEP values reported by \citet{Grimm2018}. The SEP instrument team found that the in situ loss tangent was below 0.05 and could be as low as 0.002 \citep{SEPFinal1975}. The 95th percentile of the refitted loss tangent is 0.001--0.004 for circular, stationary and range-stratified blocks, but 0.011--0.022 for common blocks. Since the inferred upper limit depends on how correlations between channels are treated, we adopt the more conservative bound, $\tan\delta\lesssim0.02$. This constraint applies to the graded profile and should not be taken as a general bound for other regolith structures.

The two-layer fit gives a different result. It has $\tan\delta=0.008$ even when measurements within 100~m of the transmitter are removed, but reaches 0.001 when the exclusion extends to 200 or 300~m. The fitted loss tangent also depends on the sharpness of the permittivity rise, parameterized by the scale $\zeta$ in Sec.~\ref{sec:sharpness}. It decreases from 0.025 at $\zeta=0.05$~m to the lower bound, 0.001, for $\zeta\geq3.3$~m. A sharp interface produces a stronger reflection and therefore requires greater loss to reduce it to the measured level.

Finally, refitting the graded model with the impedance boundary gives a poorer fit and a substantially different regolith profile. Its $\ell_{\rm eff}$ is 11.5 lower than that obtained with the full layered-medium calculation, but 3.4 higher than the fixed-parameter graded IBC result in Table~\ref{tab:sep}. The inferred permittivity rises only from 2.33 at the surface to 2.6 at 50~m. The deep permittivity and grading depth reach the limits of the search, $\varepsilon'_\infty=3.0$ and $z_g=100$~m. Applying the impedance boundary to these measurements would therefore lead to a substantially different interpretation of the subsurface structure.

\subsection{Predictions at held-out frequencies}
\label{sec:sepholdout}

\begin{table*}[t]
  \caption{Predictions at the held-out frequencies of 8.1, 16.0 and 32.1~MHz, using the profile parameters fitted at 2.1 and 4.0~MHz (Table~\ref{tab:sep}). Only the offset of each channel is refitted. The first two columns give \(\ell_{\rm eff}\) at 8.1 and 16.0~MHz, followed by the shortfall from the best description in their combined score and in the OU log-likelihood. The last two columns give the corresponding shortfalls at 32.1~MHz, evaluated over the first 1~km of the traverse. Higher scores indicate better agreement. IBC denotes the impedance-boundary approximation. The continuous graded and Grimm-like profiles are represented by 320 and 240 layers, respectively, which are sufficient for convergence at these frequencies (Appendix~\ref{app:robust}).}
  \label{tab:holdout}
  \begin{ruledtabular}
  \begin{tabular}{lcccccc}
  & \multicolumn{4}{c}{8.1 and 16.0~MHz} & \multicolumn{2}{c}{32.1~MHz} \\
  Description & 8.1~MHz & 16.0~MHz & Behind best & Behind best, OU & Behind best & Behind best, OU \\
  \hline
  graded & $-161.2$ & $-85.7$ & 3.5 & 4.6 & 0.1 & 0.5 \\
  graded, IBC & $-161.4$ & $-85.6$ & 3.7 & 6.7 & 0.3 & 2.5 \\
  Grimm-like & $-158.2$ & $-85.1$ & 0 & 0 & 0.1 & 0 \\
  Grimm-like, IBC & $-158.5$ & $-85.0$ & 0.2 & 1.7 & 0.3 & 1.6 \\
  half-space & $-160.6$ & $-88.4$ & 5.7 & 23.3 & 0 & 0.4 \\
  two-layer & $-193.7$ & $-120.6$ & 70.9 & 788.9 & 9.2 & 631.0 \\
  \end{tabular}
  \end{ruledtabular}
\end{table*}

This subsection asks whether the descriptions fitted at 2.1 and 4.0~MHz predict the measurements at 8.1--32.1~MHz, with their parameters fixed and only the offsets of the channels refitted. The predicted frequencies were recorded on the same traverse as the fitted ones, so the test checks prediction across frequency at one site; it is not a test against spatially independent data. Two measures are used. The score $\ell_{\rm eff}$ is descriptive: it ranks the descriptions. The interference ripple along the traverse is a direct observable: a buried interface predicts it, and the measured fluctuations bound it. Both test the descriptions, not the actual profile, which Sec.~\ref{sec:sharpness} discusses.

We predict 8.1 and 16.0~MHz in 5~m bins and 32.1~MHz within 1~km in 2.5~m bins (Table~\ref{tab:holdout}, Fig.~\ref{fig:holdout}). The test separates the fitted descriptions. The graded profile predicts 8.1--16.0~MHz within 3.5 in $\ell_{\rm eff}$ of the best description, the Grimm-like profile, and 32.1~MHz within 0.1. The two-layer description falls behind by 70.9 at 8.1--16.0~MHz and 9.2 at 32.1~MHz, and under the covariance model by 789 and 631. Its interface, at 17.9~m and with the contrast fitted at 2.1--4.0~MHz, from 4.27 to 6.03, returns a coherent reflection, whose predicted interference ripple exceeds the measured fluctuation in five of the six broadside channels (below). The final report of the experiment found that no model of plane layers fitted all frequencies accurately \citep{SEPFinal1975}; this test quantifies that finding for an interface fitted at the low frequencies. The report also rejected a smoothly varying permittivity, because a 20-layer model of that kind produced no interference peaks and troughs. In the present calculation the graded profile follows the smooth decline of the fields at 8.1--16.0~MHz, where the measurements show little interference, and produces interference minima at 1.0~MHz, where the measurements show them (below). Every description without such an interface, the two continuous profiles and the lossy half-space, predicts these frequencies within 5.7 of the best. With the parameters fixed and the measurements resampled, the two-layer description stays behind the graded profile at 8.1--16.0~MHz in 96--100\% of the replicates under every scheme, and at 32.1~MHz in all of them (Table~\ref{tab:seprobust}). With the samples within 100--300~m removed and every model refitted, the two-layer description falls behind the graded profile at 8.1--16.0~MHz by 90--201 and at 32.1~MHz by 4.3--44.

The missing reflection can also be seen without the score, in the ripple that it would produce along the traverse. We remove a smooth trend from each binned series, its convolution with a Gaussian of 25~m standard deviation in range, and compare the rms of the remainder over 100--800~m, with each prediction evaluated at the ranges of the raw samples and binned like the measurements (Appendix~\ref{app:ripple}). At 16.0~MHz the two-layer description predicts a ripple of 4.0 and 3.7~dB rms in the broadside vertical field and horizontal power, with a period of about 19~m, while the measurements fluctuate there by 1.9 and 1.4~dB rms, scattering and noise included [Figs.~\ref{fig:holdout}(g) and \ref{fig:holdout}(h)]. If the ripple were present and the other fluctuations independent of it, the measured fluctuation would be expected to be at least as large as the ripple. The predicted ripple also exceeds the measured fluctuation in the broadside horizontal power at 8.1~MHz and in both broadside channels at 32.1~MHz, by factors of 1.4--2.2, and at 16.0~MHz it does so by factors of 1.4--2.6 for trend widths of 25--60~m and over 50--1500~m (Appendix~\ref{app:ripple}). Only in the broadside vertical field at 8.1~MHz is it smaller than the measured fluctuation, 2.8 against 3.1~dB, so there the measured fluctuations could hide it. The test does not depend on accurate ranges: errors of scale, shifts and smooth drifts of the archived ranges, and even independent errors of 6~m rms at every position, leave the predicted ripple above the measured fluctuation in the same five channels (Appendix~\ref{app:ripple}). The graded profile predicts a ripple of at most 0.2~dB at 4.0--32.1~MHz. At 2.1 and 4.0~MHz, in contrast, the measured fluctuations of the broadside channels, 2.0--2.9~dB, exceed the ripple of both descriptions, at most 1.6~dB, so the fit there cannot detect the interface.

The loss tangents fitted at 2.1--4.0~MHz, 0.001 for the graded profile and 0.008 for the two-layer description, lie below the laboratory expectation for the composition of the site, 0.011--0.034 (Sec.~\ref{sec:regolith}), and a higher loss weakens the reflection from a buried interface. We therefore repeated the test with the loss tangent of both descriptions fixed at 0.012, the in situ value of \citet{Grimm2018}, and at 0.020, 0.028 and 0.040 (Appendix~\ref{app:lossfixed}). At every value the measurements at 2.1--4.0~MHz select the same interface, at 17.9--18.7~m, ahead of the graded fit by 18.8--20.6, while the graded predictions at 8.1--16.0~MHz change by less than 0.6. The shortfall of the two-layer description at 8.1--16.0~MHz shrinks as the loss rises, to 12.1 at 0.040, and the predicted ripple at 16.0~MHz exceeds the measured fluctuation up to a loss tangent of 0.028. The failure at the held-out frequencies therefore persists for every loss tangent that the laboratory relation allows, with less margin as the loss rises.

The reverse test (Appendix~\ref{app:reverse}) shows what each frequency band constrains. When fitted to 8.1--16.0~MHz alone, the two-layer model places the interface at a shallower depth of 10.5~m, with \(\varepsilon'=3.46\) above it and \(\varepsilon'=6.99\) below it, and gives a loss tangent of 0.023. Both models overestimate the deep permittivity because the higher frequencies are only weakly sensitive to it. The fitted values are 6.28 for the graded model and 6.99 for the two-layer model, compared with 6.01--6.03 from the low-frequency fits. When extrapolated back to 2.1--4.0~MHz, the two models score 38.9 and 47.2 below their respective fits at those frequencies. In the forward test, by contrast, the graded profile fitted at 2.1--4.0~MHz predicts the 8.1--16.0~MHz measurements within 3.5 of the score obtained by fitting those frequencies directly, whereas the two-layer prediction falls 92.3 below its own fit. The low frequencies therefore constrain the deep permittivity, while the higher frequencies are more sensitive to the sharpness of the transition and the loss. At the low frequencies, the measured fluctuations can hide the interference ripple produced by the buried interface. The fit consequently accepts an interface whose predicted ripple is inconsistent with the higher-frequency measurements.

A graded profile fitted at 2.1--4.0~MHz thus predicts the measurements across a factor of 15 in frequency, whereas the two-layer profile fitted over the same band does not. The distinction comes mainly from the 8.1--16.0~MHz measurements. At 32.1~MHz, the two-layer description falls 9.2 behind the best score, while all other descriptions lie within 0.3 of the best. The evidence against the fitted two-layer profile is therefore its failure at the held-out frequencies, both in the comparison scores and in the interference ripple it predicts beyond the measured fluctuations. This is in contrast to the fitted band, where the two-layer model leads the graded model by 18.2 in \(\ell_{\rm eff}\). The test demonstrates the predictive success of the graded parameterization, but does not establish that the regolith is actually graded (Sec.~\ref{sec:sharpness}). Nor does it exclude two-layer structures in general. When fitted jointly over 2.1--16~MHz, a two-layer model gives a better fit than the graded model by 11.4 in \(\ell_{\rm eff}\).

The 1.0~MHz measurements, which are also held out, are more difficult to interpret. The traverse spans only 0.2--5 wavelengths at this frequency. The full 70~m transmitter wire radiates with a current distribution that is not documented in detail \citep{Simmons1972}, and the receiver height is only approximately known. Changing the assumed height from 1.5 to 3.5~m changes \(\ell_{\rm eff}\) by 10. The broadside fields show interference minima [Figs.~\ref{fig:holdout}(a) and \ref{fig:holdout}(b)], produced by the beating of the wave guided in the regolith with the wave propagating above it. The exact calculation predicts these minima, although not at their measured positions, which depend on the transmitter current distribution and receiver height. The impedance-boundary approximation does not include a wave guided within the regolith and therefore predicts no such minima. Its curves pass between the measured minima, giving the graded profile a score 48--60 higher than with the exact calculation. The two-layer description falls 35 behind the best score at 1.0~MHz, about 18 less than the graded profile.

\subsection{How sharply the permittivity rises}
\label{sec:sharpness}

The SEP measurements constrain the depth at which the permittivity increases, but they do not establish how sharply this increase occurs. The inferred shape depends on which measurements are included. In particular, the preference of the 8.1--16.0~MHz data for a sharp rise comes almost entirely from the broadside vertical field measured within 300~m of the transmitter. It nearly disappears when these measurements are excluded. We examine the evidence for the depth and sharpness of the transition below.

\begin{table*}[t]
  \caption{Taurus-Littrow regolith profiles fitted to the SEP measurements. The graded and two-layer models are fitted either at 2.1 and 4.0~MHz or jointly over 2.1--16.0~MHz. The last two rows give the best sharp and gradual solutions of the transition model, with \(\zeta\leq1\)~m and \(\zeta\geq2\)~m, respectively, fitted over the same combined band. The columns give \(\ell_{\rm eff}\) for the 2.1 and 4.0~MHz data, for the 8.1 and 16.0~MHz data, and for their sum. The score at the held-out 32.1~MHz is evaluated over the first 1~km of the traverse. Higher scores indicate better agreement. Also shown are the depth \(z_5\) at which the permittivity reaches \(\varepsilon'=5\) and the ratio of the largest to smallest zenith directivity of a dipole lying on the regolith over 5--45~MHz (Sec.~\ref{sec:surfacearrays}). The profile parameters are listed as permittivities, depths in meters, and the loss tangent.\label{tab:fits}}
  \begin{ruledtabular}
  \begin{tabular}{llcccccc}
  Description & Parameters & 2.1 + 4.0 & 8.1 + 16.0 & Sum & 32.1 & $z_5$ & Directivity ratio \\
  \hline
  graded, 2.1--4.0 & $2.18\to6.01$, $z_g=11.3$, 0.001 & $-395.4$ & $-246.8$ & $-642.3$ & $-51.6$ & 15.1 & 1.02 \\
  two-layer, 2.1--4.0 & 4.27 over 6.03 at 17.9, 0.008 & $-377.2$ & $-314.3$ & $-691.5$ & $-60.8$ & 17.9 & 1.63 \\
  graded, 2.1--16 & $2.55\to6.02$, $z_g=12.0$, 0.001 & $-395.9$ & $-245.8$ & $-641.6$ & $-51.7$ & 14.7 & 1.02 \\
  two-layer, 2.1--16 & 3.37 over 5.92 at 11.5, 0.025 & $-405.4$ & $-224.9$ & $-630.2$ & $-53.1$ & 11.5 & 1.89 \\
  sharp transition & $3.37\to5.92$ at 11.5, $\zeta=0.05$, 0.025 & $-405.3$ & $-224.9$ & $-630.2$ & $-53.0$ & 11.5 & 1.87 \\
  gradual transition & $3.88\to6.04$ at 18.1, $\zeta=3.3$, 0.001 & $-384.3$ & $-242.1$ & $-626.4$ & $-51.7$ & 18.3 & 1.01 \\
  \end{tabular}
  \end{ruledtabular}
\end{table*}

The distinction between a gradual rise and a sharp interface can be tested only by fitting the 2.1--16.0~MHz data together. At 2.1--4.0~MHz, considered alone, the two-layer model gives the best fit. However, the interface fitted at these frequencies fails to predict the 8.1--16.0~MHz measurements. When fitted jointly over 2.1--16.0~MHz, the two-layer model reaches \(\ell_{\rm eff}=-630.2\), with an 11.5~m layer of \(\varepsilon'=3.37\) above a substrate with \(\varepsilon'=5.92\) and a single loss tangent of 0.025. The graded model reaches \(-641.6\), rising from \(\varepsilon'=2.55\) at the surface to 6.02 at depth, with \(z_g=12.0\)~m (Table~\ref{tab:fits}). The joint fits also have secondary maxima. For the two-layer model, these occur at interface depths of 12.7, 15.0 and 17.3~m, with scores 7.1--34.1 below the best maximum. For the graded model, a secondary maximum occurs at \(\varepsilon'_0=4.16\), with a score 9.9 below the best (Appendix~\ref{app:robust}).

\begin{figure*}[t]
  \centering
  \figorbox{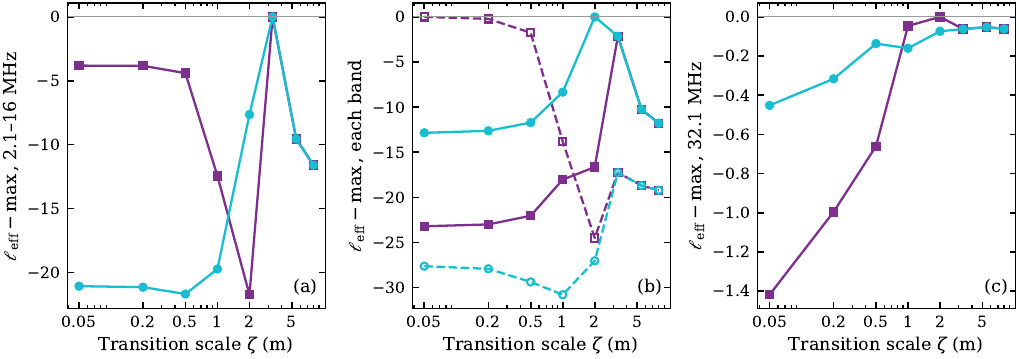}{\textwidth}
  \caption{The comparison score $\ell_{\rm eff}$ against the scale $\zeta$ of the transition, Eq.~(\ref{eq:transitionfamily}), fitted jointly to 2.1--16~MHz with the other four parameters refitted at each $\zeta$. Purple squares follow the branch that starts from a sharp solution and cyan circles the branch that starts from a gradual one. (a) The joint $\ell_{\rm eff}$. (b) The two bands at the same solutions, 2.1 + 4.0~MHz (solid lines, filled markers) and 8.1 + 16.0~MHz (dashed lines, open markers). (c) 32.1~MHz within 1~km, held out and predicted. Each curve is relative to its largest value.}
  \label{fig:sharpness}
\end{figure*}

To compare the sharpness of the transition directly, we use the transition model of Eq.~(\ref{eq:transitionfamily}), which includes both sharp and gradual profiles. Its scale parameter \(\zeta\) describes the width of the permittivity increase. Since \(\ell_{\rm eff}\) has separate maxima for sharp and gradual rises, we evaluate it over \(0.05\leq\zeta\leq8\)~m. At each fixed value of \(\zeta\), we refit the other four parameters along two branches, one continued from a sharp solution and the other from a gradual solution (Fig.~\ref{fig:sharpness}). We also perform fits with all five parameters free, starting from four points in a Latin-hypercube design. These fits end 8.0--150 below the highest point on the fixed-\(\zeta\) curve, and none finds a higher maximum.

The curve has two distinct maxima [Fig.~\ref{fig:sharpness}(a)]. The higher corresponds to a gradual rise with \(\zeta=3.3\)~m, centered at 18.1~m, from \(\varepsilon'=3.88\) to 6.04. The permittivity covers 10--90\% of the increase between 10.8 and 25.3~m, and the loss tangent reaches its lower bound. The second maximum is only 3.8 lower and corresponds to a sharp rise with \(\zeta=0.05\)--0.5~m, centered at 11.5--11.8~m, from \(\varepsilon'=3.4\) to 5.9, with \(\tan\delta=0.017\)--0.025. Intermediate rises with \(\zeta=1\)--2~m score 7.6--12.4 below the maximum. Broader rises with \(\zeta=5.5\)--8~m score 9.6--11.6 below it.

The two frequency bands favor different profiles [Fig.~\ref{fig:sharpness}(b)]. At 2.1--4.0~MHz, the gradual solution leads the sharp solution by 21.1. At 8.1--16.0~MHz, the sharp solution leads by 17.2. The small overall preference for the gradual rise is therefore the difference between two much larger and opposing preferences. The two solutions also differ in the depth of the transition, at 18.1~m for the gradual rise and 11.5~m for the sharp rise. Thus, the preference of the low-frequency data is a preference for the depth of the increase, not for its width. Indeed, when fitted to 2.1--4.0~MHz alone, a sharp rise at 17.9~m gives an even better fit (Sec.~\ref{sec:sepfits}). At the held-out frequency of 32.1~MHz, all values of \(\zeta\) predict the measurements within 1.4 of one another [Fig.~\ref{fig:sharpness}(c)].

The preference of the 8.1--16.0~MHz data for a sharp rise is dominated by a single measurement channel. The broadside vertical field within 300~m of the transmitter favors the sharp solution by 16.9 of the total 17.2. This channel is also particularly sensitive to scattering by rocks and to lateral variations in the regolith near the transmitter. The remaining channels nearly cancel one another. We test the effect of this channel and of possible model discrepancies by fixing \(\zeta\) at 0.5 and 3.3~m for the sharp and gradual solutions, respectively. The gradual solution then leads by 4.4. Removing the affected channel within 300~m at 8.1 and 16.0~MHz increases its lead to 22.6. Assigning a 4~dB scatter to that channel gives a lead of 17.9, while allowing for model discrepancy in every channel gives a lead of 5.4. In all these tests, the fitted depths and permittivities change by no more than 0.6~m and 0.04, respectively. When the parameters are held fixed and the measurements are resampled, the sharp solution is preferred in only 35--40\% of the replicates across the bootstrap schemes in Table~\ref{tab:seprobust}. Removing the samples within 100, 200 or 300~m of the transmitter reduces the preference of the 8.1--16.0~MHz data for the sharp rise to a range of \(-1.7\) to 1.4. For the combined 2.1--16~MHz data, the gradual rise is then preferred by 6.5--18.2.

Although the sharpness remains uncertain, the depth of the permittivity increase is much better constrained. Across the tested profile families, the measurements require a substantial increase at Taurus-Littrow, with \(\varepsilon'\) reaching about 5 at depths of 12--18~m and about 6 at greater depths. Every profile that fits the combined 2.1--16~MHz data reaches \(\varepsilon'=5\) at \(z_5=11.5\)--18.3~m. Above this depth, the average permittivity is 3.4--4.2, while at 100~m it is 5.9--6.0 (Table~\ref{tab:fits} and Fig.~\ref{fig:tlfits}). The instrument team concluded that the permittivity rises from about 3.5 near the surface to 6--8 at a depth of about 100~m, although no single model fitted all the measured frequencies \citep{SEPFinal1975}. Our fits are consistent with that conclusion and further constrain the depth of the increase.

This depth determination relies largely on measurements made within a few hundred meters of the transmitter. When the 2.1--4.0~MHz data are fitted without samples within 300~m, the graded and two-layer models reach \(\varepsilon'=5\) at 9.5 and 22.2~m, respectively. The preferred transition scale of \(\zeta=3.3\)~m reflects the balance between the low frequencies, which favor an increase near 18~m, and the high frequencies, which favor a sharper increase near 11.5~m. The distinction matters for antennas lying on the regolith. Their beams are nearly achromatic over a gradual rise but become chromatic when the transition is sharp (Sec.~\ref{sec:surfacearrays}).

The SEP measurements cannot distinguish a gradual subsurface transition from a rough or undulating interface. Our layered-medium calculation describes the coherent field, its average over the variations of the regolith along the traverse. The rest of the field, scattered by these variations, contributes to the measured fluctuations. For an interface whose height follows a Gaussian distribution with rms \(\sigma_h\), the coherent reflection is equivalent to that from a laterally averaged medium with an error-function transition of width \(\sigma_h\) \citep{NevotCroce1980}. Such an interface weakens the reflection at 8.1--16.0~MHz while leaving it nearly unchanged at 2.1--4.0~MHz, as does a gradual increase in permittivity. The fitted value of \(\zeta\) therefore describes the vertical extent of the transition, but does not by itself establish its physical origin. The logistic profile with \(\zeta=3.3\)~m has a spread equivalent to a Gaussian standard deviation of about 6~m.

The fitted permittivities can also be interpreted in terms of bulk density using Eq.~(\ref{eq:epsdensity}), which \citet{Olhoeft1975} established for the upper 100~m of the Moon. The average permittivity of 3.4--4.2 above \(z_5\) corresponds to a density of 1.9--2.2~g\,cm\(^{-3}\), consistent with compacted regolith. The value of \(\varepsilon'=5.9\)--6.0 at 100~m corresponds to 2.7--2.75~g\,cm\(^{-3}\), or a porosity of about 20\% for the grain density of Apollo~17 high-Ti basalt, 3.46~g\,cm\(^{-3}\) \citep{Kiefer2012}. The inferred depth \(z_5\) lies below the estimated regolith base at about 8~m, derived from crater counts \citep{Bugiolacchi2016}. The Lunar Seismic Profiling Experiment at the site measured seismic velocities of about 100~m\,s\(^{-1}\) in the upper 4~m and 327~m\,s\(^{-1}\) down to 32~m, and identified the material below the regolith as fractured \citep{Cooper1974}. Reanalyses found velocities of about 300~m\,s\(^{-1}\) down to approximately 100~m \citep{Heffels2017,Heffels2021}. \citet{Grimm2018} found that porosity decreases steeply over the upper 20--30~m. Within the adopted permittivity--density relation, the radio measurements are therefore consistent with compaction of the fractured material beneath the regolith. This inference follows from the radio measurements and provides an independent complement to the seismic results.

\subsection{LuSEE-Night profiles and the impedance boundary}
\label{sec:sepluseeibc}

We also test the two-layer structures of the LuSEE-Night subsurface inference \citep{Yousuf2026}, $\varepsilon'_{\rm top}=3.2$--4.2 over $\varepsilon'_{\rm bottom}=3.8$--4.8 with a top layer 0.5--3~m thick and $\tan\delta=0.01$, with only the offsets free. All 27 structures of a $3\times3\times3$ grid spanning this range fall behind the graded profile by at least 328 in $\ell_{\rm eff}$ at 2.1--4.0~MHz, and a continuous search from eight starting points finds nothing better. With the loss tangent free as well, the best structure takes the largest loss tangent we allow, 0.05, which damps the waves returning from depth, and still falls behind by 111. At 2.1--4.0~MHz the waves sample tens of meters, where the fitted descriptions have $\varepsilon'\approx6$, above the deepest permittivity of the range, 4.8. At 8.1--16.0~MHz, which sample shallower material, the best structure of the grid comes within 3.1 of the graded profile. Across the analysis variants of Appendix~\ref{app:robust} the grid falls behind by at least 145 at 2.1--4.0~MHz. The tested LuSEE-Night range is thus disfavored by the Apollo~17 measurements at Taurus-Littrow, already at the frequencies of the fit, while more flexible two-layer structures are not excluded.

The measurements do not discriminate the exact coefficients from the impedance boundary in a consistent way. The exact coefficients are preferred at 2.1--4.0~MHz by 11.5 with each description refitted, the two predict 8.1--32.1~MHz about equally well, and the impedance boundary fits 1.0~MHz better, where the exact result depends on the undocumented loading of the transmitter. This is as expected for a receiver hundreds of meters from the transmitter, beyond the reach of the evanescent waves through which a low antenna couples to the ground. The difference matters for the radiometer quantities of Sec.~\ref{sec:radiometer}, where the antenna is less than one wavelength above the surface.

The Apollo~17 comparison thus rests on three tests that do not treat $\ell_{\rm eff}$ as a likelihood: prediction at frequencies withheld from the fit but recorded on the same traverse, the stability of the ranking under five resampling schemes and without the samples near the transmitter, and the interference ripple of the interface fitted at 2.1--4.0~MHz, which exceeds the measured fluctuations in five of the six broadside channels at 8.1--32.1~MHz. The first holds for loss tangents fixed anywhere in the laboratory range for the site and the third up to 0.028, and the third does not depend on errors in the archived ranges. Together they place at Taurus-Littrow, for the tested profile families, a rise of the permittivity to about 5 at 11.5--18.3~m and about 6 below, disfavor the two-layer range of the LuSEE-Night inference, and show that an analysis under the impedance boundary would misread the structure. They do not determine the sharpness of the rise, and they do not measure the far-side subsurface.

% =====================================================================
\section{A far-side radiometer}
\label{sec:radiometer}
% =====================================================================

\begin{figure*}[t]
  \centering
  \figorbox{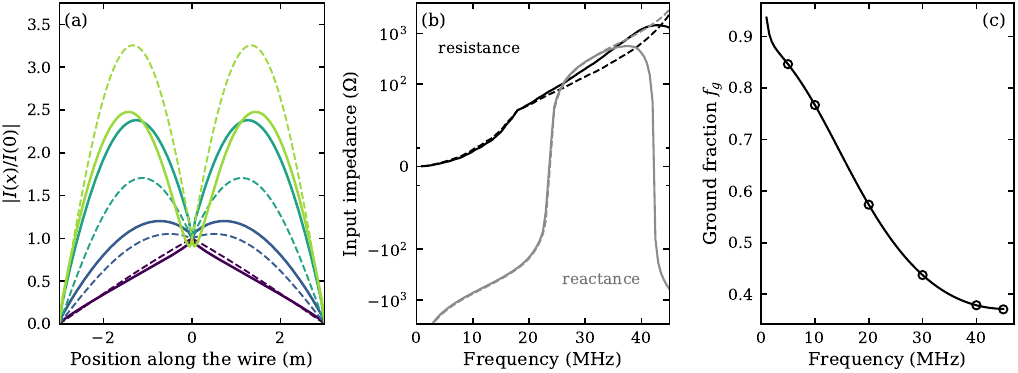}{\textwidth}
  \caption{Current, input impedance and ground fraction for a stand-alone 6~m dipole, 2.5~m above the Chang'E-4 reference profile. (a) Current magnitude along the wire, normalized to its value at the feed, at 10~MHz (dark purple), 30~MHz (blue), 40~MHz (green) and 45~MHz (yellow-green). Solid and dashed curves show the method-of-moments solution and the sinusoidal approximation, respectively; the feed is at $x=0$ and the tips at $x=\pm3$~m. (b) Input resistance $R_{\rm in}$ (black) and reactance $X_{\rm in}$ (gray) for the same two current distributions. The reactance follows the convention of Appendix~\ref{app:impedance}, with $Z_{\rm in}=R_{\rm in}-iX_{\rm in}$ for time dependence $e^{-i\omega t}$. The impedance scale is linear between $-30$ and $30~\Omega$ and logarithmic outside this range. (c) Ground fraction for the half-space reduction of the reference profile, calculated using the method of moments (line) and NEC-2 (circles).}
  \label{fig:mom}
\end{figure*}

The regolith reaches the measurement of a far-side radiometer through three quantities of the antenna. The input impedance sets the coupling to the receiver, the beam weights the sky, and the ground fraction weights the thermal emission of the ground. We compute them with the calculation verified in Sec.~\ref{sec:verification} and tested against Apollo~17 in Sec.~\ref{sec:sep}, ask how well the approximate descriptions of the regolith reproduce them, and add the lander. Sections~\ref{sec:calibrator}--\ref{sec:bias} build on these results.

We evaluate Eqs.~(\ref{eq:Zr})--(\ref{eq:fg}), with the antenna current obtained from Eq.~(\ref{eq:mom}), for a 6~m dipole of radius 1~cm above the Chang'E-4 reference profile. This stand-alone dipole represents the two 3~m monopoles of a LuSEE-Night pseudo-dipole, without the lander \citep{Bale2023,BaleSlosar2024}. The receiver is represented by a high-impedance load of 35~pF in parallel with 1~M$\Omega$ \citep{Rotermund2026}. Published descriptions place the 0.7~m instrument on the top deck of the lander but do not specify the antenna height \citep{Bale2023,Saliwanchik2024}. We therefore adopt 2.5~m above the regolith as the reference antenna height. We examine heights of 1.5--3.5~m in Sec.~\ref{sec:inputs} and 0.3--3~m in Sec.~\ref{sec:ibcantenna}.

The results are predictions for a stand-alone LuSEE-Night-like antenna over the adopted Chang'E-4 reference profile, not for the LuSEE-Night site, whose profile is unknown. Section~\ref{sec:ensemble} tests them on 72 constructed profiles that share the top unit, and on the radar values of that unit, and Sec.~\ref{sec:lander} adds an illustrative model of the lander.

\subsection{The antenna current}
\label{sec:current}

We compare the self-consistent current calculated in Sec.~\ref{sec:mom} with the sinusoidal current assumed in the induced-EMF method (Fig.~\ref{fig:mom}). Figure~\ref{fig:mom}(a) shows the current magnitude along the wire, normalized to its value at the feed. All curves equal unity at the feed and vanish at the tips. At 10~MHz, the dipole is 0.2 free-space wavelengths long, and the two current distributions are nearly identical, falling almost linearly from the feed. At 30--45~MHz, each arm carries a distinct current lobe, and the two distributions differ in their peak amplitudes relative to the feed current.

Figure~\ref{fig:mom}(b) shows the corresponding input impedances. The two results differ by 4--5\% up to 20~MHz, and both pass through the series resonance near 24~MHz, where the reactance changes sign. The difference grows to 18\% at 30~MHz and becomes much larger near the antiresonance at 40--45~MHz. At these frequencies, the 6~m dipole is 0.8--0.9 free-space wavelengths long. As the frequency approaches 50~MHz, where the dipole is one wavelength long, the sinusoidal feed current approaches zero. In the induced-EMF expression, the impedance is inversely proportional to the square of the feed current for a fixed current distribution along the rest of the wire. It therefore rises rapidly as the assumed feed current approaches zero. The self-consistent current, by contrast, remains nonzero at the feed, and its input impedance stays finite through this frequency range. Its resistance peaks at about 1.5~k$\Omega$ near 43~MHz, while its reactance crosses zero again near 42~MHz.

The beam and ground-related quantities are less sensitive to the choice of current because they depend mainly on its normalized distribution along the wire, rather than on its value at the feed. When each current distribution is normalized to its own peak, the two differ by at most 0.08 of the peak for distances greater than 0.5~m from the feed. The resulting ground fraction changes by at most 0.0013, the effective ground temperature defined in Sec.~\ref{sec:thermal} by at most 0.001~K, and the normalized beam by at most 0.023 of its peak. The beam-weighted sky of Sec.~\ref{sec:anisotropic} changes by at most 0.3\% at a fixed ecliptic longitude $\lambda_z$ of the zenith. The errors introduced by the approximate regolith descriptions are also essentially unchanged.

Figure~\ref{fig:mom}(c) compares the ground fraction calculated with the self-consistent current against NEC-2 for the half-space reduction of the reference profile, the case that NEC-2 can treat. The two calculations agree to within $9\times10^{-4}$ over 5--45~MHz (Table~\ref{tab:verification}). Across this band, the ground fraction decreases from 0.85 at 5~MHz to 0.37 at 45~MHz.

The self-consistent current is nevertheless needed to calculate the receiver match and calibrator response. Above 25~MHz, the sinusoidal approximation gives an increasingly inaccurate input impedance and hence an inaccurate receiver match. After removing the gain, the calibrator response calculated with the sinusoidal current differs from that calculated with the self-consistent current by 0.13~dB at 30~MHz and by 0.4--0.6~dB at 40--45~MHz. These differences exceed the per-sample precision adopted in Sec.~\ref{sec:calibrator}. We therefore use the self-consistent current throughout.

\subsection{Beam and ground fraction}
\label{sec:reductions}

We calculate the input impedance, beam and ground fraction of the stand-alone dipole over the full Chang'E-4 reference profile, using the self-consistent current of Sec.~\ref{sec:current} and the exact coupling of the plane waves to all layers. Previous beam calculations for LuSEE-Night represent the subsurface through a surface impedance (Table~\ref{tab:prior}). Our calculation provides a reference for the antenna response over a layered regolith. To identify which features depend on the subsurface structure, we repeat the calculation for the two-layer and half-space reductions of Sec.~\ref{sec:farside}.

\begin{figure*}[t]
  \centering
  \figorbox{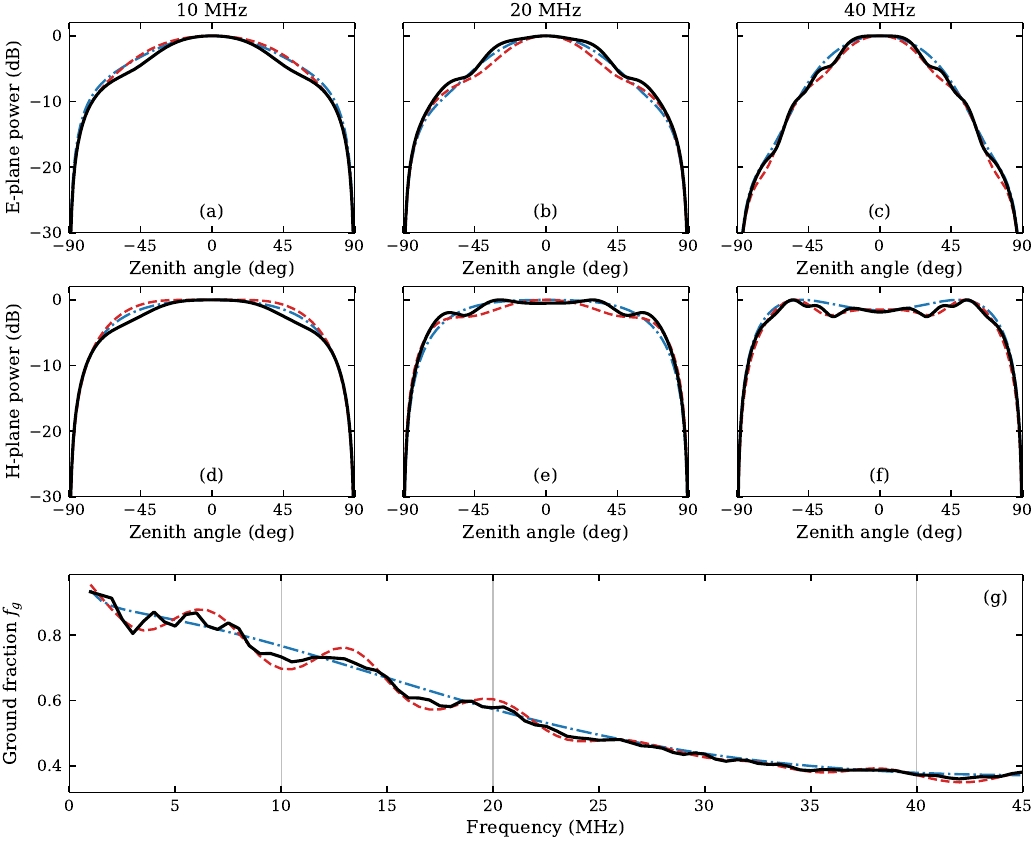}{\textwidth}
  \caption{Beam and ground fraction of a stand-alone 6~m dipole, 2.5~m above the Chang'E-4 reference profile (black, solid), its two-layer reduction (red, dashed) and its half-space reduction (blue, dash-dotted). Panels (a)--(c) show the normalized beam power, Eq.~(\ref{eq:pattern}), in the E-plane at 10, 20 and 40~MHz; panels (d)--(f) show the corresponding H-plane beams. Negative zenith angles indicate directions on the opposite side of the zenith in the same plane. Panel (g) shows the ground fraction, Eq.~(\ref{eq:fg}), from 1 to 45~MHz. Vertical lines mark the frequencies shown in panels (a)--(f).}
  \label{fig:beam}
\end{figure*}

Figures~\ref{fig:beam}(a)--\ref{fig:beam}(c) show the beam in the E-plane, which contains the dipole, at 10, 20 and 40~MHz. Figures~\ref{fig:beam}(d)--\ref{fig:beam}(f) show the corresponding beam in the H-plane, perpendicular to the dipole. The three regolith descriptions give similar beams near the zenith but differ substantially at off-zenith angles of 17--66$^\circ$. Relative to the reference profile, the two-layer reduction changes the normalized beam power by up to 0.26 of its peak in the E-plane and 0.28 in the H-plane. The corresponding differences for the half-space reduction reach 0.17 and 0.30. The subsurface structure therefore shapes the beam, and neither simplified description reproduces it.

Figure~\ref{fig:beam}(g) shows the ground fraction from 1 to 45~MHz. For the reference profile, it decreases from 0.93 at 1~MHz to 0.37 at 40~MHz. At 20~MHz, about 58\% of the antenna power couples into the ground, consistent with \citet{Yousuf2026}. Relative to the reference profile, the two-layer and half-space reductions change the ground fraction by up to 0.053 and 0.067, respectively, with the largest differences near 3--4~MHz. The corresponding changes in input impedance are at most 2.0\% and 2.1\%. The ground fraction oscillates with frequency for the reference profile and its two-layer reduction because of reflections from buried interfaces (Sec.~\ref{sec:ripple}); it varies smoothly for the half-space. The reductions therefore reproduce the input impedance to about 2\% and the ground fraction to within 0.07, but do not reproduce the beam.

\subsection{Errors of the impedance boundary}
\label{sec:ibcantenna}

\begin{figure*}[t]
  \centering
  \figorbox{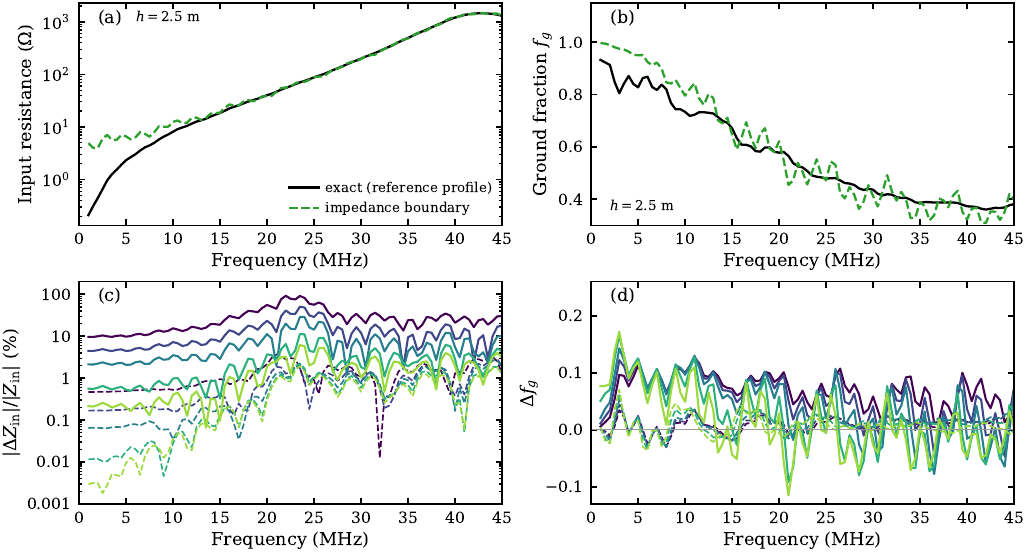}{\textwidth}
  \caption{Impedance-boundary approximation compared with the exact calculation for a stand-alone 6~m dipole above the Chang'E-4 reference profile. Panels (a) and (b) show the input resistance and ground fraction, respectively, at $h=2.5$~m, exact (black, solid) and with the impedance boundary (green, dashed). Panels (c) and (d) show the relative input-impedance error, $|\Delta Z_{\rm in}|/|Z_{\rm in}|$, and the absolute ground-fraction error, $\Delta f_g$, with the latter defined as the approximate value minus the exact value. Solid and dashed curves show the impedance-boundary and half-space results, respectively. The colors indicate the antenna height, from $h=0.3$~m (dark purple) through 0.6~m (blue), 1~m (teal) and 2~m (green) to 3~m (yellow-green).}
  \label{fig:ibc}
\end{figure*}

Figures~\ref{fig:ibc}(a) and \ref{fig:ibc}(b) compare the exact calculation for the Chang'E-4 reference profile with the impedance-boundary approximation of Eq.~(\ref{eq:ibc}), at an antenna height of 2.5~m. Across 1--45~MHz, the approximation gives an input impedance that differs from the exact result by up to 8.1\%, with a median difference of 1.6\%. Below 10~MHz, the input impedance is dominated by its reactance, yet the approximation overestimates the input resistance by factors of 1.4--23 [Fig.~\ref{fig:ibc}(a)]. It also overestimates the ground fraction below 13~MHz, with an absolute error of up to 0.17 [Fig.~\ref{fig:ibc}(b)]. The normalized beam differs from the exact result by up to 0.17 in the E-plane and 0.30 in the H-plane. The errors in input impedance and ground fraction are several times larger than those of the two reductions (Sec.~\ref{sec:reductions}). The beam errors are comparable to those of the two reductions. None of these approximations reproduces the exact beam.

Figures~\ref{fig:ibc}(c) and \ref{fig:ibc}(d) show how the errors change with antenna height, from 0.3 to 3~m. They also illustrate the two distinct limitations of the impedance boundary discussed in Sec.~\ref{sec:formalism}. The relative error in input impedance decreases from 92\% at 0.3~m to 29\% at 1~m and 6.2\% at 3~m [solid curves in Fig.~\ref{fig:ibc}(c)]. Raising the antenna suppresses the evanescent plane-wave contributions through the factor $e^{-2h(\kappa^2-k^2)^{1/2}}$ in Eq.~(\ref{eq:Zr}). This is consistent with the rapid decrease of surface effects with height found by \citet{Rotermund2026}. The error in ground fraction, however, does not decrease: it is 0.13 at 0.3~m and 0.17 at 3~m [solid curves in Fig.~\ref{fig:ibc}(d)].

This difference arises because the ground fraction depends on propagating plane waves at all angles, including those close to grazing incidence. For these waves, the impedance boundary is inaccurate because the refractive index of the upper regolith, $n\approx1.5$--1.8, does not satisfy the large-index condition $|n|\gg1$. Raising the antenna therefore reduces the error associated with evanescent fields but does not remove the error in the beam or ground fraction. By comparison, the half-space description gives relative impedance errors of 1.9--3.7\% and absolute ground-fraction errors of 0.035--0.070 over the same range of heights [dashed curves in Figs.~\ref{fig:ibc}(c) and \ref{fig:ibc}(d)].

\subsection{Effect of the lander}
\label{sec:lander}

LuSEE-Night is planned to fly to the lunar far side on Firefly's Blue Ghost Mission~2 \citep{Firefly2023,FireflyBGM2}. Its four 3~m monopoles will be mounted on a turntable on the top deck of the lander \citep{Bale2023,FireflyBGM2}. The Blue Ghost lander of Mission~1, which landed on the near side in 2025, was 2~m tall, with a span of 3.5~m across its legs \citep{BlueGhost}. We use these dimensions only to set the scale of the model below. The lander body and legs are conducting structures within a wavelength of both the antennas and the regolith, and can change the antenna currents and fields. Since the detailed shapes of the lander body, payload structure and turntable have not been published, we construct an illustrative wire-grid model with representative dimensions (Appendix~\ref{app:mom}). It is not intended to represent the flight hardware.

The model has four 2.9~m monopoles mounted on a hub 2.7~m above the surface, with their tips 3.0~m from the center. It also includes a $1.0\times1.0\times0.6$~m payload cage on the deck, a $2\times2$~m lander body extending from 0.7 to 2.0~m above the surface, and four legs ending in footpads 0.1~m above the surface. The conducting surfaces are represented by wire grids with 0.5~m spacing.

A pseudo-dipole is formed by taking the difference between the signals from two opposite monopoles. The monopoles are physically separate and each is referenced to the lander structure. Their differential response is dipole-like, but they do not form a continuous center-fed wire. We represent this configuration in the moment-method calculation by driving the selected pair with equal and opposite voltages relative to the structure, while leaving the other two monopoles open-circuited. We solve for the currents on the complete antenna and lander structure above the layered ground. Table~\ref{tab:lander} compares this pseudo-dipole with a stand-alone 6~m dipole at the same height.

\begin{table}[t]
  \caption{Antenna response with the illustrative wire-grid lander and for a stand-alone 6~m dipole, both at 2.7~m above the Chang'E-4 reference profile. The input-impedance entries are $(R_{\rm in},X_{\rm in})$ in ohms. The beam column gives the largest difference in normalized power over the upper hemisphere between the two configurations. The lower part lists the maximum errors over 5--45~MHz in ground fraction and input impedance for the impedance-boundary (IBC), two-layer and half-space descriptions, for both configurations.}
  \label{tab:lander}
  \begin{ruledtabular}
  \setlength{\tabcolsep}{3pt}
  \begin{tabular}{rccccc}
  $\nu$ & \multicolumn{2}{c}{$(R_{\rm in},X_{\rm in})$} & \multicolumn{2}{c}{$f_g$} & Beam \\
  (MHz) & lander & alone & lander & alone & \\
  \hline
  5 & $1.5,\,-1599$ & $2.3,\,-1648$ & 0.822 & 0.824 & \\
  10 & $5.6,\,-722$ & $8.0,\,-732$ & 0.723 & 0.718 & 0.014 \\
  20 & $31.7,\,-183$ & $40.4,\,-146$ & 0.612 & 0.553 & 0.043 \\
  30 & $126,\,127$ & $201,\,269$ & 0.606 & 0.418 & 0.16 \\
  40 & $586,\,677$ & $1168,\,423$ & 0.666 & 0.368 & 0.31 \\
  45 & $1679,\,-451$ & $1323,\,-593$ & 0.577 & 0.384 & \\
  \hline
  Error of & \multicolumn{2}{c}{$|\Delta f_g|$} & \multicolumn{2}{c}{$|\Delta Z|/|Z|$ (\%)} & \\
  & lander & alone & lander & alone & \\
  IBC & 0.119 & 0.122 & 14.6 & 6.8 & \\
  two-layer & 0.034 & 0.036 & 3.0 & 1.9 & \\
  half-space & 0.042 & 0.044 & 2.2 & 1.3 & \\
  \end{tabular}
  \end{ruledtabular}
\end{table}

At 5--20~MHz, including the lander changes the input impedance by 1--25\%, the ground fraction by at most 0.06 and the beam by at most 0.043 of its peak. Above 20~MHz, the lander becomes a substantial fraction of a wavelength and has a much larger effect. At 30~MHz, the ground fraction increases from 0.42 to 0.61, and at 40~MHz from 0.37 to 0.67. The input impedance changes by up to 88\%, while the beam changes by 0.16--0.31.

The errors from the approximate regolith descriptions, however, change little when the lander is included. Across 5--45~MHz, the impedance boundary gives an absolute ground-fraction error of up to 0.119 on the lander and 0.122 for the stand-alone dipole. Its normalized beam differs from the exact result by 0.17--0.32 on the lander and 0.16--0.28 for the stand-alone dipole. Including the lander changes the ground-fraction errors of the two-layer and half-space descriptions by at most 0.005 and 0.014, respectively.

The antenna response also depends on how the lander is represented (Appendix~\ref{app:mom}). We test four wire grids with spacings of 0.5 and 0.25~m and wire radii of 2--8~cm. These radii include those obtained from two forms of the equal-area rule for wire grids \citep{Ludwig1987,Rubinstein2005}. Relative to the fiducial grid, the input impedance changes by at most 2.8\% and the ground fraction by at most 0.010 at 10--20~MHz. At 30--40~MHz, these changes reach 11.5\% and 0.030. The grid-related changes in ground fraction are much smaller than the changes caused by including the lander itself, which are 0.18--0.21 at 30~MHz and 0.27--0.31 at 40~MHz. The error in ground fraction from the impedance boundary changes by at most 0.007 across the tested grids. The ground fraction is stable under the grid variations considered here, while the input impedance near 30--40~MHz remains sensitive to the grid and the assumed lander shape.

These comparisons show that the errors of the approximate regolith descriptions persist when the stand-alone dipole is replaced by the antenna on the lander model. The layered ground must therefore be retained when calculating the beam, ground fraction and regolith emission. The lander, in turn, must be included to predict the response of the deployed instrument. In particular, the input impedance and receiver match, as well as the beam and ground fraction above about 20~MHz, depend on the lander structure. The absolute calibrator response and the spectral features introduced by the lander also depend on its detailed geometry. Except where the illustrative wire-grid lander is explicitly included, the results in this paper are predictions for a stand-alone LuSEE-Night-like dipole. A calculation of the flight response will require the deployed antenna and lander geometry, including the stacers, cables and turntable, to be modeled over the layered regolith using the method developed here.

\subsection{Ripple from buried interfaces}
\label{sec:ripple}

A wave reflected from a buried interface acquires an additional phase on its round trip through the overlying material. For an interface at depth $d$ beneath material of refractive index $n$, this phase is $2\xi\simeq2knd$ at near-normal incidence, as described by the factor $e^{2i\xi_j}$ in Eq.~(\ref{eq:parratt}). As the frequency changes, the reflected and direct waves move in and out of phase, producing an oscillation in the antenna coupling. The approximate frequency spacing between successive cycles is
\begin{equation}
  \Delta\nu_d \simeq \frac{c}{2\,n\,d},
  \label{eq:ripple}
\end{equation}
where $c$ is the speed of light.

\begin{figure}[t]
  \centering
  \figorbox{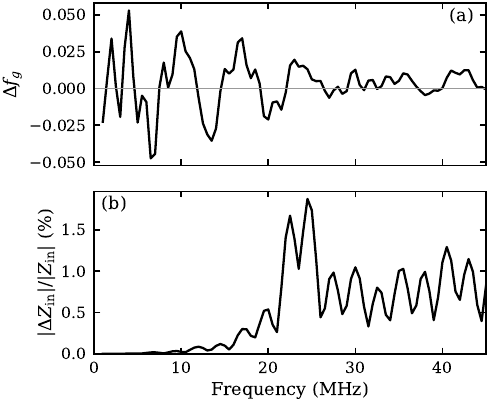}{\columnwidth}
  \caption{Effect of inserting the two buried units at depths of 12--40~m into the Chang'E-4 reference profile, for a stand-alone dipole 2.5~m above the surface. Panels (a) and (b) show the resulting changes in ground fraction and input impedance, respectively.}
  \label{fig:unitsripple}
\end{figure}

For the units mapped by Chang'E-4 beneath the upper 12~m of regolith, taking $n\approx1.80$, interfaces at depths of 12, 24 and 40~m give ripple periods of approximately 7.0, 3.5 and 2.1~MHz, respectively. These periods lie within the frequency range and spectral width relevant to the Dark Ages absorption feature. Including the buried units changes the ground fraction by up to 0.053 and the input impedance by up to 1.9\%. The resulting frequency dependence is visible in Figs.~\ref{fig:unitsripple}(a) and \ref{fig:unitsripple}(b). Above 10~MHz, the ripple has a period of about 2.0~MHz, consistent with reflection from the deepest interface at 40~m.

The buried interfaces produce a measurable frequency structure in the antenna response, even though their effect on the overall ground fraction and input impedance is relatively small. A uniform-layer reduction does not preserve the distinct interface depths, while a single effective surface impedance cannot capture their full contribution to the layered-ground response. Retaining the buried structure is therefore important when calculating the frequency-dependent ground coupling of a far-side radiometer.

\subsection{Depth sensitivity of the antenna response}
\label{sec:depths}

We examine which parts of the regolith contribute most to the antenna response by changing the permittivity of each layer by 1\% in turn, while keeping the other layers fixed. We then sum the absolute changes in each observable over all layers, starting at the surface and moving downward. For the stand-alone dipole at 2.5~m above the Chang'E-4 reference profile, half of this total sensitivity comes from material above depths of 7.4, 5.4 and 4.3~m at 10, 20 and 40~MHz, respectively, for the input impedance. The corresponding depths for the ground fraction are 6.7, 5.7 and 2.8~m. At this height both quantities are mainly sensitive to the upper few meters of the regolith. The evanescent plane-wave contributions are suppressed with height by the factor $e^{-2h(\kappa^2-k^2)^{1/2}}$, limiting the contribution from the shallowest material. When the antenna is within a meter of the surface, the input impedance becomes much more sensitive to the upper decimeters, where the evanescent fields are strongest. These depths indicate where the antenna response is most sensitive to changes in permittivity. They do not represent the vertical resolution of the subsurface structure.

Over the Chang'E-4 reference profile the two reductions thus reproduce the impedance to about 2\% and the ground fraction to within 0.07, but not the beam. The impedance boundary errs by 0.13--0.17 in the ground fraction at every height tested, the buried units add ripples with periods of 2--7~MHz, and the lander changes the response above 20~MHz without removing these errors. The response depends mostly on the top few meters of the regolith, which the passes of the orbiting calibrator could probe.

% =====================================================================
\section{The orbiting calibrator}
\label{sec:calibrator}
% =====================================================================

LuSEE-Night is planned to use a transmitter on the Elytra Dark orbiter of the same mission for calibration \citep{Firefly2023}. The orbiter is required to make at least 30 passes over the landing site within 50 Earth days, with its transmitted flux known to about 1\% \citep{Bale2023}. We examine two possible uses of these observations: calibrating the antenna beam and constraining the properties of the local regolith. We also test how the lander affects the regolith constraints.

As a baseline, we calculate the power received by the stand-alone dipole, using Eq.~(\ref{eq:calib}), along a fiducial pass at 300~km altitude whose ground track passes 10$^\circ$ from the lander. The direction of the calibrator, $(\theta_c,\phi_c)$, is given in the frame of the antenna (Sec.~\ref{sec:geometry}). The zenith angle $\theta_c$ is $90^\circ$ minus the elevation of the calibrator, and the azimuth $\phi_c$ is measured from the dipole axis. We take the calibrator signal to be polarized along $\hat{\bm\theta}$, so that $\hat{\bm e}=\hat{\bm\theta}$ in Eq.~(\ref{eq:calib}), and call this the $\theta$ polarization. Its electric field lies in the vertical plane that contains the antenna and the calibrator, as for the $p$ polarization of Sec.~\ref{sec:basis_and_amp}. By Eq.~(\ref{eq:pattern}), the received power, proportional to $|E_\theta|^2$, then contains the wave reflected by the regolith only through $\rP$. We consider the Chang'E-4 reference profile, its two-layer and half-space reductions, and the impedance-boundary approximation. The orbit and polarization of the calibrator have not been published, so this pass is illustrative. Section~\ref{sec:inputs} examines altitudes of 100--1000~km, ground-track offsets of 0--30$^\circ$, and two other polarizations. These are the $\phi$ polarization, $\hat{\bm e}=\hat{\bm\phi}$, which is horizontal and contains the reflected wave only through $\rS$, and the circular polarization $\hat{\bm e}=(\hat{\bm\theta}-i\hat{\bm\phi})/\sqrt2$, which contains it through both.

\begin{figure*}[t]
  \centering
  \figorbox{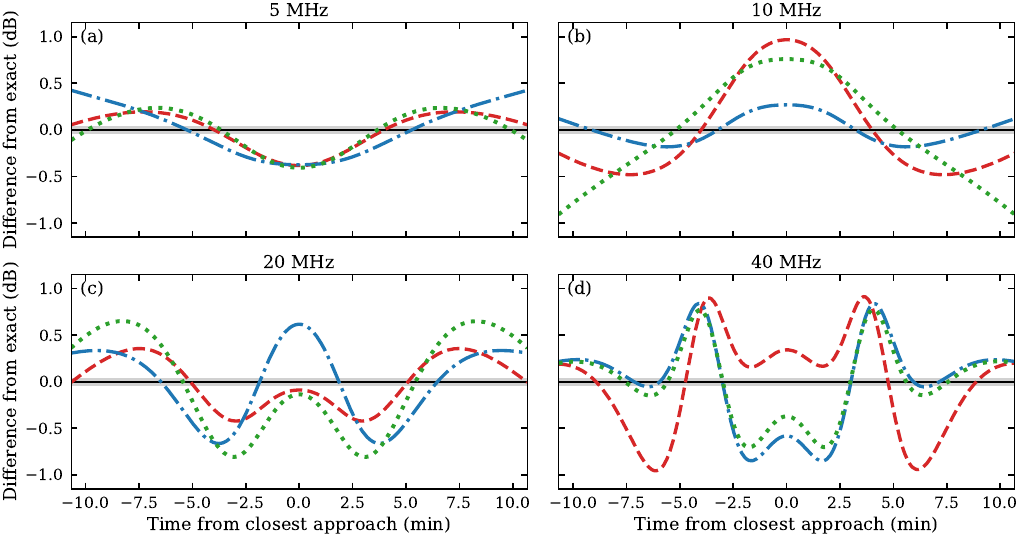}{\textwidth}
  \caption{Calibrator passes over the stand-alone dipole 2.5~m above the Chang'E-4 reference profile. The pass is at 300~km altitude, its ground track passes 10$^\circ$ from the lander, and the calibrator is polarized along $\hat{\bm\theta}$. It stays above 2$^\circ$ elevation for 21~min and reaches 37$^\circ$ at closest approach. Each panel shows, at one frequency, the received power computed with an approximate description of the regolith minus the received power computed exactly, after a constant has been removed from each difference: two-layer reduction (red, dashed), half-space reduction (blue, dash-dotted) and impedance boundary (green, dotted). The black line at zero marks agreement with the exact calculation, and the gray band, $\pm0.04$~dB, the 1\% uncertainty of the calibrator flux.}
  \label{fig:calibrator}
\end{figure*}

The absolute received power includes an unknown gain. We therefore remove a constant offset from each pass at each frequency before comparing the variation in received power along the track. Figure~\ref{fig:calibrator} shows the result at 5, 10, 20 and 40~MHz, the frequencies of the Fisher analysis below. The residual differences vary with calibrator direction because the regolith descriptions give different angular responses. Across these frequencies the largest differences from the exact calculation are 0.4--1.0~dB for the two-layer description, 0.3--0.8~dB for the half-space, and 0.4--0.9~dB for the impedance boundary applied to the reference profile. They come from the reflection coefficient $\rP$, and the change of the current between descriptions adds less than 0.01~dB rms. At closest approach the calibrator is 53$^\circ$ from the zenith, near the Brewster angle of the top unit, 55--61$^\circ$, so the surface reflects the $p$ polarization only weakly and the reflections from the buried units weigh heavily in $\rP$. These add with phases that change with frequency. The exact $|\rP|$ there is 0.01, 0.11, 0.07 and 0.18 at 5, 10, 20 and 40~MHz, while the half-space gives 0.10 at every frequency. Each approximate description therefore errs by a different amount, and with a different sign, at each frequency, which is why the curves change shape from one panel of Fig.~\ref{fig:calibrator} to the next. For comparison, a 1\% uncertainty in the calibrator flux corresponds to about 0.04~dB, shown by the gray band. The differences caused by the approximate regolith descriptions are therefore much larger than the flux-knowledge uncertainty. Accurate knowledge of the calibrator flux alone is not sufficient for beam calibration; the layered-ground response must also be modeled.

The same passes could also constrain the regolith. We estimate this sensitivity using a Fisher-matrix calculation for the four parameters of the fiducial graded profile (Sec.~\ref{sec:farside}), $\varepsilon'_0$, $\varepsilon'_\infty$, $z_g$ and $\tan\delta$. The Fisher matrix gives the uncertainties that a fit to the passes would reach for small Gaussian errors, from the derivatives of the predicted power alone, without simulating and fitting noisy data. Let $P_m$ be the predicted received power in dB for sample $m$, corresponding to a particular calibrator direction, specified by its elevation and azimuth, at a given frequency. We write $\bm\Psi$ for the four regolith parameters and $\bm\Xi$ for the vector of all fitted parameters, $\bm\Psi$ together with the unknown gains, with $\Xi_i$ and $\Xi_j$ any two of its components. For a measurement precision $\sigma_P$ per sample, the Fisher matrix is
\begin{equation}
  F_{ij}=\sum_m\frac{1}{\sigma_P^2}
  \frac{\partial P_m}{\partial\Xi_i}
  \frac{\partial P_m}{\partial\Xi_j},
  \label{eq:fishercal}
\end{equation}
where the sum runs over all samples. A diagonal element $F_{ii}$ measures how strongly the received power responds to $\Xi_i$ compared with the precision per sample, and an off-diagonal element $F_{ij}$ measures how similar the responses to $\Xi_i$ and $\Xi_j$ are. For Gaussian errors and small changes of the parameters, the inverse matrix $F^{-1}$ is the covariance matrix of the fitted parameters (Appendix~\ref{app:stats}). Its diagonal element $(F^{-1})_{ii}$ is the variance of $\Xi_i$ when all other parameters, including the gains, are fitted at the same time. The marginalized $1\sigma$ uncertainty in $\Xi_i$ is therefore $[(F^{-1})_{ii}]^{1/2}$.

These estimates assume that the calibrator flux, position and polarization are known, the gain remains constant during each pass, and the samples are independent. They also assume that the antenna response, apart from the effect of the regolith, is known. Under these idealized conditions, we take one pass with measurements at 5, 10, 20 and 40~MHz, four frequencies an octave apart, each sampled at 241 directions between rising and setting at 2$^\circ$ elevation, with a precision of $\sigma_P=0.1$~dB per sample. This pass gives uncertainties of $\pm0.07$ in the surface permittivity, $\pm0.30$ in the deep permittivity, $\pm1.5$~m in the grading depth and $\pm0.014$ in the loss tangent. These uncertainties scale linearly with $\sigma_P$ and inversely with the square root of the number of samples along the track. In each orbit examined in Sec.~\ref{sec:inputs}, the $\theta$ polarization gives the largest of the three uncertainties in $\varepsilon'_0$, so the fiducial choice is the conservative one. Combining 30 passes whose closest approaches are spaced by 12$^\circ$ in azimuth, and allowing an independent gain for each pass and frequency, reduces the uncertainties to $\pm0.013$, $\pm0.054$, $\pm0.27$~m and $\pm0.0025$, respectively, close to the factor $\sqrt{30}$ expected for 30 independent passes. Passes from different directions could therefore provide useful constraints on the regolith, provided the antenna response is known. These estimates apply to the stand-alone dipole and are not forecasts for the complete instrument.

The assumption that the antenna response is known is important. When the antenna is mounted on the lander, the body, mount and turntable change its response. If these effects are omitted, their contribution can be mistaken for a change in the regolith. We first test whether this unknown response can be represented by a smooth angular correction. We add a correction to the received power in dB that is fixed in the antenna frame and depends on the direction $(\theta_c,\phi_c)$ of the calibrator. The lander structure does not move relative to the antenna, so the correction is a fixed function of $(\theta_c,\phi_c)$, while different passes cross the sky of the antenna along different tracks. We expand it in real spherical harmonics up to degree 2 or 3, which keep only broad structure (Appendix~\ref{app:calharm}). Broad structure is what we expect at 5--20~MHz, where the lander, a few meters across, is small compared with the wavelengths of 15--60~m (Sec.~\ref{sec:lander}). Each harmonic has a constant coefficient and a coefficient of $\ln(\nu/20~{\rm MHz})$, 18 or 32 coefficients in all. A correction in dB that is linear in $\ln\nu$ multiplies the received power in each direction by a power law in frequency, the simplest smooth change with frequency beyond a constant. We write it about 20~MHz, the reference frequency of the gain and foreground models of Sec.~\ref{sec:bias}, so that the constant coefficients give the correction at 20~MHz. The same coefficients apply to every pass, for antennas held at one turntable angle. We fit these coefficients together with the regolith parameters and the gains.

For a single pass, this angular correction is strongly correlated with the regolith parameters. One pass samples the antenna frame only along one track, and along it a smooth function of direction can mimic the change in received power that a change of the regolith causes. Including harmonics up to degree 2 or 3 increases the parameter uncertainties by factors of 2.3--2.8. With 30 passes at different azimuths, the calibrator samples a wider range of directions in the antenna frame, reducing this degeneracy. Including harmonics up to degree 3 gives uncertainties of $\pm0.031$ in the surface permittivity, $\pm0.12$ in the deep permittivity, $\pm0.61$~m in the grading depth and $\pm0.0059$ in the loss tangent. These are 2.2--2.4 times the corresponding uncertainties without the angular correction. Multiple passes therefore help separate a smooth, antenna-fixed response from the regolith contribution, although the resulting constraints remain dependent on how well the angular correction represents the actual antenna response.

We test this representation using the illustrative lander model of Sec.~\ref{sec:lander}, with the fiducial graded profile, against a stand-alone dipole at the same height, 2.7~m. Along the same 30 passes, after fitting an independent gain for each pass and frequency, the received power for the antenna on the lander differs from that for the stand-alone dipole by 0.03~dB rms at 5~MHz, 0.11~dB at 10~MHz, 0.58~dB at 20~MHz and 6.2~dB at 40~MHz. Up to 20~MHz, the angular variation is smooth: harmonics up to degree 3, fitted separately at each frequency, leave a residual of at most 0.014~dB. At 40~MHz, even harmonics up to degree 6 leave a residual of 0.36~dB. The lander response also grows with frequency much faster than the constant-plus-logarithmic form can describe. That form changes by the same amount in each direction from one octave to the next, whereas the rms difference grows by factors of 4, 5 and 11 from 5 to 10, from 10 to 20 and from 20 to 40~MHz.

When we restrict the analysis to 5--20~MHz and fit the lander response with the same harmonic representation, together with the regolith parameters and the gains, the inferred regolith parameters are biased by 4--27 times their formal uncertainties (Appendix~\ref{app:calharm}). Allowing independent harmonic coefficients at every frequency, in place of the constant and logarithmic terms, can absorb the lander response, but it also removes much of the information needed to constrain the regolith. With harmonics up to degree 3 the biases fall to at most about half of the uncertainties, but the uncertainty in surface permittivity increases to $\pm1.7$. A more flexible angular correction can therefore reproduce the lander's effect, but cannot reliably distinguish it from the regolith response.

The calibrator can thus constrain the local regolith only if its observations are analyzed with an electromagnetic model of the deployed antenna and lander over the layered ground. A generic smooth correction to the beam is not sufficient. In Sec.~\ref{sec:timeforecast}, we use the 1\% calibrator-flux uncertainty and the Fisher constraints for the stand-alone dipole as idealized priors in the global-signal analysis. These provide a useful baseline, but should not be interpreted as expected calibration uncertainties for the flight instrument until its antenna and lander response are included.

% =====================================================================
\section{Thermal emission from the regolith}
\label{sec:thermal}
% =====================================================================

The regolith contributes thermal emission to the antenna temperature through the term $f_gT_{\rm eff}$ in Eq.~(\ref{eq:TA}). This contribution can amount to tens or hundreds of kelvin across the observing band, so its frequency dependence must be modeled accurately. During the lunar night, the surface cools to about 90--100~K, while the regolith below about 0.4~m remains near 250~K \citep{Hayne2017}. Temperature gradients of several kelvin per centimeter have also been measured in situ in the upper 10~cm \citep{ChaSTE2025}. We calculate the effective ground temperature for such a cooled regolith and examine how neglecting the surface cooling biases the recovered Dark Ages signal.

\begin{figure*}[t]
  \centering
  \figorbox{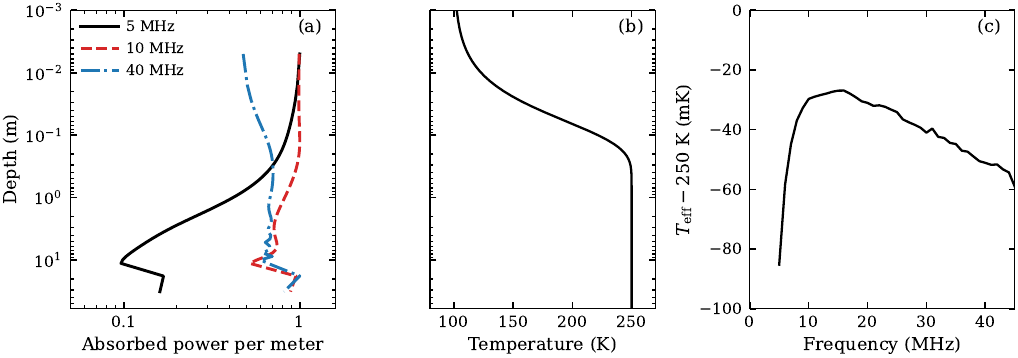}{\textwidth}
  \caption{Thermal weighting for a stand-alone dipole 2.5~m above the Chang'E-4 reference profile. (a) Power absorbed per unit depth, normalized to its maximum, at 5~MHz (black, solid), 10~MHz (red, dashed) and 40~MHz (blue, dash-dotted). (b) Adopted nighttime temperature profile, Eq.~(\ref{eq:Tprofile}). (c) Effective ground temperature relative to the deep temperature of 250~K, over 5--45~MHz.}
  \label{fig:thermal}
\end{figure*}

By Eq.~(\ref{eq:Teff}), $T_{\rm eff}$ is the temperature profile averaged over the fractions $f_j$ of the ground power absorbed in each layer. We calculate these fractions using the method of Sec.~\ref{sec:absorption}, for a stand-alone dipole 2.5~m above the Chang'E-4 reference profile. Evanescent waves deposit their power close to the surface, while propagating waves reach greater depths, particularly at lower frequencies. The depth distribution of absorbed power therefore changes across the band [Fig.~\ref{fig:thermal}(a)].

We adopt a nighttime temperature profile
\begin{equation}
  T(z)=T_{\rm deep}-(T_{\rm deep}-T_{\rm surf})\,e^{-|z|/z_T},
  \label{eq:Tprofile}
\end{equation}
with $T_{\rm surf}=100$~K, the value adopted for the LuSEE-Night site \citep{Saliwanchik2024} (95~K at the equator, \citet{Williams2017}), $T_{\rm deep}=250$~K and an e-folding depth $z_T=6$~cm. This depth lies within the 4--7~cm diurnal layer inferred by \citet{Hayne2017} [Fig.~\ref{fig:thermal}(b)]. With the deeper substrate at $T_{\rm deep}$, Eq.~(\ref{eq:Teff}) gives
\[
  T_{\rm eff}=T_{\rm deep}-(T_{\rm deep}-T_{\rm surf})\,W,
  \qquad
  W=\sum_j f_j\,e^{-|z_j|/z_T},
\]
where $W$ measures the contribution of the cold near-surface layers to the absorbed ground power. For this antenna, 64--93\% of the ground power is absorbed below 40~m, while only 0.2--0.4\% is absorbed in the upper 0.5~m over 5--45~MHz. Consequently, $W$ is only $(1.8$--$5.7)\times10^{-4}$. The effective ground temperature is therefore just 27--85~mK below the deep temperature and varies by 0.06~K across the band [Fig.~\ref{fig:thermal}(c)].

\begin{figure}[t]
  \centering
  \figorbox{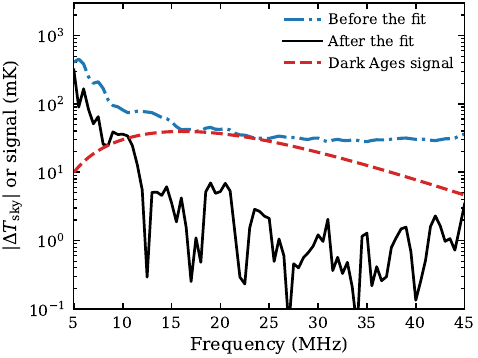}{\columnwidth}
  \caption{Sky-temperature error caused by assuming a uniform ground temperature of 250~K. The blue dash-dotted curve shows the magnitude of the error before fitting the foreground and signal. The black curve shows the magnitude of the residual after the fit. The red dashed curve is the input Dark Ages signal, with an absorption depth of 39.6~mK. The calculation uses the sky spectrum of \citet{NovacoBrown1978}.}
  \label{fig:thermalbias}
\end{figure}

Although this temperature difference is small, its effect on the inferred sky temperature is amplified by the ground fraction. If the analysis assumes a uniform ground temperature of 250~K, the ground contribution in Eq.~(\ref{eq:TA}) is in error by $f_g(T_{\rm eff}-250~{\rm K})$. The corresponding error in the inferred sky temperature is
\[
  \Delta T_{\rm sky}
  =\frac{f_g}{1-f_g}(T_{\rm eff}-250~{\rm K}).
\]
Across 5--45~MHz, the factor $f_g/(1-f_g)$ ranges from 0.6 to 6.6 and is largest at the low frequencies. It amplifies the temperature difference into a smooth, negative spectral error with a magnitude of 28--450~mK (Fig.~\ref{fig:thermalbias}).

We estimate the effect on signal recovery by adding this error to the sky spectrum of \citet{NovacoBrown1978} and an input $\Lambda$CDM Dark Ages signal with an absorption depth of 39.6~mK. We then fit the foreground and signal amplitude as in Sec.~\ref{sec:isotropic}. A five-term foreground model represents the input foreground spectrum exactly in this calculation. The foreground fit removes most of the smooth thermal error, but the remaining spectral structure resembles the Dark Ages absorption feature and biases its fitted amplitude. The recovered amplitude is 202~mK, rather than the input value of 39.6~mK. If the analysis instead allows a single, frequency-independent ground temperature to be fitted, it absorbs the overall offset but not the frequency dependence of $T_{\rm eff}$. The recovered amplitude is then 95~mK.

We also vary the surface temperature over 95--120~K, the deep temperature over 230--250~K, and the e-folding depth over 4--7~cm. The recovered amplitude ranges from 148 to 238~mK when the ground temperature is fixed at 250~K, and from 76 to 107~mK when a constant ground temperature is fitted. The bias increases with the temperature difference between the surface and the deeper regolith.

Thus, the cooling of the upper centimeters must be included in the thermal model, even though it changes $T_{\rm eff}$ by less than 0.1~K. The absorption weights follow from the same electromagnetic calculation used for the input impedance and beam; the additional input is the temperature profile. Orbital thermal measurements could help constrain this profile at the landing site \citep{Williams2017}. In Sec.~\ref{sec:timeforecast}, we include it in the time-resolved analysis and fit the thermal profile together with the foreground, signal, regolith and receiver gain.

% =====================================================================
\section{Recovering the Dark Ages signal}
\label{sec:bias}
% =====================================================================

The regolith contributes to the measured spectrum through the antenna impedance, beam, ground fraction and thermal emission. Errors in these quantities can therefore produce spectral structure that is mistaken for the Dark Ages signal or absorbed into the fitted foreground. We examine how these errors affect the recovered signal amplitude and how accurately the regolith must be modeled to avoid bias. This is an illustrative study under specified models of the sky, receiver gain and regolith, not a sensitivity forecast for the flight instrument. The assumptions are stated with each calculation, and all numerical results are conditional on them.

We report three distinct quantities. Residuals from fits to noiseless spectra measure the systematic mismatch between the simulated data and the analysis model. Statistical uncertainties in the signal amplitude are calculated from Fisher matrices using the radiometer noise. Signal biases measure the shift in the recovered amplitude caused by a model mismatch. We calculate them either to first order using the Fisher matrix or by refitting the simulated spectra.

A radiometer with a high-impedance voltage readout \citep{Bale2023} measures the voltage spectrum
\begin{multline}
  S_V(\nu) = 4k_B\,\mathrm{Re}\,Z_{\rm in}
  \left|\frac{Z_{\rm rx}}{Z_{\rm in}+Z_{\rm rx}}\right|^2\\
  \times\big[(1-f_g)\,T_{\rm sky}+f_g\,T_{\rm eff}\big]+S_{\rm rx},
  \label{eq:SV}
\end{multline}
where $k_B$ is the Boltzmann constant, $Z_{\rm rx}$ is the receiver impedance and $S_{\rm rx}$ is the receiver noise. The quantity in square brackets is the antenna temperature of Eq.~(\ref{eq:TA}), with $T_{\rm sky}$ denoting the sky temperature weighted by the antenna beam.

The regolith affects the measured spectrum in three ways: through the input impedance $Z_{\rm in}$, the ground fraction $f_g$ and the effective ground temperature $T_{\rm eff}$. When the sky is anisotropic, it also affects the beam and hence the sky temperature seen by the antenna. We separate these effects in three steps. Section~\ref{sec:isotropic} considers an isotropic sky, which removes the dependence on the beam and isolates errors in the impedance and ground contribution. Section~\ref{sec:anisotropic} introduces an anisotropic sky and examines the additional effect of an incorrect antenna beam. Finally, Sec.~\ref{sec:timeforecast} uses the time variation of the sky over the lunar sidereal month to examine which errors can be constrained from the changing sky and which remain in the measured spectrum.

\subsection{Isotropic sky}
\label{sec:isotropic}

We first isolate the effect of the regolith by considering an isotropic sky, for which the antenna beam does not affect the measured sky temperature. We calculate the voltage spectrum $S_V$ of Eq.~(\ref{eq:SV}) for the Chang'E-4 reference profile, with a $\Lambda$CDM Dark Ages absorption signal that reaches $-39.6$~mK at 16.1~MHz (redshift 87; Appendix~\ref{app:signal}). We then analyze this synthetic spectrum using several approximate regolith descriptions: the impedance boundary applied to the correct profile, the two-layer description, the two-layer description combined with the impedance boundary, as used in current LuSEE-Night simulations \citep{Yousuf2026}, and the half-space description. In each case, we fit the foreground and signal amplitude together with the regolith parameters when these are allowed to vary. The shape of the input signal is fixed, and only its amplitude $A$ is fitted.

The sky spectrum is taken from \citet{NovacoBrown1978}. In brightness temperature, it is
\begin{equation}
  T_{\rm fg}(\nu)=T_0\Big(\frac{\nu}{\rm 1~MHz}\Big)^{-2.76}
  \exp\!\Big[-3.28\Big(\frac{\nu}{\rm 1~MHz}\Big)^{-0.64}\Big],
  \label{eq:novaco}
\end{equation}
where $T_0=4.49\times10^8$~K. This normalization follows from their unabsorbed specific intensity of $1.38\times10^{-19}$~W\,m$^{-2}$\,Hz$^{-1}$\,sr$^{-1}$ at 1~MHz, converted using the Rayleigh--Jeans relation. The exponential term describes free-free absorption. The spectrum has a brightness temperature of $7.1\times10^4$~K at 20~MHz and was used for the first spectra obtained from the lunar surface \citep{Hibbard2025}.

We fit the foreground with a five-term polynomial in $\log(\nu/20~{\rm MHz})$ for $\log T_{\rm sky}$. Such a smooth form is commonly used in single-spectrum analyses \citep{Bowman2018,Tauscher2020}. Synchrotron emission gives an approximately power-law spectrum, represented by the leading terms, while the higher-order terms describe its curvature, including the turnover due to free-free absorption. For the controlled tests below, we simulate the spectrum using the same five-term representation of the input foreground. This ensures that the biases reported here arise from the regolith model, not from a mismatch between the input foreground and its fitted form. The five-term representation differs from the original Novaco--Brown spectrum by $1.2\times10^{-4}$ of the sky temperature, about 350 times the signal amplitude. Six and seven terms reduce this difference to $5.8\times10^{-6}$ and $2.7\times10^{-7}$, respectively (Appendix~\ref{app:signal}). The effects of foreground-model errors are considered separately in Sec.~\ref{sec:timeforecast}.

The spectra used for these comparisons are noiseless, allowing us to isolate systematic errors from the regolith. To keep the joint refits computationally manageable, we use the sinusoidal antenna current in both the simulated data and the analysis. This approximation changes the ground fraction by at most 0.0013, whereas the approximate regolith descriptions considered here change it by 0.05--0.17.

With the correct regolith profile and exact reflection coefficients, the analysis recovers the input signal amplitude, as expected. The impedance boundary applied to the correct profile leaves a residual of 12\% of the sky temperature, about $10^5$ times the signal amplitude. This mismatch is too large for a meaningful recovery of the signal. The two-layer and half-space descriptions, initially held at their depth-averaged parameters, leave residuals of 6.4--6.6\% of the sky. Allowing their parameters to vary together with the foreground and signal reduces these residuals, but does not remove them: the two-layer description, with or without the impedance boundary, leaves 4.0--4.5\% of the sky, while the half-space leaves 6.3\%. For comparison, when the deeper units are removed from the reference profile, a refitted two-layer description leaves a smaller residual of 0.19\% of the sky, still thousands of times larger than the signal. These results show that adjusting the parameters of an inadequate regolith description cannot compensate for subsurface structure that the model does not represent.

We next use a Fisher analysis to estimate how accurately the regolith parameters would need to be known if they were held fixed in the fit. We use the fiducial graded profile of Sec.~\ref{sec:farside}, with the same sky, signal and antenna. Let $\bm\Theta$ denote the five foreground coefficients and the signal amplitude $A$, and let $\bm\Psi$ denote the four regolith parameters. For an observation time $t_{\rm obs}=1000$~h and frequency channels of width $\Delta\nu=0.5$~MHz, we adopt the radiometer noise
\[
  \sigma_\nu=\frac{S_V}{(\Delta\nu\,t_{\rm obs})^{1/2}}.
\]
The Fisher matrix for the full parameter set $\bm\Xi=(\bm\Theta,\bm\Psi)$ is
\begin{equation}
  F_{ij}=\sum_\nu\frac{1}{\sigma_\nu^2}\,
  \frac{\partial S_V}{\partial\Xi_i}\,
  \frac{\partial S_V}{\partial\Xi_j}.
  \label{eq:fisher}
\end{equation}
If the regolith parameters are fixed in the analysis at incorrect values $\bm\Psi+\Delta\bm\Psi$, the best-fit foreground coefficients and signal amplitude shift. To first order, this shift is
\begin{equation}
  \Delta\bm\Theta=-F_{\Theta\Theta}^{-1}
  F_{\Theta\Psi}\,\Delta\bm\Psi,
  \label{eq:fisherbias}
\end{equation}
where $F_{\Theta\Theta}$ and $F_{\Theta\Psi}$ are blocks of the full Fisher matrix (Appendix~\ref{app:stats}). The signal-amplitude bias calculated this way agrees with direct refits to within 1\%.

To keep the bias in $A$ below 10\% of the input signal, the regolith parameters would have to be known to within $1.6\times10^{-7}$ in surface permittivity, $1.2\times10^{-6}$ in deep permittivity, 4~$\mu$m in grading depth and $4.5\times10^{-8}$ in loss tangent. These requirements are extremely stringent. Changing the number of foreground terms from three to six leaves the requirement on surface permittivity between $1.6\times10^{-7}$ and $6\times10^{-6}$. Thus, fixing the regolith parameters to externally specified values would require unrealistically precise knowledge of the profile.

In practice, the regolith parameters should be fitted along with the foreground and signal. Allowing them to vary increases the marginalized uncertainty in the signal amplitude, $\sigma_A=[(F^{-1})_{AA}]^{1/2}$, from 267 to 331~mK, a 24\% increase. This result is similar with and without the calibrator priors because the spectral imprint of the regolith itself provides substantial information on its parameters. For three to six foreground terms, the increase in $\sigma_A$ ranges from 6 to 51\%. Even with the regolith held fixed at its correct value, the 267~mK uncertainty for five foreground terms and 1000~h is 6.7 times the input signal amplitude. When the regolith is fitted as well, the uncertainty rises to 331~mK, or 8.4 times the signal. A single spectrum therefore has limited ability to distinguish the broad Dark Ages absorption feature from a smooth foreground.

The main requirement is consequently on the form of the regolith model, not on knowing its parameters to extreme precision beforehand. If the model can represent the actual subsurface profile, its parameters can be fitted together with the foreground and signal, at a moderate increase in the signal uncertainty. If the model cannot represent the actual structure, as for the two-layer description applied to the full Chang'E-4 reference profile, refitting its parameters still leaves a residual of about 4\% of the sky temperature. Such a mismatch overwhelms the Dark Ages signal.

\subsection{Anisotropic sky}
\label{sec:anisotropic}

The Galactic foreground is not uniform across the sky, and its angular distribution changes with frequency. Below about 20~MHz, free-free absorption by ionized gas suppresses emission from the Galactic plane. The synchrotron spectral index also varies across the sky \citep{Cong2021}. The antenna therefore receives a foreground spectrum shaped by both the sky pattern and the frequency-dependent beam. Since the regolith affects the beam, an incorrect regolith model can also lead to an incorrect estimate of the beam-weighted sky.

We write the sky temperature in direction $\hat{\bm n}$ as
\[
  T_{\rm sky}(\nu,\hat{\bm n})
  =T_{\rm fg}(\nu)\,\mathcal M(\nu,\hat{\bm n}),
\]
where $T_{\rm fg}$ is the sky-averaged spectrum and $\mathcal M$ describes its angular variation, normalized to have unit mean over the sphere. Our fiducial pattern joins the Ultralong-wavelength Sky Model with Absorption (ULSA) at and below 10~MHz \citep{Cong2021}, which includes free-free absorption, to the Low Frequency Sky Model (LFSM) for the first station of the Long Wavelength Array, LWA1, at and above 22~MHz \citep{Dowell2017}. We interpolate between the two models in each direction (Appendix~\ref{app:signal}).

For this sky, the antenna temperature in Eq.~(\ref{eq:SV}) becomes
\[
  (1-f_g)\big[T_{\rm fg}B+T_{21}\big]+f_gT_{\rm eff},
\]
where $T_{21}$ is the isotropic global signal and
\begin{equation}
  B(\nu,\lambda_z)=
  \frac{\int (dP/d\Omega)\,\mathcal M\,d\Omega}
       {\int (dP/d\Omega)\,d\Omega}
  \label{eq:Bsky}
\end{equation}
is the beam-weighted angular factor. Both integrals extend over the upper hemisphere. Thus, $B$ gives the sky temperature seen by the antenna in units of the sky average. It depends on frequency through both the sky pattern and the antenna beam.

The Moon's spin axis is close to the ecliptic pole, so the sky above a fixed lunar site rotates about the ecliptic pole once per lunar sidereal month. We describe this motion by the ecliptic longitude $\lambda_z$ of the zenith, which plays the role of local sidereal time. For our reference calculation, we take a site at selenographic latitude zero, on the far side opposite the Earth, and average the beams of two crossed dipoles. Each is a stand-alone LuSEE-Night-like dipole at 2.5~m, with its current calculated by the method of moments. The planned LuSEE-Night site is at $23.8^\circ$S and $182.3^\circ$E \citep{Bale2023}, about $24^\circ$ south of our reference site. We also examine sites at latitudes of $30^\circ$S and $30^\circ$N.

We use seven angular patterns in total (Appendix~\ref{app:signal}), all normalized to the same sky-averaged spectrum of Eq.~(\ref{eq:novaco}). They represent alternative descriptions of the angular distribution, not independent measurements of the absolute sky temperature. In the first set of comparisons, we use the same pattern to simulate and analyze the data. This isolates errors caused by the regolith model from those caused by an incorrect sky map. As in Sec.~\ref{sec:isotropic}, the residuals are those of fits to noiseless spectra, and the statistical errors come from the Fisher matrix with the radiometer noise defined there.

\begin{figure*}[t]
  \centering
  \figorbox{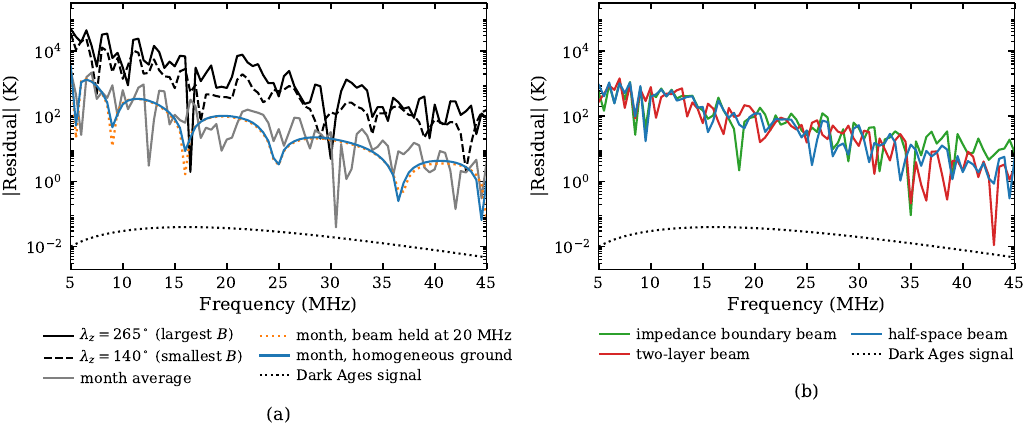}{\textwidth}
  \caption{Residuals after fitting a five-term foreground to spectra simulated over the Chang'E-4 reference profile with the fiducial sky pattern and the Novaco--Brown spectrum, for two crossed stand-alone dipoles at 2.5~m and latitude zero. (a) Residuals when the analysis assumes an isotropic sky ($B=1$), shown at the longitudes of maximum and minimum $B$ and for the spectrum averaged over the lunar month. For the month-averaged spectrum, we also show results with the beam fixed to its 20~MHz shape and with the data simulated over homogeneous ground. (b) Month-averaged residuals when the analysis uses the exact impedance and ground fraction but an approximate beam. The dotted curve shows the input Dark Ages signal.}
  \label{fig:skyresid}
\end{figure*}

At 20~MHz, the beam-weighted sky varies over the lunar month from 0.68 to 1.61 times the sky average. It reaches its maximum at $\lambda_z=265^\circ$, when the Galactic center passes close to the zenith, and its minimum at $\lambda_z=140^\circ$. At 5~MHz, free-free absorption suppresses emission from the Galactic plane, reducing the variation to 0.85--1.22 times the sky average. Across 5--45~MHz, $B$ varies by $-25\%$ to $+31\%$ relative to its value at 20~MHz. Most of this frequency dependence comes from changes in the sky pattern, while the chromatic beam contributes a variation of $-10\%$ to $+8\%$. The choice of regolith model also affects $B$. At a given $\lambda_z$, the values of $B$ obtained with the impedance-boundary and two-layer beams differ from those with the exact beam over the Chang'E-4 reference profile by up to 6.2\%, and those with the half-space beam by up to 5.1\%. These differences are smaller when averaged over the lunar month, ranging from 0.26\% to 0.36\%.

We first examine what happens if the analysis ignores the beam altogether and assumes an isotropic sky, $B=1$. With the correct regolith and beam, the five-term foreground fit recovers the input signal amplitude at every $\lambda_z$. Without the beam, the rms residuals range from $1.4\times10^3$ to $1.1\times10^4$~K at individual longitudes and are 469~K for the spectrum averaged over the month [Fig.~\ref{fig:skyresid}(a)]. These residuals are about $10^4$ times the Dark Ages signal. For the fiducial sky pattern, most of the error comes from its frequency-dependent angular structure. The buried interfaces add further frequency structure through the ripple described in Sec.~\ref{sec:ripple}. The sky pattern must therefore be combined with the frequency-dependent antenna beam when calculating the received foreground, as is also done in ground-based analyses that correct for beam chromaticity \citep{Bowman2018}.

We next isolate the error caused by an approximate beam. We retain the exact input impedance and ground fraction in the analysis, but replace the exact beam with that of each approximate regolith description. For the month-averaged spectrum, the resulting residuals are 296~K for the impedance-boundary beam, 306~K for the two-layer beam and 308~K for the half-space beam [Fig.~\ref{fig:skyresid}(b)]. These are again about $10^4$ times the signal. At individual values of $\lambda_z$, the residuals range from 320 to 9100~K. For sites at latitudes of $\pm30^\circ$, the month-averaged residuals increase to 520--760~K. The differences arise because the approximate descriptions give different reflection coefficients, particularly at oblique incidence, and hence different angular beams.

If the same approximate regolith description is used for the input impedance, ground fraction and beam, the residuals are 12.3\% of the sky for the impedance boundary, 6.7\% for the two-layer description and 6.4\% for the half-space. These values are close to those found for an isotropic sky in Sec.~\ref{sec:isotropic}, because the error in the ground fraction dominates the total mismatch.

We also allow the four parameters of the fiducial graded profile to vary together with the foreground and signal amplitude. For the month-averaged spectrum, this increases the marginalized uncertainty in the signal amplitude from 281 to 345~mK, or by 23\%, similar to the increase found for the isotropic sky. If the regolith is instead held fixed, keeping the bias below 10\% of the signal again requires the surface permittivity to be known to $1.6\times10^{-7}$.

The main results are not specific to the fiducial sky pattern. We repeat the analysis with six alternatives (Appendix~\ref{app:signal}): three other reconstructions based on sky surveys \citep{deOliveiraCosta2008,Polisensky2007,Zheng2017}, a sky calculated from a Galactic model \citep{lowsky2026}, ULSA with a direction-independent spectral index, and the 22~MHz LFSM pattern held fixed across the band. For these patterns, the beam-only residuals range from 290 to 700~K. The requirement on the fixed surface permittivity remains $1.6\times10^{-7}$, and jointly fitting the regolith increases the signal-amplitude uncertainty by 23--24\%. For patterns that vary smoothly with frequency, the residual from ignoring the beam is still 520--710~K. In these cases it is produced mainly by the chromatic beam, particularly the effects of the buried interfaces.

We also test other sky-averaged spectra, random angular errors at the 15\% level on $10^\circ$ scales, representative of differences among current maps, and cases in which the simulated and fitted patterns are different. These changes affect the residuals, with the last set giving 62--1165~K, but do not alter the conclusions (Appendix~\ref{app:signal}).

The beam error from an approximate regolith model and the error from an uncertain sky pattern are therefore comparable, both leaving residuals of several hundred kelvin, about $10^4$ times the signal. A month-averaged spectrum requires the sky pattern and the regolith response to be modeled together with much greater accuracy than is currently available. An accurate regolith model is essential, but is not sufficient on its own: the angular distribution of the Galactic foreground must also be known.

\subsection{Time-resolved analysis}
\label{sec:timeforecast}

A spectrum averaged over the lunar month loses the variation of the beam-weighted Galactic foreground as the Moon rotates. This variation provides a way to distinguish the anisotropic foreground from the isotropic Dark Ages signal \citep{Bale2023,Tauscher2020,Anstey2023}. We therefore simulate the voltage spectra $S_V(\nu,\lambda_z)$ in bins of the ecliptic longitude $\lambda_z$ of the zenith, including the radiometer noise for the observing time in each bin. We use the fiducial graded regolith profile of Sec.~\ref{sec:farside}, represented by 80 layers with a minimum thickness of 5~mm, and the stand-alone crossed dipoles of Sec.~\ref{sec:anisotropic}, with their beam assumed to be known. The simulation is described in Appendix~\ref{app:signal}.

Retaining the time variation improves the statistical precision on the signal amplitude by about an order of magnitude compared with fitting a month-averaged spectrum. It also helps identify errors in the angular distribution of the sky: if the true sky pattern is outside the range allowed by the fitted model, the fit leaves large residuals and a high $\chi^2$. An incorrect sky-averaged spectrum is more difficult to identify. It can be absorbed by the foreground and other fitted parameters, leaving a $\chi^2$ per degree of freedom close to unity while biasing the signal amplitude by hundreds of times its true value.

\begin{table*}[t]
  \caption{Statistical error $\sigma_A$ and bias in the recovered Dark Ages amplitude for the observing scenarios considered here, for the stand-alone crossed dipoles at 2.5~m over the fiducial graded profile. P0--P3 use five foreground terms; the final two columns give $\sigma_A$ and the bias for P4 with seven foreground terms, for data containing the full spectrum of Eq.~(\ref{eq:novaco}). The ranges span the 12 starting seasons for the scheduled observations and the 24 possible bins for the single-bin reference case. Biases are for the median season. P4 with five foreground terms gives results within 1\% of P3. All errors and biases are in mK; the input signal amplitude is 39.6~mK.}
  \label{tab:time}
  \begin{ruledtabular}
  \begin{tabular}{lccccccccc}
  & & & & \multicolumn{4}{c}{Five foreground terms} & \multicolumn{2}{c}{P4, seven terms} \\
  Scenario & Duration & Observing & $\lambda_z$ covered & P0 & P1 & P2 & P3 & $\sigma_A$ & Bias \\
  \hline
  One night & 14.8~d & 354~h & $195^\circ$ & 34--118 & 37--124 & 235--338 & 252--387 & 567--840 & $-183$ \\
  Three nights & 74~d & 1063~h & $253^\circ$ & 20--46 & 21--47 & 135--169 & 144--186 & 327--434 & $-227$ \\
  Three nights, averaged & 74~d & 1063~h & $253^\circ$ & 162--379 & 227--448 & 563--1377 & 2130--3692 & 5561--13125 & $-1520$ \\
  Twelve nights & 340~d & 4253~h & $360^\circ$ & 11 & 12 & 70 & 73 & 171 & $-235$ \\
  Twelve nights, averaged & 340~d & 4253~h & $360^\circ$ & 136 & 174 & 309 & 1654 & 4359 & $-1330$ \\
  \hline
  \multicolumn{10}{l}{Reference integrations of 1000~h, not observing schedules} \\
  One $15^\circ$ bin & & 1000~h & $15^\circ$ & 116--599 & 167--611 & 502--2470 & 1291--10756 & 2874--22257 & $-1030$ \\
  24 bins, equal time & $\geq7$ nights & 1000~h & $360^\circ$ & 23 & 24 & 145 & 150 & 352 & $-235$ \\
  Month average & & 1000~h & $360^\circ$ & 280 & 359 & 632 & 3431 & 9127 & $-1330$ \\
  \end{tabular}
  \end{ruledtabular}
\end{table*}

\paragraph{Observing coverage.}
At a site on the lunar equator, each lunar night lasts approximately 14.8 Earth days, or 354~h. During this interval, the zenith sweeps through \(194.6^\circ\) in \(\lambda_z\). The next lunar night begins \(29.1^\circ\) farther along. We divide each night into 13 bins, each \(15^\circ\) wide (Table~\ref{tab:time}), so that one night samples approximately \(195^\circ\) in zenith longitude. Three consecutive lunar nights span 74~days and cover \(253^\circ\). We repeat this three-night schedule for 12 different starting seasons, which cover different longitudes. We also consider a longer schedule of 12 consecutive lunar nights. Spread over 340~days, these observations cover the full \(360^\circ\), with a total observing time of 4253~h.

We consider three additional reference integrations, each with a total observing time of 1000~h. These are idealized reference cases rather than observing schedules. In the first, all observing time is assigned to a single \(15^\circ\) bin. In the second, the spectrum is averaged over a full lunar month. In the third, the observing time is divided equally among 24 bins spanning \(360^\circ\). This last case represents repeated observations across the full range of zenith longitudes. It cannot be completed within a single lunar sidereal month, which lasts 656~h. Covering all 24 bins requires observations over at least seven lunar nights, since each night spans \(194.6^\circ\) and the starting longitude advances by \(29.1^\circ\) between successive nights.

\paragraph{Parameters fitted in the analysis.}
We introduce the parameter sets P0--P4 in stages to examine how additional uncertainties affect signal recovery. Each set defines a more complete analysis model applied to the same synthetic data. They do not represent different physical scenarios. P0 fits the foreground coefficients and the signal amplitude \(A\). P1 also fits the four parameters describing the regolith profile. P2 adds a smooth, frequency-dependent receiver gain that multiplies the spectrum,
\[
  G(\nu)=\exp\!\left[
  g_0+g_1\ln(\nu/\nu_0)+g_2\ln^2(\nu/\nu_0)
  \right],
\]
where \(\nu_0=20\)~MHz. The three coefficients describe the gain normalization, slope and curvature.

P3 adds eight parameters to account for uncertainties in the angular distribution of the sky. We divide the sky into five regions and allow independent changes in amplitude and spectral index in four of them. These are the inner and outer Galactic plane, at Galactic latitudes within \(10^\circ\) and longitudes within and beyond \(60^\circ\) of the Galactic center, respectively; the intermediate latitudes, \(10^\circ\)--\(30^\circ\) north and south together; and the northern cap, at latitudes above \(30^\circ\). The southern cap is the fifth region and has no separate parameters. In each of the four fitted regions, the change is described by a factor that is linear in \(\ln(\nu/\nu_0)\) and multiplies the fiducial sky pattern. A change common to all five regions is already captured by the foreground spectrum.

P4 adds the three parameters describing the thermal profile in Eq.~(\ref{eq:Tprofile}), namely the surface temperature \(T_{\rm surf}\), the deep temperature \(T_{\rm deep}\) and the e-folding depth \(z_T\). For the fiducial graded profile, the surface-weighting factor \(W\), defined in Sec.~\ref{sec:thermal}, is 0.0013, 0.0005 and 0.0011 at 5, 20 and 45~MHz, respectively. Because the gain is free, its frequency-independent normalization is degenerate with the overall amplitudes of the foreground, signal and ground temperatures. The thermal profile must therefore be constrained by informative priors unless the gain is calibrated in absolute terms. We adopt priors of 10~K on each temperature and 1.5~cm on \(z_T\). We also examine priors twice as wide, as well as profiles in which \(T_{\rm surf}\) and \(z_T\) vary linearly during each night. For these time-varying profiles, we use priors of 20~K and 3~cm, respectively.

We assume that the receiver-noise contribution \(S_{\rm rx}\) in Eq.~(\ref{eq:SV}) is either subtracted, for example by switching to a reference load, or fitted separately. We therefore omit it from the spectral fit. This approach requires the receiver noise to be controlled in the units of the measured data. An independent measurement of the receiver noise in absolute input units would also provide an absolute calibration of the gain.

\paragraph{What time resolution gains.}
We first consider the case in which the sky pattern, receiver gain and regolith are known, so that only the foreground coefficients and signal amplitude are fitted (P0). For three nights of observations, the statistical uncertainty in \(A\) is 20--46~mK, depending on the season and the range of \(\lambda_z\) covered. Averaging the same observations into a single spectrum increases the uncertainty to 162--379~mK. For 12 nights covering the full circle, the uncertainty falls to 11~mK, compared with 136~mK for the month-averaged spectrum. The improvement therefore comes from preserving the time dependence of the sky response, rather than simply from accumulating more observing time.

Allowing the four regolith parameters to vary (P1) increases the uncertainty by only 2--6\% for the time-resolved three-night observations. For the month-averaged spectra, the increase is about 28\%. Retaining the time dependence thus helps separate the effects of the regolith from both the isotropic signal and the anisotropic foreground.

The thermal profile is harder to constrain. When its parameters are fitted without priors, but the receiver gain is assumed to be known, the signal uncertainty for three nights increases by a factor of three to five, reaching 109--155~mK. Changes in the surface temperature produce a smooth variation in \(T_{\rm eff}\) across the observing band, which can be partly confused with the signal. The effect of an incorrect thermal model is even more direct. If the thermal profile is ignored and the analysis assumes the deep temperature throughout the regolith, the recovered signal amplitude is biased by 245--250~mK, about six times the input signal.

\paragraph{Foreground spectral errors.}
Accurate modeling of the foreground spectrum is also essential. We test this by generating synthetic data with the full Novaco--Brown spectrum of Eq.~(\ref{eq:novaco}) and fitting them with polynomials in \(\log\nu\). When the receiver gain and sky pattern are known (P0), a five-term foreground model leaves a signal bias of \(-0.39\) to \(-0.54\)~K, or 10--14 times the input signal. With six terms, the bias falls to \(+42\)--\(+44\)~mK. Seven terms reduce it to less than 1~mK and are therefore sufficient in this case.

The result is very different when the gain is also fitted. Even with seven foreground terms, the residual difference between the polynomial and the full foreground spectrum is only \(2.7\times10^{-7}\) of the sky brightness. However, the additional freedom in the fit allows this small residual to project onto the signal, producing biases of \(-193\) to \(-302\)~mK in P2 and \(-183\) to \(-235\)~mK in P3 and P4. With fewer foreground terms, the biases are much larger. They range from \(-13\) to \(-25\)~K for five terms and from \(+2.9\) to \(+3.3\)~K for six terms. Thus, a foreground model that is adequate when the gain is known may fail when the frequency dependence of the gain is also fitted.

\paragraph{Results when all components are fitted.}
We now consider the most complete analysis model, P4, in which the foreground, signal, regolith, receiver gain, angular sky pattern and thermal profile are all fitted together. With seven foreground terms, the statistical uncertainty for three nights is 327--434~mK, depending on the season, and the median bias across seasons is \(-227\)~mK. For 12 nights, the uncertainty is 171~mK and the bias is \(-235\)~mK. This bias is about six times the input signal amplitude and exceeds the statistical uncertainty. The signal is therefore not recovered reliably in this model.

With the adopted P4 priors, fitting the thermal profile has little effect on the statistical uncertainty. Doubling the prior widths, or allowing the surface temperature and e-folding depth to vary during each night, increases the uncertainty by only about 1\%. When the gain is also free, ignoring the thermal profile produces a much smaller bias in \(A\), between 9 and 19~mK, because the gain absorbs most of the resulting change in the spectrum. This does not make the thermal emission negligible. Rather, its effect is partly degenerate with the gain in this particular analysis.

\paragraph{What the calibrator can constrain.}
We next include the calibration information discussed in Sec.~\ref{sec:calibrator}. As a simple representation of a calibrator with a flux known to 1\%, we initially impose priors of 0.01 on each of the three gain coefficients. This is an assumption of the model. An uncertainty in the calibrator flux that is common to all frequencies affects only the gain normalization \(g_0\). Comparable constraints on the slope \(g_1\) and curvature \(g_2\) also require the frequency dependence of the calibrator flux to be known. We include the regolith constraints inferred from 30 calibrator passes for the stand-alone dipole. For 12 nights, these priors give a statistical uncertainty of 170~mK and a bias of \(-234\)~mK, essentially unchanged from the result without them. The regolith priors have little effect because the time-resolved spectra already constrain the regolith. A prior on \(g_0\) alone also has no effect, even when its width is reduced from \(10^{-2}\) to \(10^{-7}\).

The frequency dependence of the gain is the main limitation. Tightening the priors on all three gain coefficients to \(10^{-3}\) reduces the uncertainty and bias for 12 nights to 144 and \(-170\)~mK, respectively. With priors of \(10^{-4}\), they fall to 65 and \(-17\)~mK. The bias is then smaller than the input signal amplitude of 39.6~mK, although the statistical uncertainty remains larger. At \(10^{-5}\) or tighter, the uncertainty and bias reach 58 and \(-8\)~mK, respectively. Imposing the same priors on \(g_1\) and \(g_2\) alone gives uncertainties and biases that differ by less than 0.3~mK from those obtained when all three coefficients are constrained. The gain normalization is therefore unimportant in this fit, while the slope and curvature must be constrained to about \(10^{-4}\) for the bias to fall below the input signal amplitude. These priors are 100 times narrower than the 0.01 priors adopted to represent the calibrator. They should be understood as a requirement of this illustrative model, not as a demonstrated capability of the calibrator or the instrument. They apply to the smooth gain coefficients and do not imply that every frequency channel must be calibrated independently.

\paragraph{Errors in the angular and sky-averaged foreground.}
The Fisher estimates assume that the true sky is adequately described by the fitted model. We test this assumption by generating additional spectra with alternative sky patterns and fitting them with a nonlinear Gauss--Newton method. We use seven foreground terms and test two sky models. The first has the eight angular parameters of P3. The second is more flexible, with an independent amplitude and spectral index in 11 of 12 equal-area sky regions (Appendix~\ref{app:signal}). We carry out these tests for three nights observed in the season with the median statistical uncertainty and for the twelve-night schedule.

For the fiducial sky pattern and the full Novaco--Brown spectrum, the nonlinear fits give biases consistent with the linearized estimates, \(-216\)~mK for three nights and \(-236\)~mK for twelve nights. The alternative angular patterns give a different result. Spectra generated with any of the other six sky patterns, or with random angular errors, additional spectral curvature or an incorrect free-free optical depth near the Galactic plane, cannot be adequately fitted by either sky model. The resulting \(\chi^2\) per degree of freedom ranges from \(10^3\) to \(10^7\), and remains above \(10^3\) even for the more flexible model. Thus, the time variation makes these angular mismatches apparent rather than allowing them to be hidden in an apparently acceptable fit. However, it does not recover the input signal when the assumed sky pattern is incorrect.

Errors in the sky-averaged spectrum are much harder to identify. A power-law spectrum is reproduced exactly by the seven-term foreground model. Changing the strength of free-free absorption in the Novaco--Brown spectrum produces signal biases of \(-132\) to \(-416\)~mK. A more severe example is the spectrum of \citet{Cane1979}. The seven-term foreground model fits it with \(\chi^2\) per degree of freedom of 0.73--0.99, even though the input spectra contain no noise. Yet the recovered signal amplitude is biased by \(-10.7\) to \(-12.2\)~K, about 300 times the input signal. Time resolution can therefore reveal errors in the angular distribution of the foreground, but cannot by itself identify an incorrect sky-averaged spectrum.

\paragraph{Carbon recombination lines.}
Carbon recombination lines are not included in the smooth foreground model or in the sensitivity analysis above. We examine their effect separately as a screening test (Appendix~\ref{app:signal}). The 569 C\(\alpha\) lines between 5 and 45~MHz produce fluctuations in the absorption, with an rms variation of 7--10\% from one 0.5~MHz channel to the next. This frequency structure cannot be absorbed by a smooth foreground model. With five foreground terms, the lines bias the recovered signal amplitude by \(-5\) to \(-590\)~mK. They must therefore be modeled \citep{Yousuf2026CRRL}, or identified and removed using data with sufficient frequency resolution before the spectrum is averaged.

These tests show that time resolution helps separate an isotropic signal from an anisotropic foreground, but it does not guarantee an unbiased signal measurement. The angular structure of the sky must be modeled accurately, and the sky-averaged spectrum must also be correct. The regolith model must account for the relevant subsurface structure, while the thermal calculation must include emission from the cooled upper centimeters. In the illustrative P4 model, the slope and curvature of the receiver gain must be constrained to about \(10^{-4}\) for the signal bias to fall below the input amplitude. An acceptable fit to time-resolved data is not, by itself, evidence that the sky-averaged foreground has been modeled correctly.

These conclusions are conditional on the fiducial graded regolith profile, the assumed \(\Lambda\)CDM signal shape and the beam of the stand-alone crossed dipoles being known. The lander changes the antenna beam above about 20~MHz (Sec.~\ref{sec:lander}). Analyzing flight data will therefore require an electromagnetic model of the deployed antenna and lander above the layered regolith. The sky model must reproduce both the angular distribution and the sky-averaged spectrum to the accuracy required by the foreground fit. Other effects arising in the receiver and in the data, rather than in the regolith, are discussed in Sec.~\ref{sec:limitations}.

%%%%%%%%%%%%%%%%%%%%%%%%%%%%%%%%%%%%%%%
\section{Discussion}
\label{sec:discussion}

The results show that the electromagnetic response of an antenna above lunar regolith depends on the subsurface structure, the antenna current distribution and the coupling between the antenna and the ground. A description based only on a surface impedance or on the reflection coefficient at normal incidence does not, in general, capture this response. The calculation developed here treats the antenna current self-consistently above a lossy layered medium and provides a common framework for predicting the input impedance, beam, ground fraction and thermal emission. Its numerical consistency is tested against analytical limits, energy conservation and independent calculations. More importantly, the model is tested against the Apollo~17 SEP measurements. A graded profile fitted at 2.1 and 4.0~MHz predicts the measurements at higher frequencies, which were not used in the fit. At Taurus-Littrow, the graded profile describes these held-out data better than the best-fitting two-layer profile. This is a frequency-held-out test along a single traverse. It does not uniquely determine the subsurface profile or rule out every possible two-layer structure.

These results have implications beyond the interpretation of the SEP measurements. They establish a forward model that can be used to study how the regolith affects antenna calibration and low-frequency radiometry, and they identify the accuracy required of the subsurface, sky and instrumental models for recovering a global 21-cm signal. We discuss the implications for LuSEE-Night, examine the sensitivity to inputs that are not specified in published descriptions and to the far-side structure, consider orbiting and surface experiments, and state the limitations of the calculation and what is still needed for predictions of the flight instrument.

\subsection{Implications for LuSEE-Night}

For LuSEE-Night, the results for the stand-alone dipole provide a theoretical baseline. The calculations with the illustrative lander model show how strongly the antenna response can be affected by coupling to the lander (Sec.~\ref{sec:lander}). Neither calculation should be interpreted as a prediction for the flight instrument. Such a prediction, including the use of calibrator passes to constrain the regolith at the landing site (Sec.~\ref{sec:calibrator}), requires the deployed antenna geometry, feed, cables, turntable and lander structure. These details are not available in the published descriptions, and we do not assume them in this work. The calibrator uncertainties in Sec.~\ref{sec:calibrator} are idealized benchmarks for the stand-alone dipole. The harmonic correction examined there is a diagnostic of how an unmodeled lander response can mix with the response of the regolith. It is not a model of the flight lander. Similarly, the gain priors used to represent the calibrator in Sec.~\ref{sec:timeforecast} are adopted assumptions, not demonstrated calibration uncertainties.

The regolith should be included in the analysis as a fitted, parameterized model whose allowed range encompasses the actual subsurface profile (Sec.~\ref{sec:bias}). At Taurus-Littrow, the only site for which we have tested the profile families against Apollo~17 SEP measurements, the fitted profiles place the rise in permittivity at depths of 11.5--18.3~m. The data do not determine how sharply this transition occurs. A suitable model must therefore allow both its depth and sharpness to vary, as the transition model of Eq.~(\ref{eq:transitionfamily}) does (Sec.~\ref{sec:sep}). This result should guide the choice of parameterization, rather than be taken as evidence that the far-side regolith has the same profile.

The calculation is also practical for use in parameter estimation. Evaluating the impedance, beam and ground fraction with the self-consistently solved antenna current takes 0.23~s per frequency on a single processor core. A sweep from 1 to 45~MHz takes about 20~s for the Chang'E-4 reference profile. This makes the calculation suitable for use as a regolith forward model in a joint fit to calibrator observations, the sky spectrum and the global signal.

One conclusion from Sec.~\ref{sec:timeforecast} applies to global-signal experiments more generally. As the sky drifts through the antenna beam, its angular structure changes the measured spectrum with time. This variation can help distinguish an isotropic global signal from an anisotropic foreground. It does not, however, provide an independent check of the sky-averaged foreground spectrum. If an incorrect spectrum lies within the range of the adopted foreground model, it can bias the recovered signal without producing a poor fit. An experiment that uses sky rotation to separate the foreground from the global signal must therefore model both the angular distribution and the sky-averaged spectrum accurately.

For a laterally uniform regolith, the impedance, ground fraction and effective temperature of a stand-alone horizontal dipole are independent of its azimuth. Its beam simply rotates with the antenna. Rotating the antennas on the turntable therefore provides additional information about the regolith only through the changing angular response to the anisotropic sky, or through lateral variations in the regolith that make the impedance azimuth-dependent. On the lander, rotation also changes the positions of the antennas relative to the lander body. Separating these effects requires an electromagnetic model of the deployed antenna and lander.

\subsection{Sensitivity to unpublished inputs}
\label{sec:inputs}

Several inputs needed for quantitative predictions are not specified in the published descriptions. We therefore vary the principal uncertain parameters over ranges that cover plausible configurations. These tests establish which conclusions are insensitive to the missing information and which numerical predictions must be revisited once the instrument and site parameters are known.

First, we vary the height of the stand-alone LuSEE-Night-like dipole from 1.5 to 3.5~m. At 20~MHz, increasing the height lowers the ground fraction from 0.71 to 0.47 and reduces the impedance error of the impedance-boundary approximation from 17\% to 5\%. The corresponding error in ground fraction remains at 0.16--0.17 for the impedance-boundary approximation and at 0.05--0.06 for the two-layer description. Across 7--46~MHz and these antenna heights, the ground fraction ranges from 0.36 to 0.88. For the fiducial graded profile, one calibrator pass then constrains the surface permittivity \(\varepsilon'_0\) to \(\pm 0.04\)--0.07.

The lander, payload structure and turntable are represented by the illustrative model described in Sec.~\ref{sec:lander}. This model is intended to demonstrate the importance of antenna--lander coupling, not to reproduce the flight configuration. Quantitative predictions for the deployed instrument will require its actual geometry, including the antenna supports, cables, feed and surrounding structure.

We also vary the assumed calibrator orbit and observing geometry. For orbital heights of 100--1000~km, ground tracks 0--30$^\circ$ from the lander, and the \(\theta\), \(\phi\) and circular polarizations considered in Sec.~\ref{sec:calibrator}, a single pass measured to 0.1~dB constrains \(\varepsilon'_0\) to between \(\pm 0.02\) and \(\pm 0.11\). The weakest constraints come from the \(\theta\) polarization at each orbital height. Over these configurations, the impedance-boundary approximation misestimates the calibrator response by 0.6--1.2~dB.

The assumed permittivities of the deeper Chang'E-4 units also affect the results. We vary the permittivities of the two deeper units and the substrate in three steps, from 3.5, 3.5 and 5.5 to 5.5, 5.0 and 7.5, respectively. Over this range, the impedance-boundary approximation has errors of 7.5--11\% in impedance and 0.14--0.17 in ground fraction. The disc-averaged reflectivity of the smooth surface ranges from 0.10 to 0.25 across 5--50~MHz.

Finally, we vary the height of the SEP receiving antenna from 1.5 to 3.5~m. This change does not alter the relative ranking of the Taurus-Littrow profile descriptions. In the held-out 8.1--16~MHz comparison, the two-layer description remains behind the graded profile by 36--135 (Appendix~\ref{app:robust}).

Taken together, these variations leave the main conclusions of Secs.~\ref{sec:radiometer}, \ref{sec:calibrator} and \ref{sec:orbit} unchanged over the ranges examined. The precise numerical predictions for LuSEE-Night will depend on the final antenna height and deployed geometry, the orbital observing configuration, and the subsurface profile at the landing site. Once these inputs are specified, the same calculation can be used to evaluate the response of the flight instrument and to include the regolith consistently in the analysis of its calibration and radiometric measurements.

\subsection{Other far-side profiles}
\label{sec:ensemble}

The Chang'E-4 reference profile represents only one possible structure. Radar measurements show that the subsurface changes along the rover's path \citep{Li2020,Chen2022CE4}. To examine how these variations affect the antenna response, we construct 72 profiles in six structural classes, with 12 profiles in each class. All profiles share the top unit of the reference profile. The underlying regolith is 5.7--15.6~m thick \citep{Chen2022CE4}, with a loss tangent of 0.003--0.006.

The six classes represent different types of subsurface structure. The weak-interface profiles contain one to three units whose permittivities differ by less than 0.6. The strong-interface profiles contain two to four units that alternate between \(\varepsilon'=3.5\)--4.2 and 5.5--6.5. The thin-unit profiles contain three to five layers, each 2--5~m thick, as might arise from ejecta deposits. The smooth profiles have a gradual increase in permittivity without distinct interfaces. The sharp profiles have a single sharp interface at the base of the regolith. Finally, the lateral profiles vary the reference structure by up to 2~m in depth and 0.3 in permittivity, comparable to the differences between adjacent radar profiles. The substrate has \(\varepsilon'=5.5\)--7.5. In the weak-interface class, its permittivity is 0.2--0.8 higher than that of the deepest unit.

\begin{table*}[t]
  \caption{Sensitivity of the antenna response to the far-side structure. Each entry gives the range across the 12 constructed profiles in a class for a stand-alone LuSEE-Night-like dipole at 2.5~m height over 5--45~MHz. The columns give the largest impedance error of the impedance-boundary approximation (IBC $|\Delta Z|$), its largest ground-fraction error (IBC $|\Delta f_g|$), and the largest ground-fraction errors of the two-layer (2L) and half-space (HS) reductions. Ripple is the rms variation of the ground fraction about a smooth fit. Reflectivity is the disc-averaged smooth-surface reflectivity at 20~MHz. The directivity ratio is the ratio of the largest to the smallest zenith directivity of a dipole lying on the regolith, evaluated on a 0.1~MHz grid. The beam-only residual is the rms residual of a five-term fit to the month-averaged spectrum when the exact impedance and ground fraction are combined with the beam from the IBC, two-layer or half-space description. The ensemble uses an angular grid of $48\times96$ nodes, which reproduces the beam-only residuals of the reference profile in Sec.~\ref{sec:anisotropic} to 1\%.\label{tab:ensemble}}
  \begin{ruledtabular}
  \setlength{\tabcolsep}{4pt}
  \begin{tabular}{lcccccccc}
  Class & IBC & IBC & 2L & HS & Ripple & Reflectivity & Directivity & Beam only \\
  & $|\Delta Z|$ (\%) & $|\Delta f_g|$ & $|\Delta f_g|$ & $|\Delta f_g|$ & & & ratio & (K) \\
  \hline
  Reference & 8.1 & 0.12 & 0.048 & 0.041 & 0.016 & 0.12 & 2.8 & 297--310 \\
  Weak & 3.5--5.7 & 0.09--0.12 & 0.014--0.045 & 0.014--0.028 & 0.008--0.011 & 0.13--0.15 & 1.3--1.7 & 30--150 \\
  Strong & 7.4--15.3 & 0.14--0.20 & 0.032--0.081 & 0.069--0.116 & 0.023--0.034 & 0.12--0.24 & 3.8--10.6 & 208--477 \\
  Thin & 3.3--10.6 & 0.11--0.20 & 0.040--0.108 & 0.053--0.149 & 0.016--0.035 & 0.11--0.21 & 2.5--6.2 & 65--409 \\
  Smooth & 2.4--3.8 & 0.09--0.12 & 0.050--0.087 & 0.011--0.021 & 0.006--0.010 & 0.12--0.14 & 1.1--1.7 & 8--194 \\
  Sharp & 3.3--9.9 & 0.12--0.14 & 0.005--0.006 & 0.059--0.082 & 0.023--0.033 & 0.09--0.20 & 2.2--3.2 & 2--217 \\
  Lateral & 6.2--9.3 & 0.11--0.14 & 0.040--0.065 & 0.035--0.067 & 0.014--0.019 & 0.11--0.17 & 2.5--3.3 & 255--358 \\
  \end{tabular}
  \end{ruledtabular}
\end{table*}

These profiles are constructed sensitivity cases, not a probability-weighted ensemble drawn from the radar uncertainties. They show how the antenna response changes across the structures considered, but do not indicate how likely any of these structures is at a particular site. Since all 72 profiles share the same top unit, the ensemble also does not capture the full uncertainty in the shallow regolith. This is important because the upper few meters have a strong influence on both the impedance and the ground fraction. For a dipole at 2.5~m, the depth containing half of the sensitivity to either quantity is 2.8--7.4~m (Sec.~\ref{sec:depths}). The ensemble therefore probes variations below the shared top unit rather than the full range of possible far-side structures. A measurement of the upper meters at the landing site, for example through the calibrator observations discussed in Sec.~\ref{sec:calibrator}, would constrain the antenna response more directly than further refinement of the deeper structure. Table~\ref{tab:ensemble} summarizes the range of each quantity across the six classes.

We examine the shallow material separately by varying the top unit of the reference profile over the radar-derived values discussed in Sec.~\ref{sec:farside}. For the Chang'E-4 values, changing the permittivity of the top 30~cm to 2.35 or 2.78 changes the impedance by at most 1.3\% and the ground fraction by at most 0.012. Changing the permittivity of the entire top unit to 2.64 or 3.85 produces changes of up to 3.9\% and 0.068, respectively. We also consider the Chang'E-6 measurements as an effective description of the shallow material at that site. Changing the permittivity of the top 3~m to 2.22, with \(\tan\delta=0.010\), changes the impedance and ground fraction by up to 5.3\% and 0.10, respectively. Across the quoted permittivity range of 1.50--2.94 over the top 1.7 or 3~m, the changes range from 1.3--12\% in impedance and 0.03--0.20 in ground fraction. The loss alone has little effect. Assigning a loss tangent of 0.010 to the entire top unit changes the impedance by only 0.2\% and the ground fraction by 0.002. The attenuation length given in Sec.~\ref{sec:density} is 60--640~m across 5--45~MHz, much greater than the thickness of this unit. For all these variants, the impedance-boundary approximation misstates the impedance by 6.8--12.6\% and the ground fraction by 0.10--0.22 over 5--45~MHz, compared with 8.1\% and 0.12 for the reference profile. The uncertainty in the permittivity of the upper meters, over the range of radar measurements at the two sites, therefore affects the antenna response at least as strongly as the deeper structure.

Some conclusions are common to all six structural classes. Across the 72 profiles, the impedance-boundary approximation misstates the ground fraction by 0.09--0.20 over 5--45~MHz, while the ground fraction at 20~MHz remains between 0.51 and 0.62. Even when the exact impedance and ground fraction are used, the beam from each approximate description leaves a residual of at least 1.8~K in the month-averaged spectrum, more than 40 times the input signal. Other aspects of the response depend more strongly on the detailed structure, particularly on whether the permittivity changes gradually or across sharp interfaces. The impedance error of the boundary approximation ranges from 2.4\% for a smooth increase to 15\% for strong interfaces. The two-layer reduction is nearly exact for a single sharp interface, but its ground-fraction error reaches 0.11 for profiles with thin units. The rms ground-fraction ripple ranges from 0.006--0.011 for weak interfaces and smooth profiles to 0.016--0.035 for strong, thin or sharp structures. The beam-only residuals range from about 2~K, when the approximate description matches the structure, to nearly 480~K. Thus, the size of the spectral ripple and the chromaticity of surface elements depend on the actual subsurface profile. By contrast, the failure of the impedance-boundary approximation and the need for an accurate beam calculation persist across the full set of profiles tested.

\subsection{Orbiting and surface experiments}
\label{sec:secondary}
\label{sec:orbit}
\label{sec:surfacearrays}

The same reflected field affects lunar experiments beyond a radiometer deployed on the surface (Appendix~\ref{app:secondary}). We consider its implications for orbiting radiometers and for arrays of antennas lying on the lunar surface.

\paragraph{Orbiting experiments.}
An orbiter observes the lunar surface over a wide range of incidence angles. Its received reflected sky emission therefore depends on the reflectivity averaged over angle, rather than on the reflectivity at normal incidence alone. For a uniform, smooth regolith with \(\varepsilon'=3.0\), the normal-incidence reflectivity is 7.2\%, consistent with the radar value. However, the reflectivity defined in Eq.~(\ref{eq:TB}) rises toward unity at grazing incidence. Its average is therefore 12.5\% over the lunar disc and 14.0\% over the surface visible from an altitude of 300~km. Across the 72 constructed far-side profiles of Sec.~\ref{sec:ensemble}, the disc-averaged reflectivity at 20~MHz ranges from 0.09 to 0.24.

Below about 90~MHz, the reflected Galactic emission exceeds the Moon's thermal emission, so the lunar brightness depends almost linearly on the reflectivity. Earlier orbiter sensitivity studies used a constant reflectivity of 7\%. This value is based on radar measurements \citep{Evans1969} and already includes the effect of surface roughness. Replacing it with the angle-averaged smooth-surface reflectivity calculated here changes the antenna temperature of a broad beam at 300~km by about \(0.02\,T_{\rm sky}\), or \(7\times10^3\)~K at 10~MHz. These calculations provide a smooth-surface baseline for assessing reflected Galactic emission. They are not a measurement of the reflectivity of the real Moon.

\paragraph{Surface arrays.}
The SEP experiment measured the electromagnetic coupling between two antennas lying on the lunar surface. This is the same coupling that determines interactions between elements of a surface array, such as FARSIDE, FarView or DEX \citep{Burns2021FARSIDE,DAPowerSpec2026,Ghosh2026DEX}. Simulations for DEX show that element displacements caused by deployment and local topography can bias the 21-cm power spectrum by 10--30\% \citep{Ghosh2026DEX}. The layered regolith introduces a separate effect through the beam of each element.

For a wire lying on the surface, the regolith enters the far-field pattern through the factors in Eq.~(\ref{eq:pattern}),
\[
\cos\theta\,(1-\rP{}e^{2ikh\cos\theta})\quad{\rm and}\quad 1+\rS{}e^{2ikh\cos\theta}.
\]
Over homogeneous ground, these factors are frequency-independent and the element beam is achromatic. A layered regolith makes them frequency-dependent. For a dipole 2~cm above the Chang'E-4 reference profile, the zenith directivity relative to the power radiated into the sky varies from 2.4 to 6.7 over 5--45~MHz. For the Taurus-Littrow profiles that fit the SEP measurements, this directivity varies by 1.3--2.5\% when the permittivity rises gradually, but by a factor of 1.6--1.9 when the rise occurs within a meter (Table~\ref{tab:fits}). The chromaticity of a surface-array element therefore depends strongly on the sharpness of the subsurface transition, which the Apollo~17 measurements do not determine (Sec.~\ref{sec:sharpness}).

\subsection{Limitations}
\label{sec:limitations}

The calculation assumes a laterally uniform regolith with planar layers and a smooth surface. Lunar curvature enters only through the geometry. Along the SEP traverse, the diffraction parameter defined in Sec.~\ref{sec:sepdata} is at most 0.12, well within the planar regime. For an orbiter, each point on the lunar disc is treated in its local tangent plane (Sec.~\ref{sec:absorption}). Since the lunar radius exceeds \(10^4\) wavelengths, this approximation fails only within about a degree of grazing emission at the limb, where the contribution to the disc average vanishes. For a surface array with baselines of about 10~km, the diffraction parameter reaches about 0.5 at 40~MHz. Calculating the coupling between widely separated elements in this regime would require a model that includes the curvature and layered structure of the Moon.

Surface roughness provides another limitation. For a surface whose heights follow a Gaussian distribution with rms \(\sigma_h\), the coherent reflection coefficient is reduced by a factor \(\exp[-2(k\sigma_h\cos\theta)^2]\), and the coherent reflectivity by \(\exp[-4(k\sigma_h\cos\theta)^2]\) \citep{Ament1953,Beckmann1963}. At normal incidence, the reflectivity decreases by 0.7\% at 20~MHz and 3.5\% at 45~MHz for \(\sigma_h=0.1\)~m. For \(\sigma_h=0.3\)~m, the corresponding reductions are 6\% and 27\%. The reductions are smaller at oblique incidence. Power removed from the specular direction is redistributed into diffuse reflection and partly enters the regolith, where some of it is absorbed. The factor above describes only the coherent component. Our results therefore provide a smooth-interface baseline. A complete prediction of the angular response, including diffuse and volume scattering, requires a rough-surface scattering model. The 1.1--4.4~dB rms residuals of the SEP fits provide a bound on the combined mismatch from scattering, lateral variations along the traverse and the idealized transmitter and receiver. These effects cannot be separated with the present data. The SEP attenuation measurements also show little lateral heterogeneity on scales of 10~m \citep{Grimm2018}.

We assume that the regolith is nonmagnetic, with relative permeability \(\mu_r=1\). Measurements of Apollo samples at 0.3--5~GHz find magnetic loss tangents comparable to or larger than the dielectric loss tangents below 1~GHz \citep{Linton2026}. However, we know of no measurements of the magnetic permeability or magnetic loss of lunar regolith at 5--45~MHz. A real part of \(\mu_r\) different from unity would multiply \(\varepsilon\) in the wavenumber of each layer. It would also divide the \(s\)-polarized admittance in Eq.~(\ref{eq:admittance}) by \(\mu_r\), so its effect cannot be absorbed exactly into the permittivity. To first order in the losses, the magnetic loss tangent adds to the dielectric loss tangent in determining the attenuation. The upper limit on the loss tangent at Taurus-Littrow (Sec.~\ref{sec:sepfits}) therefore applies, to this order, to their sum. For the far-side calculations, the loss is an adopted input. Increasing the loss tangent of the top unit by a factor of 2.5 has little effect on the response (Sec.~\ref{sec:ensemble}).

The antenna is represented by a thin wire with a delta-gap feed. The feed region, the hub joining the monopoles and their stacer construction are idealized. The feed is represented by a capacitance in parallel with the antenna terminals, and its value depends on how the gap is modeled. Differences between thin-wire codes shift the antiresonance near 43~MHz by about 1~MHz and change the reactance there. They do not significantly change the conductance, ground fraction, normalized beam or change in admittance caused by the ground (Appendix~\ref{app:impedance}). The large changes in the upper-band response caused by the regolith and the lander are therefore robust. However, predicting the complex input impedance near the antiresonance requires the capacitance of the actual feed. For the lander model, the wire-grid representation introduces an uncertainty of about 10\% in the response (Appendix~\ref{app:mom}). The impedances at 40--45~MHz in Table~\ref{tab:lander} should therefore not be interpreted as precise predictions for the flight antenna.

The lander itself is represented by a wire-grid model of representative shape. It does not include the cables, electronics or turntable mechanism. This model identifies which aspects of the antenna response are affected by the lander (Sec.~\ref{sec:lander}), but it does not reproduce the flight hardware. That requires an electromagnetic calculation using the actual deployed structure. The polarization response is also calculated without a polarized sky.

The layered-medium calculation is checked against analytical limits, energy conservation, an alternative integration path in the complex plane and the independent code \texttt{empymod}. These checks cover the SEP fields and the reflected field along the antenna. The antenna current above the layered medium is checked through power balance, while the independent comparison with NEC-2 is limited to a homogeneous half-space. A further test against an independent full-wave calculation of an antenna above layered ground, for example using a volumetric solver, would provide an additional check.

The current distribution of the SEP transmitter is prescribed rather than solved self-consistently. The transmitter is modeled as a half-wave element tuned for a permittivity of 3.3. The channel offsets absorb its overall amplitude but not its distribution along the wire. We test nine current distributions, ranging from a point dipole and a uniform current to elements 20\% shorter or longer than the nominal length, as well as the method-of-moments current calculated over each profile. The ranking of the profile descriptions at both the fitted and held-out frequencies does not change (Sec.~\ref{sec:sepfits} and Appendix~\ref{app:robust}). A more complete model of the transmitter traps and feed would provide a further test.

The signal-recovery study in Sec.~\ref{sec:bias} accounts for errors associated with the regolith, its thermal emission, the sky and a smooth receiver gain. It does not include effects arising in the receiver chain or in the data, such as internal reflections, channel-to-channel gain structure, radio-frequency interference and data gaps. These depend on the design and operation of each experiment and must be addressed in its data analysis \citep{Bowman2018,Singh2022}. To first order, the signal bias is linear in the modeling error. For a fixed set of fitted parameters, these additional effects therefore contribute their own biases on top of those calculated in Sec.~\ref{sec:bias}. Systematic errors in the orbiting calibrator enter through the resulting priors on the gain and regolith. At the widths adopted to represent the calibrator, these priors leave the signal uncertainty and bias essentially unchanged (Sec.~\ref{sec:timeforecast}). We also assume an unpolarized sky and a laterally uniform temperature profile, consistent with the assumed regolith structure.

Finally, the space above the regolith is treated as vacuum. Chandrayaan-3 measured daytime electron densities of 380--600~cm\(^{-3}\) just above the lunar surface \citep{Manju2025}. Even if this plasma filled the entire space above the surface, it would lower the permittivity there by at most 0.1\% at 7~MHz and 1.1\% at 2.1~MHz.

The calculation is not restricted to lunar radiometers. It applies to antennas less than one wavelength above stratified ground, including an antenna floating on a lake above its bed \citep{Raghunathan2021,Singh2022}. Since the Green's function also describes a source within the layered medium, the same framework can be applied to radio pulses from cosmic-ray cascades in the regolith \citep{Costello2025,ARIA2025}. Both applications require appropriate site-specific information and are beyond the scope of this work.

% =====================================================================
\section{Summary and conclusions}
\label{sec:summary}

We have developed a self-consistent calculation of the electromagnetic response of an antenna less than one wavelength above a lossy, layered regolith. The layered-medium Green's function includes both propagating and evanescent fields, and the antenna current is solved over the same layered ground. Within the assumptions of a planar, laterally uniform stack, this provides a common framework for calculating the input impedance, beam, ground fraction and thermal emission of the antenna. The antenna is represented by a thin wire with an idealized feed. We verify the calculation against analytical limits and energy conservation, compare its layered-medium fields with the independent code \texttt{empymod}, and check the antenna calculation against NEC-2 for a homogeneous half-space.

A central result is that the model can be tested against radio measurements made on the Moon. We fit the Apollo~17 SEP fields at 2.1 and 4.0~MHz and compare the resulting predictions with measurements at the held-out frequencies from 8.1 to 32.1~MHz along the same traverse. The best-fitting two-layer profile gives a better fit at the two fitted frequencies. However, its buried interface predicts an interference ripple at the held-out frequencies, most clearly at 8.1 and 16.0~MHz, whose rms is up to 2.6 times that of the measured fluctuations. The graded profile gives a better description of these higher-frequency measurements. This result persists for all tested transmitter current distributions and across the laboratory range of loss tangents considered for the site. It tests the particular two-layer profile selected by the low-frequency fit. It does not rule out all possible two-layer structures.

Across the profile families tested at Taurus-Littrow, the permittivity rises to about 5 at depths of 12--18~m, below the base of the regolith. The two-layer range considered in the LuSEE-Night inference does not reproduce the SEP measurements even at 2.1 and 4.0~MHz, because its deepest permittivity, 4.8, is below the value of about 6 required by the fits. The measurements do not determine whether the rise is gradual or sharp, and the profile at a far-side landing site remains unknown. The Apollo~17 comparison therefore demonstrates why a subsurface model should not be accepted solely because it fits a limited frequency range. Its predictions must also be tested against measurements outside the fitted band.

For a stand-alone LuSEE-Night-like dipole above the radar-informed Chang'E-4 reference profile, the impedance-boundary approximation produces absolute errors of 0.13--0.17 in the ground fraction across antenna heights of 0.3--3.5~m. The beam and ground fraction depend on oblique plane-wave components, whereas the impedance boundary is fixed by the response at normal incidence. It is therefore inadequate for the relatively small refractive indices of 1.5--1.8 in the upper meters of the reference profile. This error persists across the constructed far-side profiles, including variations based on the Chang'E-6 measurements of the top unit, and remains important in the illustrative lander calculation. The beam must therefore be calculated using the layered-medium response rather than inferred from a normal-incidence boundary condition. Quantitative predictions for the flight instrument, including the use of calibrator passes to constrain the regolith at the landing site, require an electromagnetic model of the deployed antenna and lander above the layered ground.

Our signal-recovery study illustrates how errors in the regolith, thermal emission, sky and receiver gain can affect a global 21-cm measurement. It is not a forecast of the flight instrument. Within the adopted models, reliable recovery requires a regolith parameterization that includes the actual subsurface profile and a thermal model that accounts for the cooling of the upper centimeters. Even with 12 lunar nights of time-resolved observations, the most complete fit gives a statistical uncertainty of 171~mK and a signal bias of about \(-235\)~mK, compared with an input signal of 39.6~mK. In this illustrative model, imposing priors of \(10^{-4}\) on the slope and curvature of the smooth receiver gain reduces the bias below the input signal, but does not bring the statistical uncertainty below it. The gain normalization has little effect. These priors are 100 times narrower than the 0.01 priors adopted to represent a calibrator flux known to 1\%. They are a requirement of this particular model, not a demonstrated calibration capability. Time-resolved observations help expose errors in the angular sky pattern, but cannot by themselves identify an incorrect sky-averaged foreground spectrum.

The broader conclusion is that the regolith is an active part of a low-frequency lunar antenna system, not simply a boundary beneath it. Its layered structure affects the antenna current, input impedance, beam, coupling to the ground and thermal emission. These effects must therefore be treated together when interpreting measurements and designing lunar experiments to detect the global 21-cm signal. By combining a self-consistent antenna calculation with a test against Apollo~17 measurements, this work provides a unified framework for addressing these effects in future experiments. It also identifies the measurements and instrument models needed to move from a theoretical baseline to reliable predictions for a deployed lunar radiometer.

\section*{Data availability}
The Apollo~17 SEP measurements analyzed in this article are openly available from the NASA Planetary Data System \citep{SEPdata2021}.

\begin{acknowledgments}
The author acknowledges the Apollo~17 crew and the Surface Electrical Properties (SEP) experiment team for acquiring and documenting the lunar measurements analyzed in this work, and the NASA Planetary Data System for restoring, preserving and providing access to the archived SEP data and associated documentation. This work was supported by the Center for Cosmology and AstroParticle Physics (CCAPP) at The Ohio State University.
\end{acknowledgments}

% =====================================================================
\appendix
% =====================================================================

\section{Numerical integration}
\label{app:numerics}
\label{app:verification}

\paragraph{Integration contours and quadrature.}
All integrals are evaluated along the contour \(C\) defined in Eq.~(\ref{eq:weyl}). On its real segment, we use \(\alpha\) as the integration variable. The integration measure then becomes
\[
\frac{\kappa\,d\kappa}{\kzo}=k\sin\alpha\,d\alpha,
\]
which removes the singularity of \(1/\kzo\) at \(\kappa=k\). On the evanescent segment, the integrand of Eq.~(\ref{eq:Zr}) decays as
\[
e^{-2h(\kappa^2-k^2)^{1/2}}
\]
\citep{Michalski1998}. We terminate the integration when this factor falls to \(e^{-45}\).

For the SEP integrals in Eq.~(\ref{eq:sep}), we divide the integration into four intervals. The first covers \(\kappa<k\), and the second covers the region immediately above \(\kappa=k\). The third extends to \(1.25\,k\max_j(\varepsilon'_j)^{1/2}\). It contains the guided-wave poles of the low-loss regolith, which we sample with a spacing much smaller than their characteristic width of order \(k\tan\delta\). The fourth interval covers the remaining high-\(\kappa\) tail, where the integrand decays as \(e^{-|\kzo|(h_t+h_r)}\). We use composite Gauss--Legendre quadrature, with each panel narrower than half a period of \(J_m(\kappa r_{\max})\), where \(r_{\max}\) is the largest separation considered \citep{Michalski1998,MichalskiMosig1997}.

We test numerical convergence for the 6~m dipole at a height of 2.5~m above the Chang'E-4 reference profile at 5, 20 and 45~MHz. Doubling the number of spectral nodes or the number of tabulated kernel separations changes the impedance and ground fraction by less than \(10^{-12}\). Extending the evanescent integration until the decay factor reaches \(e^{-60}\), or doubling the far-field angular grid, changes these quantities by less than \(10^{-15}\).

\paragraph{Discretization of continuous profiles.}
A continuous regolith profile must be represented by a sufficiently fine set of layers. For the surface-antenna calculations, we represent the Taurus-Littrow profile using 240 layers. Increasing this number to 960 changes the fraction of sky power received within \(45^\circ\) of the zenith by only \(10^{-4}\). The corresponding convergence test for the SEP comparison scores is given in Appendix~\ref{app:robust}.

\paragraph{Comparison with an independent field calculation.}
We compare the SEP fields with those calculated using the independent code \texttt{empymod} \citep{Werthmuller2017}. The comparison includes a lossy half-space and the two-layer and graded profiles fitted at 2.1--4.0~MHz, for all three measured channels.

When both calculations use our integration nodes, the field obtained from \texttt{empymod} agrees with ours to \(1.4\times10^{-8}\) over distances of 50--1500~m and frequencies of 2.1--32.1~MHz. In this comparison, the two codes retain their own reflection coefficients, source and receiver terms, and field assembly. We also compare the results using the independent Hankel transform implemented in \texttt{empymod}, so that the integration and field evaluation are independent as well. Where this transform converges, the agreement is \(4.7\times10^{-4}\) over 50--600~m at 2.1--8.1~MHz and for the graded profile at all frequencies. The difference decreases as the quadrature is refined.

This comparison requires the direct field to be included inside the wavenumber integral. At grazing incidence, the reflected field from a transmitter on the surface nearly cancels its direct field. At 16--32.1~MHz, the Hankel transform in \texttt{empymod} fails to converge at some distances for the half-space and two-layer profiles, regardless of the settings used. The resulting differences reach 8\%. However, the wavenumber-domain field from \texttt{empymod}, evaluated on our integration nodes, still agrees with our calculation. This indicates that the discrepancy arises in the Hankel transform rather than in the wavenumber-domain field calculation.

\paragraph{Independent integration contour.}
We further check the integrals using a second contour. It follows a half-ellipse below the real \(\kappa\) axis from \(\kappa=0\) to \(1.5\,k\max_j|n_j|\), and then continues along the real axis. With the time dependence \(e^{-i\omega t}\), the branch points of air and the lossy substrate, together with the poles of \(\rS\) and \(\rP\), lie on or above the real axis. Deforming the integration path below the real axis therefore does not cross any singularities, and both contours give the same integral. Along the elliptical part, the integrand varies smoothly and stays a distance of order its depth,
\[
\kappa_{\rm e}=\min(0.5k,3/r),
\]
from every singularity.

We compare the two integration paths for the fitted half-space, two-layer, graded and transition profiles at 2.1--32.1~MHz and distances of 50--1500~m. With the numerical settings used in the fits, the real-axis and deformed-contour results agree to \(2.3\times10^{-9}\). With the default settings, the difference is \(2\times10^{-10}\). The result from the deformed contour changes by less than \(2\times10^{-10}\) when its depth is halved or the number of quadrature panels is doubled. The range tabulation used to interpolate the fitted results introduces a relative error of at most \(2.4\times10^{-4}\) in the integrals, which changes the fields by no more than 0.03~dB.

\paragraph{Reflected field along the antenna.}
The same deformed contour provides an independent check of the reflected field along the antenna, \(K_{\rm ref}\) in Eq.~(\ref{eq:Kr}). This field determines the regolith contribution to the moment matrix and is dominated by evanescent waves. We perform the comparison for the Chang'E-4 reference profile, its two reduced descriptions and the variant with the Chang'E-6 values in the top 3~m. We consider antenna heights of 0.3 and 2.5~m and frequencies of 5--45~MHz. The real-axis and deformed-contour calculations agree to \(1.4\times10^{-8}\), relative to the largest field value along the wire. Halving the contour depth or doubling the number of panels changes the deformed-contour result by less than \(3\times10^{-15}\).

We also compare this field with the wavenumber-domain response calculated using \texttt{empymod} on our integration nodes. For the Chang'E-4 reference profile and its two reduced descriptions, at heights of 0.3 and 2.5~m and over 5--45~MHz, the agreement is \(2.8\times10^{-7}\) relative to the largest field value along the wire. The independent Hankel transforms in \texttt{empymod} do not converge for this calculation, because the separations are shorter than the antenna height and the integrand is dominated by the air branch point at \(\kappa=k\).

\paragraph{Power balance.}
Table~\ref{tab:verification} summarizes the numerical verification tests. For a sinusoidal antenna current, the radiation resistance calculated from the far field agrees with the induced-EMF result to \(3\times10^{-8}\). The total power radiated into the sky and transmitted into the ground agrees with the input power to \(3\times10^{-7}\) over a perfect conductor and to \(6\times10^{-6}\) over a lossless dielectric with \(\varepsilon'=3\). The power imbalance decreases with increasing frequency.

For the lossless dielectric, the remaining imbalance arises from the branch point of the half-space at \(\kappa=k\sqrt{\varepsilon'}\). This branch point lies on the evanescent integration path used for \(Z_{\rm ref}\) and slows the convergence of the quadrature. For a lossy regolith, the branch point lies away from the integration path. In this case, quadrupling the number of spectral nodes changes \(Z_{\rm ref}\) by less than \(10^{-11}\).

\section{Plane waves and reflection coefficients}
\label{app:decomp}

This appendix gives the steps from the Hertz vector, Eq.~(\ref{eq:hertz}), to the incident amplitudes, Eq.~(\ref{eq:incident}), and the reflection coefficients of Eq.~(\ref{eq:parratt}).

\subsection{The Weyl identity}

The spherical wave has the two-dimensional Fourier representation \citep{Sommerfeld1909,Weyl1919,Chew1995}
\begin{equation}
  \frac{e^{ikR}}{R}=\frac{i}{2\pi}\int d\kappa_x\,d\kappa_y\,\frac{e^{i(\kappa_xx+\kappa_yy)+i\kzo|z-h|}}{\kzo},
  \label{eq:A_fourier}
\end{equation}
with $\kappa^2=\kappa_x^2+\kappa_y^2$ and $\kzo=(k^2-\kappa^2)^{1/2}$, $\mathrm{Im}\,\kzo\ge0$. Writing $\kappa_x=k\sin\alpha\cos\beta$ and $\kappa_y=k\sin\alpha\sin\beta$ gives $\kzo=k\cos\alpha$ and
\begin{equation}
  \frac{d\kappa_x\,d\kappa_y}{\kzo}=\frac{\kappa\,d\kappa\,d\beta}{\kzo}=k\sin\alpha\,d\alpha\,d\beta,
  \label{eq:A_measure}
\end{equation}
which turns Eq.~(\ref{eq:A_fourier}) into Eq.~(\ref{eq:weyl}). As $\kappa$ runs from 0 to $k$, $\alpha$ runs from 0 to $\pi/2$. As $\kappa$ runs from $k$ to $\infty$, $\alpha$ runs from $\pi/2$ to $\pi/2-i\infty$, with $\sin\alpha=\kappa/k>1$ and $\kzo=i(\kappa^2-k^2)^{1/2}$. This is the contour $C$.

\subsection{The $s$ and $p$ parts of one plane wave}

Below the dipole $|z-h|=h-z$. Inserting Eq.~(\ref{eq:weyl}) in Eq.~(\ref{eq:hertz}) writes the Hertz vector as a sum of plane waves, as in Ref.~\cite{Dasgupta2026},
\begin{equation}
  \bm\Pi=\Pi_0\,\frac{ik}{2\pi}\int_0^{2\pi}\!\!d\beta\int_C\!d\alpha\,\sin\alpha\;e^{i\kzo h}\,e^{ik\kd\cdot\bm r}\,\xhat ,
  \label{eq:A_pw}
\end{equation}
where $\bm r=(x,y,z)$ and each element of the integrand is a plane wave traveling along $\kd$, polarized along $\xhat$. For a plane wave the derivatives in Eq.~(\ref{eq:EHfromPi}) become $\nabla\to ik\kd$. The electric field of each element is then the part of $k^2\bm\Pi$ perpendicular to $\kd$, and its magnetic field follows from it,
\begin{equation}
  \bm E=k^2\big[\bm\Pi-\kd(\kd\cdot\bm\Pi)\big],\qquad \bm H=\frac{1}{\eta_0}\,\kd\times\bm E,
  \label{eq:A_transverse}
\end{equation}
and the same relations hold for an upgoing wave with $\ku$ in place of $\kd$. The vectors of Eq.~(\ref{eq:basis}) are perpendicular to each other and to the direction of the wave, and because $\pd=\shat\times\kd$,
\begin{equation}
  \kd\times\shat=-\pd,\qquad \kd\times\pd=\shat,
  \label{eq:A_cross}
\end{equation}
with the same two relations for $\ku$ and $\pu$. Since $\shat$ and $\pd$ are perpendicular to $\kd$, the components of the electric field on them are those of $k^2\bm\Pi$ itself, and $\shat\cdot\xhat=-\sin\beta$ and $\pd\cdot\xhat=-\cos\alpha\cos\beta$ give Eq.~(\ref{eq:incident}) with the amplitude of Eq.~(\ref{eq:amplitude}). Equations~(\ref{eq:A_transverse}) and (\ref{eq:A_cross}) show that the magnetic field of a $p$ wave is $E_p/\eta_0$ along $\shat$, so the ratio of the reflected to the incident tangential magnetic field, which defines $\rP$ in Sec.~\ref{sec:stack}, equals $E^{\rm ref}_p/E^{\rm inc}_p$ with the reflected field along $\pu$, as in Eq.~(\ref{eq:Ereflected}). With Eq.~(\ref{eq:hertz}) for $\Pi_0$,
\begin{equation}
  \frac{\Amp}{\eta_0}=-\frac{k^2\,Il}{8\pi^2}\,\sin\alpha\,e^{i\kzo h},
  \label{eq:A_prefactor}
\end{equation}
the prefactor used in Appendices~\ref{app:sep} and \ref{app:impedance}.

\subsection{The reflection coefficients of the stack}
\label{app:stack}

In layer $j$ the tangential fields of one polarization are the sum of a downgoing and an upgoing wave with the same $\kappa$ and vertical wavenumbers $\mp k_{z,j}$. As in Ref.~\cite{Dasgupta2026}, let $U$ be the tangential electric field ($s$) or the tangential magnetic field ($p$), and $V$ the other tangential field, normalized by the admittance of Eq.~(\ref{eq:admittance}). Both are continuous across every interface, and across layer $j$
\begin{equation}
  \begin{aligned}
  &\begin{pmatrix}U_{j-1}\\ V_{j-1}\end{pmatrix}=M_j\begin{pmatrix}U_{j}\\ V_{j}\end{pmatrix},\\
  &M_j = \begin{pmatrix} \cos\xi_j & -\dfrac{i}{q_j}\sin\xi_j \\[6pt] -i\,q_j\sin\xi_j & \cos\xi_j \end{pmatrix},
  \end{aligned}
  \label{eq:charmatrix}
\end{equation}
where $U_j$ and $V_j$ are the fields at the bottom of layer $j$. With $M=M_1M_2\cdots M_N$, of elements $M_{11}$, $M_{12}$, $M_{21}$ and $M_{22}$, and a single outgoing wave in the substrate, $V_N=q_{\rm sub}U_N$, the reflection coefficient of the stack is
\begin{equation}
  \rst = \frac{(M_{11}+M_{12}\,q_{\rm sub})\,q_0 - (M_{21}+M_{22}\,q_{\rm sub})}{(M_{11}+M_{12}\,q_{\rm sub})\,q_0 + (M_{21}+M_{22}\,q_{\rm sub})},
  \label{eq:rstack}
\end{equation}
with $q_0$ and $q_{\rm sub}$ the admittances of free space and the substrate \citep{BornWolf1999}. Without layers it is the single-boundary coefficient $(q_0-q_{\rm sub})/(q_0+q_{\rm sub})$. The recursion of Eq.~(\ref{eq:parratt}) is the same result evaluated from the bottom up. At the deepest interface only the substrate lies below, so $\mathcal R_N=r_{N,{\rm sub}}$. Crossing layer $j$ from its lower to its upper boundary multiplies the downgoing amplitude by $e^{i\xi_j}$ and divides the upgoing one by it, so the ratio at the top of the layer is $\mathcal R_je^{2i\xi_j}$, and matching at the interface above gives Eq.~(\ref{eq:parratt}) \citep{Parratt1954}. A complex $\varepsilon_j$ makes $k_{z,j}$ and $\xi_j$ complex and builds the attenuation across each layer into the recursion without further change. The sign of each $r_{j-1,j}$ follows from the admittances, so an arbitrary profile needs no special treatment, and for a perfect conductor the admittances give $\rS=-1$ and $\rP=+1$.

As an independent check we also solve the boundary conditions at all $N+1$ interfaces at once. In each layer we write the field as a downgoing and an upgoing wave, with amplitudes referred to the top and to the bottom of the layer, so that every propagation factor has modulus at most one. Continuity of $U$ and $V$ at each interface gives $2N+2$ linear equations for the reflection coefficient, the $2N$ amplitudes in the layers and the transmission coefficient into the substrate, which we solve directly. We compute both solutions for propagating and evanescent waves, both polarizations, 2--45~MHz and the far-side, two-layer and Taurus-Littrow profiles. The direct solution reproduces the reflection coefficient of the recursion to $8\times10^{-15}$. The power entering the ground, computed from the reflection coefficient alone, equals the power absorbed in the layers plus the power transmitted into the substrate to $2\times10^{-14}$. For $s$ waves the power absorbed in each layer also equals the volume integral $k\,\mathrm{Im}\,\varepsilon_j\int|E|^2\,dz$ of Poynting's theorem to $2\times10^{-15}$, and a lossless stack absorbs nothing to $10^{-15}$. These residuals are numerical. Energy conservation requires the exact differences, and the absorption in a lossless stack, to be zero.

\section{Absorption and emission by depth}
\label{app:emission}

For an $s$ wave, the tangential electric field at the surface is $(1+\rS)E^{\rm inc}_{s}$ and the tangential magnetic field is $(q_0/\eta_0)(1-\rS)E^{\rm inc}_{s}$. The resulting downward Poynting flux is $\tfrac12|E^{\rm inc}_{s}|^2\,\mathrm{Re}[(q_0/\eta_0)(1-\rS)(1+\rS)^{*}]$. For a $p$ wave, the tangential magnetic and electric fields are $(1+\rP)H^{\rm inc}_{s}$ and $\eta_0q_0(1-\rP)H^{\rm inc}_{s}$, respectively. The corresponding downward Poynting flux is $\tfrac12|H^{\rm inc}_{s}|^2\,\mathrm{Re}[\eta_0q_0(1-\rP)(1+\rP)^{*}]$, with $q_0=\kzo/k$ for both polarizations. For a propagating wave incident on a lossless ground, these expressions reduce to the incident flux multiplied by $1-|r^{s,p}_{\rm stack}|^2$. An evanescent wave carries no flux on its own, but its interference with the reflected wave produces a net flux when the ground is lossy. This is the mechanism by which the near field of an antenna close to the surface deposits power in the regolith.

The fields at every interface are obtained by propagating Eq.~(\ref{eq:charmatrix}) upward from the substrate, starting with $U_N=1$ and $V_N=q_{\rm sub}$. The downward flux at the bottom of layer $j$ is proportional to $P_j=\mathrm{Re}(U_jV_j^{*})$. The fraction of the power entering the ground that is absorbed in layer $j$ is therefore
\begin{equation}
  f^{s,p}_j=\frac{P_{j-1}-P_j}{P_0},\qquad f^{s,p}_{\rm sub}=\frac{P_N}{P_0}.
  \label{eq:F_frac}
\end{equation}
For an antenna, the flux associated with each plane wave is weighted by its amplitude from the current transform. Integrating the absorbed flux over $(\alpha,\beta)$ gives the fractions $f_j$ defined in Eq.~(\ref{eq:Teff}). For the Chang'E-4 reference profile made lossless, these fractions vanish to within $2\times10^{-7}$ of the power entering the ground.

By Kirchhoff's law and reciprocity, each absorbing part of the regolith emits thermal radiation with a weight equal to the fraction of the power that the antenna would deposit in that part \citep{TsangKongShin1985}. Summing these contributions gives $f_gT_{\rm eff}$. For a single plane wave leaving the surface at angle $\theta$, the emissivity is $1-|r^{s,p}_{\rm stack}|^2$, and the fraction of the emission originating in layer $j$ is $f^{s,p}_j(\theta)$. Combining this emission with the reflected sky contribution gives Eq.~(\ref{eq:TB}).

For an orbiter, the emission angle is defined in Sec.~\ref{sec:absorption}. The Moon subtends a solid angle of $2\pi(1-\cos\psi_{\rm limb})$, where $\psi_{\rm limb}$ is the nadir angle of the limb. This solid angle is 3.0~sr at an altitude of 300~km. The solid-angle average is obtained by weighting $T_B$ by $2\pi\sin\psi\,d\psi$.

\section{Derivation of the SEP fields}
\label{app:sep}

\subsection{Components of one reflected plane wave}

With $\pu$ and $\shat$ from Eq.~(\ref{eq:basis}), the Cartesian components of the integrand of Eq.~(\ref{eq:Href}) are
\begin{equation}
  \begin{aligned}
  H^{\rm ref}_{x}&=\frac{\Amp}{\eta_0}\cos\alpha\,\sin\beta\cos\beta\,(\rS+\rP),\\
  H^{\rm ref}_{y}&=\frac{\Amp}{\eta_0}\cos\alpha\,(\rS\sin^2\beta-\rP\cos^2\beta),\\
  H^{\rm ref}_{z}&=-\frac{\Amp}{\eta_0}\sin\alpha\,\sin\beta\,\rS,
  \end{aligned}
  \label{eq:C_components}
\end{equation}
each multiplied by the phase $\ph\,e^{i\kzo z}$ with which the plane wave reaches the point $(x,y,z)$. For the transmitter at height $h_t$ and the receiver at $z=h_r$, the factor $e^{i\kzo h_t}$ in $\Amp$ and $e^{i\kzo h_r}$ combine into $e^{i\kzo(h_t+h_r)}$. For a receiver at range $r$ and azimuth $\phi$ the phase along the surface is $\kappa r\cos(\beta-\phi)$. Inserting Eq.~(\ref{eq:A_prefactor}) and using $\sin\alpha\,d\alpha=\kappa\,d\kappa/(k\kzo)$, $\cos\alpha=\kzo/k$ and $\sin\alpha=\kappa/k$ gives
\begin{equation}
  \begin{aligned}
  H^{\rm ref}_{x}&=-\frac{Il}{8\pi^2}\int\!\kappa\,d\kappa\,d\beta\,\sin\beta\cos\beta\,(\rS+\rP)\,\mathcal F,\\
  H^{\rm ref}_{y}&=-\frac{Il}{8\pi^2}\int\!\kappa\,d\kappa\,d\beta\,(\rS\sin^2\beta-\rP\cos^2\beta)\,\mathcal F,\\
  H^{\rm ref}_{z}&=\frac{Il}{8\pi^2}\int\!\frac{\kappa^2}{\kzo}\,d\kappa\,d\beta\,\sin\beta\,\rS\,\mathcal F,
  \end{aligned}
  \label{eq:C_kappa}
\end{equation}
with the phase factor $\mathcal F=e^{i\kappa r\cos(\beta-\phi)}\,\mathcal E$ and $\mathcal E=e^{i\kzo(h_t+h_r)}$ as in Eq.~(\ref{eq:sep}).

\subsection{Integration over the azimuth}

From Eq.~(\ref{eq:besselid}) with $m=0,\pm1,\pm2$ and $J_{-m}=(-1)^mJ_m$,
\begin{equation}
  \begin{aligned}
  \int_0^{2\pi}\!\sin\beta\cos\beta\;\mathcal F\,d\beta&=-\pi J_2\sin2\phi,\\
  \int_0^{2\pi}\!\sin^2\beta\;\mathcal F\,d\beta&=\pi\big(J_0+J_2\cos2\phi\big),\\
  \int_0^{2\pi}\!\cos^2\beta\;\mathcal F\,d\beta&=\pi\big(J_0-J_2\cos2\phi\big),\\
  \int_0^{2\pi}\!\sin\beta\;\mathcal F\,d\beta&=2\pi i\,J_1\sin\phi,
  \end{aligned}
  \label{eq:C_azimuth}
\end{equation}
where every Bessel function has the argument $\kappa r$ and the factor $\mathcal E$ is carried along. Inserting Eq.~(\ref{eq:C_azimuth}) in Eq.~(\ref{eq:C_kappa}) gives the three lines of Eq.~(\ref{eq:sep}).

\subsection{Direct field and perfect-conductor limit}

The field that the dipole radiates directly through free space, added to Eq.~(\ref{eq:sep}), is
\begin{equation}
  \begin{aligned}
  \bm H^{\rm dir}&=\frac{Il}{4\pi}\,\nabla\Big(\frac{e^{ikR}}{R}\Big)\times\xhat\\
  &=\frac{Il}{4\pi}\Big(ik-\frac1R\Big)\frac{e^{ikR}}{R^2}\,(0,\ z-h_t,\ -y),
  \end{aligned}
  \label{eq:C_free}
\end{equation}
with $R$ the distance from the transmitter to the receiver. Over a perfect conductor Eq.~(\ref{eq:sep}) gives $H^{\rm ref}_x=0$ and
\[
H^{\rm ref}_y=\frac{Il}{4\pi}\int_0^\infty\kappa J_0(\kappa r)\,e^{i\kzo(h_t+h_r)}\,d\kappa,
\]
which by the derivative of Eq.~(\ref{eq:A_fourier}) with respect to $z$ is the field of the image dipole, $-Il\,\xhat$ at $z=-h_t$. The numerical implementation reproduces this to $1.2\times10^{-7}$ (Table~\ref{tab:verification}).

\subsection{The finite transmitter}
\label{app:wire}

The field of a finite wire is obtained by superposing the fields in Eq.~(\ref{eq:wire}) from its individual elements. We divide the wire into segments of length $\Delta x'$, using about 20 segments per wavelength along the wire. Each segment is treated as a dipole with moment $I(x'_m)\Delta x'$, and its field is evaluated using its own range $r_m$ and azimuth $\phi_m$ relative to the receiver. The three $\kappa$ integrals in Eq.~(\ref{eq:sep}) depend only on range. We therefore tabulate them once on a grid in $r$, with spacing equal to the smaller of $r/30$ and 1/24 of the free-space wavelength, and interpolate the results using cubic splines. In the SEP geometry, this procedure reproduces direct evaluation to $2.2\times10^{-3}$. A wire oriented along $\hat{\bm y}$ is treated by rotating the coordinates by $90^\circ$.

The standing-wave current vanishes at both ends of the wire. At 1~MHz, the current along the full 70~m wire is $I(x')=1-2|x'|/L$, decreasing linearly toward the tips. A uniform 0.5~m wire reproduces the field of a point dipole with the same moment to $3\times10^{-5}$.

We compare the broadside field of the transmitter in its actual configuration with that of a point dipole above the Taurus-Littrow profile, after removing a constant offset. The difference is 0.22~dB rms at 2.1~MHz, 0.02~dB at 8.1~MHz and less than 0.01~dB at 32.1~MHz. It is confined to distances within the far-field limit of the transmitter element, which is 33~m at 2.1~MHz, where it reaches 2~dB rms. Beyond 50~m, the inner limit of the fits, the difference is 0.11~dB rms at 2.1~MHz and at most 0.74~dB near the interference minima.

A current on a wire lying on the surface propagates with a wavenumber between those of free space and the regolith, $k_{\rm wire}=k[(1+\varepsilon_{\rm tune})/2]^{1/2}$, where the tuning permittivity is $\varepsilon_{\rm tune}=3.3$ \citep{Simmons1972}. A 70~m wire carrying the loaded current $(1-2|x'|/L)\,e^{ik_{\rm wire}|x'|}$ at every frequency, unlike the SEP transmitter, would change the field by up to 0.4~dB.

\section{Input impedance}
\label{app:impedance}

The reflected field produced by an antenna carrying a current $I(x)$ over $|x|\le L/2$ is obtained from Sec.~\ref{sec:formalism}, with the dipole moment replaced by the current distribution. Each current element $I(x')\,dx'$ contributes to the plane-wave spectrum in Eq.~(\ref{eq:weyl}) with an additional phase $e^{-i\kappa_xx'}$. Integrating over $x'$ therefore replaces $Il$ by $\tilde I(\kappa_x)$. From Eq.~(\ref{eq:Ereflected}), the $x$ component of the reflected field due to one plane wave is
\begin{equation}
  \begin{aligned}
  E^{\rm ref}_{x}&=\rS E^{\rm inc}_{s}(\shat\cdot\xhat)+\rP E^{\rm inc}_{p}(\pu\cdot\xhat)\\
  &=\Amp\,\big[\rS\sin^2\beta-\rP\cos^2\alpha\cos^2\beta\big],
  \end{aligned}
  \label{eq:D_Ex}
\end{equation}
where we have used Eq.~(\ref{eq:incident}) and $\pu\cdot\xhat=\cos\alpha\cos\beta$. At the antenna, the plane wave carries the phase $e^{2i\kzo h}\,e^{i\kappa_x x}$. The reaction integral of Sec.~\ref{sec:antennaobs} introduces a second current transform, $\int I(x)\,e^{i\kappa_xx}\,dx=\tilde I(-\kappa_x)$. The current thus enters through the product $\tilde I(\kappa_x)\,\tilde I(-\kappa_x)$. For a real, even current distribution, such as the sinusoidal current, $\tilde I(-\kappa_x)=\tilde I(\kappa_x)^*=\tilde I(\kappa_x)$, and the product becomes $|\tilde I(\kappa_x)|^2$. The solved current is complex. Although its transform remains even in $\kappa_x$ because the current is even in $x$, it is not real. Replacing the product by $|\tilde I|^2$ would therefore change the reflected contribution to the impedance by up to 13\% of $|Z_{\rm in}|$ at 40--45~MHz. By contrast, the power entering the ground (Appendix~\ref{app:mom}) is a flux and depends on $|\tilde I|^2$.

Replacing $Il$ by $\tilde I$ in Eq.~(\ref{eq:amplitude}) and using $\Pi_0$ from Eq.~(\ref{eq:hertz}) gives a prefactor of $k^2\eta_0/(8\pi^2I_{\rm feed}^2)$ in the reaction integral. Changing variables using $\sin\alpha\,d\alpha=\kappa\,d\kappa/(k\kzo)$ removes one power of $k$ and gives Eq.~(\ref{eq:Zr}). The sinusoidal current used in the induced-EMF method,
\[
I(x)=I_{\rm feed}\,\frac{\sin[k(L/2-|x|)]}{\sin(kL/2)},
\]
has the Fourier transform
\begin{equation}
  \tilde I(\kappa_x)=\frac{2kI_{\rm feed}}{\sin(kL/2)}\,\frac{\cos(\kappa_xL/2)-\cos(kL/2)}{k^2-\kappa_x^2},
  \label{eq:D_ft}
\end{equation}
where the limit $\kappa_x\to k$ is taken by continuity.

All calculations use the time dependence $e^{-i\omega t}$. We complex-conjugate the impedances quoted in the tables and figures, so that a capacitive antenna has a negative reactance. The receiver coupling $|Z_{\rm rx}/(Z_{\rm in}+Z_{\rm rx})|^2$ in Eq.~(\ref{eq:SV}) is evaluated with both impedances expressed using the same sign convention. Complex-conjugating both impedances leaves this quantity unchanged.

The delta-gap feed described in Sec.~\ref{sec:mom} adds a capacitance in parallel with the antenna terminals. Its value depends on how the gap is represented. In the numerical solution, the gap is set by the segment length, and the capacitance increases as the segments are shortened. It increases by 0.25~pF in our calculation and by 0.29~pF in NEC-2 when the number of segments is increased from 80 to 160. At 5--10 and 35--45~MHz, the difference between the admittances from NEC-2 and our 80-segment solution is equivalent to the susceptance of a capacitance of 0.41--0.44~pF. This difference is the same with and without the ground. A parallel capacitance changes the antenna susceptance but not its conductance. Its effect is largest near the antiresonance, where the admittance is smallest. In this region, it shifts the frequency at which the reactance changes sign from 43.5~MHz in NEC-2 to 42.5~MHz in our calculation. For this reason, Table~\ref{tab:verification} compares the impedances at 5--30~MHz and the conductances at 35--45~MHz.

The capacitance of the flight antenna is determined by its feed hardware, which is not represented in the thin-wire model. It is in parallel with the receiver input, which has a capacitance of 35~pF for LuSEE-Night (Sec.~\ref{sec:radiometer}), and should therefore be included in the receiver model. Since this capacitance does not depend on the ground, it cancels when calculating the change in admittance caused by the ground. It also has little effect on comparisons between regolith descriptions, which all use the same feed. For example, adding 0.44~pF in parallel changes the impedance error of the impedance boundary at 45~MHz from 3.5\% to 3.8\%, while leaving the ground fraction and beam unchanged.

\section{Far field by stationary phase}
\label{app:spa}

Above the antenna, at a distance $R\gg h,L$ in the direction $\hat{\bm R}=(\sin\theta\cos\phi,\sin\theta\sin\phi,\cos\theta)$, every plane wave of Eq.~(\ref{eq:Ereflected}) carries the phase
\begin{equation}
  kR\,\ku\cdot\hat{\bm R}=kR\big[\sin\alpha\sin\theta\cos(\beta-\phi)+\cos\alpha\cos\theta\big],
  \label{eq:E_phase}
\end{equation}
which is stationary at $\alpha=\theta$, $\beta=\phi$, the plane wave that travels toward the observer. There the phase is $kR$, its second derivatives are $-kR$ with respect to $\alpha$ and $-kR\sin^2\theta$ with respect to $\beta$, and the mixed derivative vanishes. The integral of a slowly varying amplitude $\mathcal A(\alpha,\beta)$ is then
\begin{equation}
  \int\!\!\int \mathcal A\,e^{ikR\,\ku\cdot\hat{\bm R}}\,d\alpha\,d\beta\simeq-\frac{2\pi i}{kR\sin\theta}\,\mathcal A(\theta,\phi)\,e^{ikR},
  \label{eq:E_spa}
\end{equation}
and the factor $\sin\alpha$ in Eq.~(\ref{eq:amplitude}) cancels the $1/\sin\theta$. Only propagating plane waves have a stationary point, so the far field contains none of the evanescent waves. The direct wave reaches the observer with phase $e^{-ikh\cos\theta}$ relative to the origin, and the reflected wave with $e^{+ikh\cos\theta}$. At the stationary point $\pu=\hat{\bm\theta}$ and $\shat=\hat{\bm\phi}$, and Eq.~(\ref{eq:incident}) gives the reflected components $-\rP\cos\theta\cos\phi$ along $\hat{\bm\theta}$ and $-\rS\sin\phi$ along $\hat{\bm\phi}$. Adding the direct field gives Eq.~(\ref{eq:pattern}). Over a perfect conductor both components are proportional to $1-e^{2ikh\cos\theta}$, the array factor of the dipole and its reversed image.

The power radiated into the sky is
\[
P_{\rm sky}=\frac{R^2}{2\eta_0}\int(|E_\theta|^2+|E_\phi|^2)\,d\Omega,
\]
integrated over the upper hemisphere, which with the input power gives Eq.~(\ref{eq:fg}). By the reciprocity theorem, the open-circuit voltage induced by a plane wave incident from $(\theta_c,\phi_c)$ with polarization $\hat{\bm e}$ is proportional to $\hat{\bm e}^{*}\cdot\bm E(\theta_c,\phi_c)$, which is Eq.~(\ref{eq:calib}).

For an extended sky, the power per unit solid angle radiated by the antenna, $dP/d\Omega=(R^2/2\eta_0)(|E_\theta|^2+|E_\phi|^2)$, integrates to $P_{\rm sky}=(1-f_g)P_{\rm in}$, with $P_{\rm in}$ the power accepted by the antenna. By reciprocity the effective area of the same antenna as a receiver is $(c/\nu)^2(dP/d\Omega)/P_{\rm in}$ \citep{Balanis2016}. The sky emits incoherently, so the powers received from different directions add. An unpolarized sky of brightness temperature $T_{\rm sky}$ has the specific intensity $2k_BT_{\rm sky}\nu^2/c^2$ in the Rayleigh--Jeans limit, with $k_B$ the Boltzmann constant, of which the antenna receives the half in its own polarization. The power per unit bandwidth received from the sky, half the integral over directions of the effective area times the specific intensity, is therefore
\begin{equation}
  P_{\nu,{\rm sky}}=\frac{k_B}{P_{\rm in}}\int \frac{dP}{d\Omega}\,T_{\rm sky}\,d\Omega .
  \label{eq:psky}
\end{equation}
The ground contributes $k_Bf_gT_{\rm eff}$ (Appendix~\ref{app:emission}), and dividing the sum by $k_B$ gives Eq.~(\ref{eq:TA}). If the sky and the ground are at one temperature $T$, then $T_A=T$, as thermodynamic equilibrium requires.

\section{The method of moments}
\label{app:mom}

\subsection{One wire}

The free-space contribution to Eq.~(\ref{eq:mom}) is the thin-wire matrix in its mixed-potential form,
\begin{equation}
  \begin{aligned}
  Z^{\rm fs}_{mn}=-ik\eta_0\int\!\!\int&\Big[\Lambda_m(x)\Lambda_n(x')\\
  &-\frac{\Lambda'_m(x)\Lambda'_n(x')}{k^2}\Big]\mathcal G\,dx\,dx',
  \end{aligned}
  \label{eq:F_zfs}
\end{equation}
where the reduced kernel is $\mathcal G=e^{ikR}/4\pi R$, with $R=[(x-x')^2+a^2]^{1/2}$ for a wire of radius $a$. Both terms depend on the matrix indices only through $x_m-x_n$. The matrix is therefore Toeplitz, so only one row needs to be computed.

The double integral in each term can be reduced to a single integral,
\begin{multline}
  \int\!\!\int \Lambda_m(x)\Lambda_n(x')\,\mathcal K(x-x')\,dx\,dx'\\
  =\int (\Lambda\star\Lambda)(u-u_{mn})\,\mathcal K(u)\,du,
  \label{eq:D_corr}
\end{multline}
where $\mathcal K$ is either of the kernels in Eq.~(\ref{eq:F_zfs}), $u_{mn}=x_m-x_n$, and $(\Lambda\star\Lambda)(u)$ is the autocorrelation of a triangular basis function of half-width $\Delta x$. This autocorrelation is a cubic B-spline,
\[
(\Lambda\star\Lambda)(u)=\Delta x\,\big[\tfrac23-(u/\Delta x)^2+\tfrac12|u/\Delta x|^3\big]
\]
for $|u|\le\Delta x$ and
\[
(\Lambda\star\Lambda)(u)=\tfrac16\,\Delta x\,(2-|u|/\Delta x)^3
\]
for $\Delta x\le|u|\le2\Delta x$. The autocorrelation of the derivatives is $-(\Lambda\star\Lambda)''(u)$.

We evaluate the integral over each polynomial interval using Gauss--Legendre quadrature. To handle the sharp peak of $\mathcal G$ at $u=0$, we use the substitution $u=a\sinh t$. Since $R=(u^2+a^2)^{1/2}=a\cosh t$, this gives $du/R=dt$ and removes the peak from the integrand. The reflected-field kernel $K_{\rm ref}$ in Eq.~(\ref{eq:Kr}) is evaluated on the spectral nodes described in Appendix~\ref{app:numerics}, at 241 separations between 0 and $L$. We then interpolate it using cubic splines. For the far field, we interpolate the transform in Eq.~(\ref{eq:momft}) at 4001 points. The interpolated result agrees with direct summation to $2.4\times10^{-15}$ of its largest value.

We verify the solution using several independent checks. First, over a perfect conductor, the reflected field must equal that of an image dipole with moment $-Il$ at depth $h$. Its kernel is
\begin{equation}
  K_{\rm ref}(u)=\frac{i\eta_0}{k}\Big(k^2+\frac{\partial^2}{\partial u^2}\Big)\frac{e^{ikR}}{4\pi R},\qquad R=(u^2+4h^2)^{1/2},
  \label{eq:D_image}
\end{equation}
which Eq.~(\ref{eq:Kr}) reproduces to $2\times10^{-15}$ at 5, 20 and 45~MHz.

We also check power conservation for the solved current. The input power, $\tfrac12\,{\rm Re}\,Z_{\rm in}|I_{\rm feed}|^2$, must equal the sum of the power radiated into the sky and the power entering the ground. We calculate the latter from the plane-wave spectrum,
\begin{multline}
  P_g=\frac{\eta_0}{32\pi^2}\int_0^\infty\!\kappa\,d\kappa\int_0^{2\pi}\!d\beta\,e^{-2\,{\rm Im}(\kzo)h}\,|\tilde I(\kappa_x)|^2\\
  \times\Big[\frac{k^2}{|\kzo|^2}\,{\rm Re}(q^{s}_{\rm in})\,|1+\rS|^2\sin^2\beta\\
  +{\rm Re}(q^{p}_{\rm in})\,|1+\rP|^2\cos^2\beta\Big],
  \label{eq:D_Pg}
\end{multline}
where the normalized input admittances of the stack are (Appendix~\ref{app:emission})
\[
q^{s,p}_{\rm in}=\frac{\kzo}{k}\,\frac{1-r^{s,p}_{\rm stack}}{1+r^{s,p}_{\rm stack}}.
\]
The input power calculated from the Galerkin matrix agrees with the sum of the far-field radiation and the power entering the ground to between $2\times10^{-7}$ and $2\times10^{-5}$. We obtain this agreement for a perfect conductor, a lossless half-space and the Chang'E-4 reference profile at 5, 20 and 45~MHz.

We further compare the free-space Galerkin solution with a Hall\'en solver that uses pulse basis functions and point matching. The impedances agree to 1--4\% at 5--35~MHz and 8--9\% at 40--45~MHz. After removing the effect of a capacitance of 0.23~pF in parallel with the feed, the agreement improves to 0.1--4\% over 5--45~MHz (Appendix~\ref{app:impedance}). For a half-wave dipole with radius $10^{-3}$ of its wavelength, the two methods give resistances of 86 and 88~$\Omega$ and reactances of 47 and 50~$\Omega$, respectively. These values are close to the classical results \citep{Balanis2016}. The comparison with NEC-2 is described in Sec.~\ref{sec:verification}.

The remaining differences near the antiresonance arise from the way the feed is represented in each method. Our delta gap is placed at the central node, NEC-2 applies the field over the central segment, and the Hall\'en solver uses a source at the central collocation point. Changing the number of segments from 80 to 40 or 160 changes the impedance by 0.7--1.9\% up to 30~MHz and by 6--11\% at 40--45~MHz. These changes correspond to a variation of 0.19--0.25~pF in the feed capacitance. In contrast, the ground fraction changes by less than $4\times10^{-5}$ and the normalized beam by less than $8\times10^{-4}$.

\subsection{Wires of arbitrary orientation}

The method can also be applied to structures made of straight wire segments joined at nodes. We expand the current in rooftop basis functions, each spanning two segments that meet at a node. At a junction shared by $N_{\rm node}$ segments, we use $N_{\rm node}-1$ basis functions. This construction ensures that the current remains continuous at every junction and vanishes at free ends.

The free-space matrix is given by Eq.~(\ref{eq:F_zfs}), with the scalar product of the unit vectors along the two segments multiplying the first term. The $1/R$ part of the inner integral can be evaluated analytically for any pair of straight segments. The reflected contribution is expressed as a sum over plane waves,
\begin{multline}
  Z^{\rm ref}_{mn}=\frac{k\eta_0}{8\pi^2}\int_0^\infty\!\frac{\kappa\,d\kappa}{\kzo}\int_0^{2\pi}\!d\beta\\
  \times\big(\rS\,\tilde\Lambda^{s+}_m\tilde\Lambda^{s-}_n+\rP\,\tilde\Lambda^{p+}_m\tilde\Lambda^{p-}_n\big),
  \label{eq:F_zr}
\end{multline}
where
\[
\tilde\Lambda^{s\pm}_m=\sum(\hat{\bm e}_s\cdot\hat{\bm u})\int\Lambda_m\,e^{\pm i\bm k^\pm\cdot\bm r}\,dl,
\]
and $\tilde\Lambda^{p\pm}_m$ is defined in the same way, with $\hat{\bm e}^\pm_p$ replacing $\hat{\bm e}_s$. These are the plane-wave transforms of basis function $m$, summed over the two segments it spans. Each segment contributes with its own unit vector $\hat{\bm u}$, and the integral is projected onto the polarization of the plane wave. Here $\bm r$ is the position along the wire and
\[
\begin{aligned}
\bm k^\pm&=(\kappa\cos\beta,\ \kappa\sin\beta,\ \pm\kzo),\\
\hat{\bm e}_s&=(-\sin\beta,\ \cos\beta,\ 0),\\
\hat{\bm e}^\pm_p&=(\pm\kzo\cos\beta,\ \pm\kzo\sin\beta,\ -\kappa)/k.
\end{aligned}
\]

For a single horizontal wire, Eq.~(\ref{eq:F_zr}) reduces to Eqs.~(\ref{eq:Zr}) and (\ref{eq:Kr}). We compare the two implementations for the 6~m dipole above the Chang'E-4 reference profile. They agree to $5.5\times10^{-6}$ in the impedance and $1.8\times10^{-9}$ in the ground fraction. A feed port is represented by a delta gap at a node, across the junction into one of the segments. The far field of the solved current, including both the direct and reflected fields, gives the beam, its polarization and the power radiated into the sky.

\subsection{The lander model}
\label{app:lander}

The lander model used in Sec.~\ref{sec:lander} consists of four monopoles, each 2.9~m long and 1~cm in radius. They extend along $\pm x$ (E, W) and $\pm y$ (N, S), and are connected to a central hub by 0.1~m arms. The hub is 2.7~m above the surface, placing the monopole tips 3.0~m from the center. A 0.1~m post connects the hub to the top of a payload cage measuring $1.0\times1.0\times0.6$~m, which stands on the deck. The lander body is represented by a $2\times2$~m box extending from 0.7 to 2.0~m above the surface, with its top face forming the deck. Four legs connect the lower corners of the body to footpads 0.1~m above the surface. The footpads are located at the corners of a square whose diagonal is 3.5~m.

The faces of the payload cage and lander body are represented by wire grids. The spacing is 0.5~m across each face, with two intervals in height, 0.3~m for the cage and 0.65~m for the body. Every wire crossing is a node. The grid wires and legs have a radius of 4~cm, while the hub and post have a radius of 2~cm. Each monopole is divided into 18 segments. The complete structure contains 177 straight wires, 253 segments and 334 basis functions.

We excite the E-W pseudo-dipole by applying $\pm\tfrac12$~V at the E and W ports, using delta gaps at the inner ends of the two monopoles. The N and S ports are left open. This excitation is odd under $x\to-x$ and even under $y\to-y$. We can therefore solve the problem within this symmetry subspace, which contains 84 basis functions, by projecting the impedance matrix onto it. The projected solution reproduces the full solution to $1.6\times10^{-14}$.

The reflected matrix is symmetric to $3\times10^{-16}$, as required by reciprocity. The free-space matrix obtained from the direct calculation, however, is symmetric only to $10^{-2}$. This asymmetry arises because the reduced thin-wire kernel uses the radius of the source segment. When two wires of different radii meet, interchanging the source and test wires therefore gives different matrix elements. For wires with the same radius, the matrix is symmetric to $3\times10^{-16}$. We restore reciprocity by using the symmetric part of the calculated matrix, $(Z+Z^{\rm T})/2$. This changes the impedance by at most $5\times10^{-4}$ and the ground fraction by at most $4\times10^{-4}$. Doubling the spectral nodes changes the impedance by at most $5\times10^{-7}$. Increasing the number of segments per monopole from 18 to 27 changes it by 0.2--0.6\%.

The wire-grid representation is the main numerical approximation in this model. For a wire grid representing a solid surface, the equal-area rule assigns the wires running in each direction the surface area of the face they represent. For a square grid with spacing $\Delta_g$, this gives a wire radius $a=\Delta_g/2\pi$ \citep{Ludwig1987,Rubinstein2005}. Alternatively, if the wires in both directions share the area of the face, the corresponding radius is $a=\Delta_g/4\pi$. We test both prescriptions at grid spacings of 0.5 and 0.25~m. The finer grid contains 798 basis functions. Our fiducial model uses $\Delta_g=0.5$~m and $a=4$~cm, following the second prescription.

We first examine the effect of halving the grid spacing. For $a=\Delta_g/4\pi$, this reduces the radius from 4 to 2~cm and changes the impedance by up to 7.0\% and the ground fraction by up to 0.016. For $a=\Delta_g/2\pi$, the radius decreases from 8 to 4~cm, giving changes of up to 11.2\% in impedance and 0.033 in ground fraction. The largest changes occur at 30--40~MHz. At $\Delta_g=0.25$~m, the two prescriptions differ by up to 7.8\% in impedance and 0.013 in ground fraction. Relative to the fiducial grid, the other three combinations change the impedance by at most 2.8\% and the ground fraction by at most 0.010 at 10--20~MHz. At 30--40~MHz, the corresponding changes reach 11.5\% and 0.030. The grid prescription therefore sets the accuracy of the impedance near 30--40~MHz to about 10\% and that of the ground fraction to about 0.03.

The grid spacing must also preserve the connections between the cage and deck and between the post and cage. A spacing of 1/3~m does not place nodes at either junction and consequently divides the model into three separate conductors. We therefore choose grid spacings that divide 0.5~m.

NEC-2 does not accurately model the four monopoles when they are fed at the hub. Its thin-wire kernel loses accuracy when a source is located next to a junction shared by several wires. Its power balance in free space is also only 0.91--0.95 for antennas placed 0.1~m above the payload. For the four monopoles in free space, however, its calculated impedance converges toward our result as the segmentation is refined. With 72 segments per monopole, the difference is 2.3\% at 10~MHz and 4.2\% at 30~MHz.

For the half-space comparison, we therefore use two simpler configurations. A vertical dipole tests the treatment of vertical currents, while the 6~m dipole placed 0.5 and 0.9~m above the payload tests scattering by the lander body and its image. The ground fractions agree to 0.0007 for the vertical dipole and to 0.003 for the dipole above the lander. The corresponding impedances for the dipole above the lander agree to 1.3--2.7\%.

\section{Reading the archived SEP data}
\label{app:sepdata}

The straightened archive table \citep{SEPdata2021} contains 217 rows and 386 comma-separated values per row, with rows numbered from 1. Its layout follows the archive bundle description and pp.~61--64 of Part~2 of the final report \citep{SEPFinal1975}. Each column corresponds to one measurement cycle, in the order in which the cycles were acquired. Row~1 contains the three-digit mode word for each cycle, with the meaning of each digit defined in the final report \citep{SEPFinal1975}. Row~2 contains a temperature.

The first digit of the mode word specifies the receiver mode. A value of 1 indicates data acquisition, while 2 indicates synchronization acquisition. The second digit specifies the frequency pair associated with the calibration and transmitter-off records in that cycle. The pairs are 32.1 and 16.0~MHz for digit 1, 8.1 and 4.0~MHz for digit 2, and 2.1 and 1.0~MHz for digit 3. The third digit is a resynchronization flag. The X, Y and Z loop measurements are stored in separate arrays. The frequency digit does not identify which loop was connected.

The 156 data-acquisition cycles cover ranges of 0--1.67~km. The remaining 230 cycles are synchronization-acquisition cycles, which contain the partial records discussed in Sec.~\ref{sec:sepdata} and cover 1.68--4.32~km. During synchronization acquisition, the receiver remains connected to a single loop \citep{SEPFinal1975}, as also indicated by the archived measurements. In these cycles, the X and Y arrays of the same samples differ by a median of 0--1.6~dB and are identical for up to 63\% of the samples. During data acquisition, the corresponding differences are 2.4--11.7~dB, with at most 2\% of the samples identical. The archive does not specify which loop was connected during the partial records, so we exclude them from the analysis. Interpreting the frequency digit of these cycles as the connected-loop identifier would not change the sample selection, since all our fits use ranges below 1.5~km.

During data acquisition, the frequency digit cycles through 1, 2 and 3 from one measurement cycle to the next. Each frequency pair therefore has 52 cycles of transmitter-off records. Rows 3--5 contain the transmitter-off levels for the X, Y and Z loops at the higher frequency of the pair, while rows 6--8 contain those at the lower frequency. When grouped by frequency digit, the median levels in each transmitter-off row differ by 3.6--15.2~dB, consistent with the frequency assignments described above.

We determine the noise floor of each receiver component from the median of its transmitter-off row, using all cycles whose frequency digit selects the corresponding frequency. The resulting floors range from $-125.8$ to $-134.9$~dBm. During synchronization acquisition, the same row records the transmitter-off level of the single connected loop. Using only the data-acquisition cycles to determine the noise floors changes them by $-2.7$ to $+8.1$~dB. The largest change occurs for the horizontal power at 16.0~MHz. The transmitter-off level of the Y loop during data acquisition is $-119.8$~dBm, 14~dB above the level in the synchronization-acquisition records. We repeat the model comparison using these alternative noise floors in Appendix~\ref{app:robust} (Table~\ref{tab:sep_robust}). Rows 9--14 contain calibration values expressed in units of the voltage-controlled oscillator and are not used in the analysis.

\begin{table*}[t]
  \caption{Robustness of the Apollo~17 comparison. The entries give the difference in $\ell_{\rm eff}$ between the exact graded profile and each alternative description. Positive values favor the graded profile. The first five columns compare the descriptions at the fitted frequencies, 2.1 + 4.0~MHz. HS denotes the half-space, 2L the two-layer description fitted at 2.1--4.0~MHz, Gr the Grimm-like profile, and Lu the best of the 27 LuSEE-Night grid structures. IBC denotes the graded profile evaluated with the impedance-boundary approximation, using the parameters from the exact fit. The last four columns give the differences for 2L and IBC at the held-out frequencies, 8.1 + 16.0~MHz and 32.1~MHz. The first three rows use different values of the scatter per sample, $\sigma_d$, and the row $r\le1000$~m uses only the samples within 1~km of the transmitter. The rows labeled weights use correlation weights calculated from the residuals of one of the six reference descriptions instead of all six (Appendix~\ref{app:stats}), weights based on the $1/e$ lag of the autocorrelation, or weights from an AR(1) fit. The final row uses noise floors determined from the transmitter-off records of the data-acquisition cycles alone. The rows labeled $h_r$ give predictions for receiver heights of 1.5 and 3.5~m, with the profile parameters fixed to those fitted at 2.5~m.\label{tab:sep_robust}}
  \begin{ruledtabular}
  \begin{tabular}{lccccccccc}
  Choice & HS & 2L & Gr & Lu & IBC & 2L, 8--16 & IBC, 8--16 & 2L, 32 & IBC, 32 \\
  \hline
  $\sigma_d=1.5$~dB & 267.1 & $-31.8$ & 33.0 & 583.0 & 25.8 & 121.3 & 0.3 & 16.3 & 0.4 \\
  $\sigma_d=2$~dB (fiducial) & 150.8 & $-18.2$ & 18.3 & 328.0 & 14.9 & 67.4 & 0.2 & 9.2 & 0.2 \\
  $\sigma_d=3$~dB & 67.5 & $-8.4$ & 7.8 & 145.6 & 7.0 & 29.1 & 0.1 & 4.1 & 0.1 \\
  $r\le1000$~m & 146.1 & $-18.1$ & 19.1 & 234.0 & 15.9 & -- & -- & -- & -- \\
  $h_r=1.5$~m & 159.7 & $-16.4$ & -- & 296.7 & 15.2 & 135.1 & 0.8 & 38.5 & 0.2 \\
  $h_r=3.5$~m & 143.4 & $-19.1$ & -- & 321.8 & 14.3 & 35.7 & $-0.4$ & 0.7 & 0.3 \\
  weights: graded & 147.8 & $-18.3$ & 17.3 & 325.5 & 14.2 & 66.8 & 0.1 & 9.4 & 0.2 \\
  weights: two-layer & 154.3 & $-17.7$ & 19.5 & 356.3 & 15.1 & 246.7 & $-1.5$ & 10.0 & 0.2 \\
  weights: half-space & 70.6 & $-13.2$ & 10.6 & 195.7 & 5.6 & 69.1 & 0.2 & 9.8 & 0.2 \\
  weights: $1/e$ lag & 164.0 & $-22.2$ & 21.8 & 401.8 & 15.6 & 124.4 & 1.2 & 23.2 & 0.5 \\
  weights: AR(1) & 97.3 & $-12.3$ & 14.5 & 259.3 & 9.5 & 86.3 & 0.7 & 50.5 & 1.1 \\
  floors: data acquisition & 149.4 & $-18.5$ & 18.1 & 328.4 & 15.5 & 81.7 & $-0.2$ & 9.2 & 0.2 \\
  \end{tabular}
  \end{ruledtabular}
\end{table*}

Rows 15--217 contain the SEP measurements. There is one block for each frequency, from 1.0 to 32.1~MHz, with the blocks beginning at rows 15, 22, 29, 43, 71 and 127. Each block contains a range array in meters, followed by six arrays of measured levels in dBm. These are the X, Y and Z loop measurements with the endfire transmitter, followed by the corresponding measurements with the broadside transmitter.

Each array spans $N_{\rm row}$ rows. The values of $N_{\rm row}$ are 1, 1, 2, 4, 8 and 13 for frequencies from 1.0 to 32.1~MHz, respectively. Thus, each measurement cycle contains more samples at higher frequencies. The rows belonging to one array are consecutive parts of the same time series and must be joined in row order to form an array of $386\,N_{\rm row}$ samples. Counting both samples and cycles from zero, sample $i$ belongs to cycle $\lfloor i/N_{\rm row}\rfloor$. Its mode word is therefore in column $\lfloor i/N_{\rm row}\rfloor+1$ of row~1.

The 4.0~MHz block provides an example. Its first row contains cycles 0--192 and its second contains cycles 193--385. The range sequence continues across the row boundary, from the end of the first row to the start of the second. Ten corrupted range values in row~29 are marked as $-888.8$ in the archive and are excluded from the analysis. At 1.0 and 2.1~MHz, the broadside vertical field exceeds the endfire field by 14~dB. This is consistent with a traverse approximately 11$^\circ$ away from the transmitter axis. At higher frequencies, the null becomes less pronounced as scattering by rocks increases \citep{Simmons1972}.

The measurements also provide an independent check of the array and orientation assignments. For a horizontal dipole, the vertical magnetic field vanishes along its axis. The endfire vertical field should therefore be weak where the broadside vertical field is strong. The horizontal power does not have this null. The archived data show precisely this pattern. Over 100--1000~m, the broadside vertical field is 4--14~dB stronger than the endfire field at every frequency, while the horizontal power is 3--14~dB stronger in the endfire orientation. Exchanging the transmitter orientations or interchanging the vertical and horizontal components would reverse these trends.

\section{Robustness of the Apollo~17 comparison}
\label{app:robust}

This appendix examines the sensitivity of the Apollo~17 model comparison to the initial conditions of the fits, the bootstrap procedure, the assumed transmitter current and other analysis choices. We also present the tests of the interference ripple, fits with the loss tangent fixed and the comparison performed in the reverse frequency direction.

\subsection{Starting points of the fits}

The comparison score has several local maxima because the interference pattern along the traverse changes with the depth and contrast of the permittivity increase. We therefore use several starting points across the allowed parameter ranges for each fit at 2.1--4.0~MHz.

For the two-layer model, the starting points comprise three points with $\varepsilon'_1=2.5$--3.3, four points with $\varepsilon'_1=3.5$--4.6 and $d_1=16$--20~m, and six points from a Latin-hypercube design over the full parameter bounds. For the graded model, we use three points with $\varepsilon'_0=2.2$--3.2 and four points with $\varepsilon'_0=3.6$--4.9 and $z_g=5$--30~m.

We also evaluate each model on a grid of 180 points spanning $\varepsilon'_1=1.75$--5.75 or $\varepsilon'_0=1.6$--4.8, depths of 3--40~m and two values of the deep permittivity. The grid search provides an independent check of the locations of the maxima found by fitting. For the two-layer model, high values of $\ell_{\rm eff}$ occur only near the maxima described below. Apart from the neighboring grid point, which lies 4.1 below the maximum, no other grid point comes within 23 of the best two-layer fit. For the graded model, all grid points within 4 of the best fit lie around that maximum.

The two-layer model has its highest maximum at the 17.9~m interface found in Sec.~\ref{sec:sepfits}. This maximum is reached from all four starting points in the second group. A second maximum, 24.1 lower, occurs for a 14.9~m layer with $\varepsilon'=2.89$ above a substrate with $\varepsilon'=5.06$, with a single loss tangent of 0.018. It is reached from two starting points in the first group and one of the Latin-hypercube points. The remaining starting points converge to fits with scores 85--391 below the highest maximum.

For the graded model, the highest maximum corresponds to the profile found in Sec.~\ref{sec:sepfits}. It is reached from the first group of starting points and from the point with $\varepsilon'_0=3.6$ and $z_g=30$~m. A second maximum, 12.7 lower, occurs at $\varepsilon'_0=4.16$, $\varepsilon'_\infty=6.09$ and $z_g=16.4$~m. This solution is close to the graded profiles obtained when samples within 200 and 300~m of the transmitter are excluded.

For the joint fits in Sec.~\ref{sec:sharpness}, we use four starting points for the two-layer model and three for the graded model. These include the maxima found in the 2.1--4.0~MHz fits and the sharp-rise solution of the transition model. In all these fits, we use the weights of the reference descriptions defined in Appendix~\ref{app:stats}.

\subsection{Bootstraps and linearization}

\paragraph{Bootstrap results.}
The bootstrap results depend on how the samples along the traverse are resampled. With circular blocks, the loss tangent of the refitted graded model remains at the lower bound of the search, 0.001, in 94\% of the replicates. The refitted half-space, Grimm-like surrogate and best structure within the LuSEE-Night range all remain behind the graded model in every replicate.

The different bootstrap schemes give similar results when the information supporting the model comparison is spread along the traverse. They differ more when a small region dominates the comparison, as in the test of the sharpness of the permittivity increase. With moving blocks, the sharp rise is preferred in only 6\% of the replicates, and the median score difference between the sharp and gradual rises is $-17.7$. This is substantially different from the value of $-3.8$ for the original measurements. The reason is that samples close to the transmitter, which favor the sharp rise, are drawn less often in this scheme. Bootstrap schemes that retain these samples with equal frequency give a preference for the sharp rise in 35--40\% of the replicates, with median score differences of $-3.3$ to $-5.7$. When the circular blocks are made twice as long, the graded model leads the two-layer model in 2 of 200 replicates. With common blocks, the surface permittivity and loss tangent are the least well constrained (Sec.~\ref{sec:sepfits}).

\paragraph{Comparison with full refits.}
The bootstrap comparisons above use linearized refits. We therefore select three sets of replicates to test how well the linearization reproduces the full calculation.

The first set contains eight moving-block replicates. In this test, the two-layer model is linearized about its lower maximum only. Across the 16 refits of the two-layer and graded models, the full and linearized calculations agree in $\ell_{\rm eff}$ to within 1.4 in 15 cases. The difference in the remaining case is 4.8. For the graded model, the full and linearized refits differ by median values of 0.16 in $\varepsilon'_0$, 0.008 in $\varepsilon'_\infty$ and 0.3~m in $z_g$.

The second and third sets contain 16 circular-block and 29 common-block replicates, respectively. In these sets, the two-layer model is refitted from both of its maxima. Each full refit starts from the corresponding linearized refit. The 29 common-block replicates comprise all 15 cases in a separate set of 80 for which the linearized surface permittivity reaches its upper bound of 5.0, four other replicates selected at random and ten additional replicates in which the linearized graded and two-layer scores differ by less than 10.

For the circular-block replicates, the better of the two linearized two-layer fits agrees with the better full refit to within 1.8 in $\ell_{\rm eff}$, with a median difference of 0.4. The graded refits agree to within 1.1. The full refit from the higher maximum gives the better result in all 16 cases, and the difference between the graded and two-layer scores has the same sign in every case.

For the 14 common-block replicates in which the linearized surface permittivity does not reach its upper bound, the difference between the graded and two-layer scores changes by at most 8.7, with a median change of 1.3. The sign of the model preference is preserved in 12 of the 14 cases. The two cases in which it changes sign are close comparisons. The linearized graded model leads by 0.95 and 1.03, respectively, whereas the full refits put the two-layer model ahead by 5.7 and 0.9. In the first case, the full graded refit moves $\varepsilon'_0$ from 4.77 to its upper bound. Every replicate of the 80 that is not refitted in full has the two models at least 10.2 apart in $\ell_{\rm eff}$.

The remaining 15 common-block replicates belong to the separate set of 80 in which the linearized surface permittivity reaches the bound. For this subset, the linearization fails. The linearized and full refits differ by up to 29 in $\ell_{\rm eff}$ for the two-layer model and by up to 6.1 for the graded model. The sign of the model preference is reversed in 10 of the 15 cases. In six of these cases the linearized comparison favors the graded model, whereas the full refits favor it in four. In seven of the ten cases with a reversal, the two models differ by less than 4 in both the linearized and full calculations. The largest reversal occurs in the case where the linearized comparison gives a lead of $-5.9$, while the full refit gives a lead of $+29.3$. This is also the only replicate in which the full graded fit retains $\varepsilon'_0$ at the upper bound, with the loss tangent at 0.044. For the other replicates at the bound, the full refits give $\varepsilon'_0=3.4$--4.4. Replicates at the bound make up 16--19\% of the common-block bootstrap.

For the separate set of 80 common-block replicates, replacing the linearized results with the full refits reduces the number of cases in which the graded model leads from 8 (10\%) to 4 (5.0\%). The latter fraction has a 95\% Wilson interval of 2.0--12.2\%. The common-block fractions of 7--12\% reported in Table~\ref{tab:seprobust} are therefore uncertain by a few percentage points and, if anything, somewhat overestimate the preference for the graded model.

\subsection{The transmitter current}

Our forward model prescribes the current distribution on the SEP transmitter (Sec.~\ref{sec:sepfields}). The fitted channel offsets absorb the overall current amplitude, but not its distribution along the wire. This distribution determines the relative amplitudes of the waves that reach different positions along the traverse. We therefore test nine current distributions. For each one, we refit the graded and two-layer models at 2.1--4.0~MHz and evaluate their predictions at the held-out frequencies using the same weights and numerical settings as in the fiducial fits.

The nine currents cover a range of possible transmitter distributions. They include a point dipole, representing the most compact source, and a uniform current over the element, representing the flattest distribution. We also use the triangular current of an electrically short dipole and sinusoidal currents with the same wavenumber on elements that are 20\% shorter or longer than the nominal length. Two further sinusoidal currents use the wavenumber of a wire over ground with permittivity 2.2 or 4.3 instead of 3.3. These values correspond to the surface permittivities of the graded and two-layer fits, respectively. Finally, we calculate the current using the method of moments for a wire of the element length, with a radius of 1~mm and a height of 5~cm above the ground. This current is solved separately over each profile as it is evaluated, with the transmitter traps represented by the open ends of the element.

The comparison changes very little across these nine current distributions. At 2.1--4.0~MHz, the two-layer model leads the graded model by 17.7--18.6. At the held-out frequencies, the two-layer model falls behind the graded profile by 65.4--68.6 at 8.1--16.0~MHz and by 8.6--8.9 at 32.1~MHz. The refitted two-layer parameters remain unchanged, while the graded parameters change by at most 0.12 in $\varepsilon'_0$ and 0.23~m in $z_g$. The method-of-moments current differs from the prescribed sinusoidal current by at most 0.10 of the feed current over both fitted profiles. Using it changes every comparison score by less than 0.1. The lower maximum of the two-layer fit also remains 22.7--25.1 below the higher maximum.

The comparison between the graded and two-layer profiles is therefore largely insensitive to the assumed transmitter current. This is expected because, apart from the first few tens of meters excluded from the fits (Sec.~\ref{sec:sepdata}), the receiver is beyond the far-field distance of the transmitter element.

\subsection{Analysis choices and layers}

Table~\ref{tab:sep_robust} summarizes the effect of the different analysis choices on the comparison between the regolith descriptions. For each choice, it gives the score difference between the exact graded profile and each alternative description. Positive values favor the graded profile. The predictions for each description are calculated once and reused across the different analysis choices. Only the comparison scores change between rows, except for the tests with different receiver heights, for which the predictions are recalculated. In those tests, the receiver height is changed to 1.5 or 3.5~m while the profile parameters remain fixed at the values fitted with a receiver height of 2.5~m.

The rows labeled weights test alternative ways of estimating the correlation weights. They use the residuals of a single reference description instead of the median residuals from all six descriptions, the lag at which the autocorrelation falls to $1/e$, or an AR(1) fit to the residuals. The row labeled floors uses noise floors determined only from the transmitter-off records during data acquisition. In these records, each array corresponds to its own loop. This replaces the fiducial choice, which uses all cycles whose mode word selects the relevant frequency. The resulting noise floors change by $-2.7$ to $+8.1$~dB (Appendix~\ref{app:sepdata}), with the largest change in the number of censored bins occurring at 16~MHz. Under this choice, the shortfall of the two-layer description relative to the graded profile at the held-out 8.1--16.0~MHz increases from 67.4 to 81.7.

The Lu column gives the smallest shortfall among the 27 structures in the LuSEE-Night grid for each analysis choice. The ranges quoted in Secs.~\ref{sec:sepfits} and \ref{sec:sepluseeibc} for the two-layer description and the LuSEE-Night grid are taken from this table. Across all choices, the half-space trails the graded profile by at least 67 at 2.1--4.0~MHz, while the Grimm-like profile trails by 7.8--33.0. The two-layer description fitted at these frequencies leads the graded profile by 8.4--31.8 in the fitted band, but trails it by at least 29.1 at 8.1--16.0~MHz and by at least 0.7 at 32.1~MHz. The smallest shortfall at 32.1~MHz occurs for a receiver height of 3.5~m. Even in this case, the two-layer description trails by 35.7 at 8.1--16.0~MHz. At the graded optimum, the exact layered-medium calculation gives scores 5.6--25.8 higher than the impedance-boundary approximation at 2.1--4.0~MHz. At the held-out frequencies, the two calculations agree to within 1.6.

Continuous profiles are represented numerically by layers whose thickness increases geometrically with depth. We use 40 layers when fitting the graded profile. Compared with a representation using 1280 layers, this gives an rms difference of 0.006~dB in the predicted fields at 2.1--4.0~MHz, with a maximum difference of 0.045~dB. The resulting change in $\ell_{\rm eff}$ is 0.08. At higher frequencies, the relatively thick layers at greater depths introduce artificial reflections. The difference increases to 0.12~dB rms, with a maximum of 2.8~dB, at 8.1--16.0~MHz and is 0.06~dB rms at 32.1~MHz. Increasing the number of layers to 320 reduces the difference to at most 0.003~dB at every frequency. We therefore use 320 layers to evaluate the graded profile and 240 layers for the Grimm-like profile.

The transition model is represented by 64 equal-thickness layers over $z_t\pm8\zeta$. When $z_t<8\zeta$, the interval begins at the surface. When the interval starts below the surface, we add six layers above it and six below it. Increasing the grid resolution by a factor of four changes the predicted levels by at most 0.023~dB at 2.1--4.0~MHz and 0.046~dB at 8.1--16.0~MHz.

\subsection{Interference ripple along the traverse}
\label{app:ripple}

The ripple test in Sec.~\ref{sec:sepholdout} compares the measured fluctuations along the traverse with those predicted by the different regolith descriptions. We first evaluate each description at the ranges of the individual measurements in a channel. The raw samples are separated by 3.5, 1.7 and 0.9~m at 8.1, 16.0 and 32.1~MHz, respectively. We then bin the predictions in exactly the same way as the measurements, using the median in each bin. This ensures that the averaging within each bin is applied to both the measured and predicted series.

To separate the ripple from the slowly varying field, we smooth each binned series by convolving it with a Gaussian in range. Unless stated otherwise, the Gaussian has a standard deviation of 25~m, which we refer to as the trend width. We define the fluctuation as the difference between the binned series and its smooth trend, and calculate the rms over the detected samples within the chosen range window. Table~\ref{tab:ripple} compares the measured rms fluctuations with those predicted by four descriptions. These are the two-layer and graded profiles fitted at 2.1--4.0~MHz, the lower maximum of the two-layer fit and the two-layer profile fitted jointly at 2.1--16.0~MHz.

The two-layer profile fitted at 2.1--4.0~MHz predicts a prominent interference pattern at the higher frequencies. Its dominant ripple period is about 40~m at 8.1~MHz, 19~m at 16.0~MHz and 9~m at 32.1~MHz. The lower maximum of the two-layer fit also produces a large ripple. At 16.0~MHz, its predicted rms fluctuations are 3.1 and 2.8~dB in the two broadside channels, both larger than the measured values.

The two-layer profile fitted jointly at 2.1--16.0~MHz behaves differently. Its interface is shallower, and its higher loss tangent of 0.025 reduces the reflected wave. It consequently predicts rms fluctuations of 2.1 and 2.0~dB at 16.0~MHz, close to the measured values of 1.9 and 1.4~dB. This explains how a two-layer profile can fit both frequency bands when its interface depth and loss are determined by the joint fit (Sec.~\ref{sec:sharpness}). The endfire horizontal power shows no corresponding interference minima. Its measured fluctuations are 0.4--0.9~dB, and none of the descriptions predicts more than 0.2~dB.

The ripple comparison does not require the archived ranges to be known with high precision, because it compares the rms fluctuation rather than the phase of the predicted ripple. We test this by perturbing the ranges used for the two-layer prediction. The tests include a scale error of $\pm3\%$, a constant shift of $\pm5$~m, and a smooth random drift with an rms of 5~m and a correlation length of 100~m. We then bin the perturbed predictions at the archived ranges. In all these cases, the predicted ripple changes by at most 0.4~dB and remains larger than the measured fluctuation in the same five channels, by factors of at least 1.36.

We also consider independent range errors of up to 6~m rms at every archived position. Such errors are a third of the ripple period at 16.0~MHz and can scramble the phase of the ripple from one sample to the next. However, most of its amplitude remains, and the predicted fluctuations still exceed the measured ones by factors of 1.18--2.18. Independent range errors of this kind are also inconsistent with the archive. The recorded ranges never decrease over the 388--842 steps between successive positions at 8.1--32.1~MHz, whereas independent errors of just 1~m rms would reverse 2--23\% of these steps.

\begin{table*}[t]
  \caption{Interference ripple along the traverse (Sec.~\ref{sec:sepholdout}). The entries give the rms fluctuation in dB about a smooth trend over 100--800~m for the broadside vertical field (BV) and broadside horizontal power (BH). The predictions are binned in the same way as the measurements. 2L denotes the two-layer fit at 2.1--4.0~MHz, 2L$'$ its lower maximum, and 2L, 2.1--16 the two-layer fit over 2.1--16~MHz. The last column gives the ratio of the 2L ripple to the measured fluctuation, evaluated for trend widths of 25, 40 and 60~m and for the range windows 100--800~m and 50--1500~m (50--1000~m at 32.1~MHz). Values above 1 indicate that the predicted ripple exceeds the measured fluctuation.\label{tab:ripple}}
  \begin{ruledtabular}
  \begin{tabular}{lcccccc}
  Channel & Measured & 2L & Graded & 2L$'$ & 2L, 2.1--16 & 2L/measured, all variants \\
  \hline
  2.1~MHz, BV & 2.90 & 1.24 & 1.14 & 0.77 & 1.40 & 0.43--0.55 \\
  2.1~MHz, BH & 2.21 & 0.73 & 0.56 & 0.54 & 0.81 & 0.33--0.49 \\
  4.0~MHz, BV & 2.02 & 0.95 & 0.16 & 0.19 & 0.24 & 0.40--0.61 \\
  4.0~MHz, BH & 1.98 & 1.56 & 0.16 & 0.16 & 0.16 & 0.74--0.79 \\
  8.1~MHz, BV & 3.08 & 2.80 & 0.17 & 1.61 & 1.20 & 0.73--0.91 \\
  8.1~MHz, BH & 1.25 & 2.79 & 0.17 & 1.71 & 0.89 & 1.19--2.23 \\
  16.0~MHz, BV & 1.87 & 3.97 & 0.12 & 3.13 & 2.06 & 1.39--2.12 \\
  16.0~MHz, BH & 1.42 & 3.69 & 0.13 & 2.81 & 2.03 & 1.44--2.60 \\
  32.1~MHz, BV & 0.95 & 1.37 & 0.15 & 0.63 & 0.32 & 1.12--1.55 \\
  32.1~MHz, BH & 1.06 & 2.22 & 0.15 & 1.08 & 0.39 & 1.54--2.08 \\
  \end{tabular}
  \end{ruledtabular}
\end{table*}

\subsection{Fixed loss tangent}
\label{app:lossfixed}

The fits at 2.1--4.0~MHz allow the loss tangent to vary freely. Both the graded and two-layer descriptions prefer values below the laboratory range expected for soils of the site. The measured range is 0.016--0.026 for the mean composition and 0.011--0.034 across the full range of compositions (Sec.~\ref{sec:regolith}). We therefore repeat the comparison with the loss tangent fixed at several values and refit the other parameters. The results are summarized in Table~\ref{tab:lossfixed}.

For the graded model, the fits at loss tangents of 0.012 and 0.020 are taken from the curve of $\ell_{\rm eff}$ against the loss tangent, with the other parameters refitted at each value (Sec.~\ref{sec:sepfits}). The fits at 0.028 and 0.040 are obtained by continuing the fit from the last point. All these fits use 40 layers to represent the profile during fitting and 320 layers for the final evaluation.

For the two-layer model, we use six starting points at each fixed loss tangent. These include the two maxima of the fit with free loss, the fit to 2.1--16~MHz and three additional starting points distributed over interface depth. The fits converge to three distinct maxima at every loss tangent. The best of these retains an interface close to that of the free-loss fit, at 17.9--18.7~m. The other two maxima have interface depths of 13.2--15.3~m. They lie 16.7--23.9 below the best two-layer fit at 2.1--4.0~MHz and also trail the graded profile at 8.1--16.0~MHz by 7.4--65.4.

The measurements at 2.1--4.0~MHz themselves favor a relatively low loss tangent. Increasing the loss tangent to 0.040 reduces the best two-layer score by 8.3 relative to its fit with free loss. The corresponding reduction for the graded profile is 8.9.

\begin{table}[!htb]
  \caption{Apollo~17 comparison with the loss tangent fixed for both descriptions. The graded and two-layer profiles are refitted at 2.1--4.0~MHz and then evaluated at 8.1--16.0 and 32.1~MHz with all profile parameters fixed. The columns give the best two-layer fit, with $\varepsilon'_1$, $\varepsilon'_2$ and $d$ in meters, the lead of the two-layer fit over the graded fit at 2.1--4.0~MHz, the lead of the graded profile at 8.1--16.0 and 32.1~MHz, and the ratio of the predicted two-layer ripple at 16.0~MHz to the measured fluctuation in the broadside vertical field and horizontal power (Table~\ref{tab:ripple}). Ratios above 1 indicate that the predicted ripple exceeds the measured fluctuation. In the first row, both loss tangents are free, with 0.001 for the graded profile and 0.008 for the two-layer profile.\label{tab:lossfixed}}
  \begin{ruledtabular}
  \begin{tabular}{lccccc}
  $\tan\delta$ & Two-layer fit & 2.1--4.0 & 8.1--16.0 & 32.1 & Ripple ratio \\
  \hline
  free & 4.27, 6.03, 17.9 & 18.2 & 67.4 & 9.2 & 2.12, 2.60 \\
  0.012 & 4.26, 6.02, 17.9 & 20.6 & 48.5 & 4.6 & 1.79, 2.35 \\
  0.020 & 4.24, 6.01, 18.1 & 20.6 & 29.5 & 1.4 & 1.16, 1.56 \\
  0.028 & 4.21, 6.00, 18.3 & 19.8 & 28.1 & 0.6 & 0.92, 1.29 \\
  0.040 & 4.16, 5.98, 18.7 & 18.8 & 12.1 & 0.3 & 0.59, 0.71 \\
  \end{tabular}
  \end{ruledtabular}
\end{table}

\subsection{Reverse test}
\label{app:reverse}

We also reverse the frequency test. We fit each model family using only the 8.1 and 16.0~MHz measurements, then predict the 2.1, 4.0 and 32.1~MHz measurements. All profile parameters are held fixed when making these predictions, and only the channel offsets are fitted.

For the graded profile, we use four starting points. Two are the profiles fitted at 2.1--4.0 and 2.1--16~MHz, and the other two are additional points in the parameter space. All four converge to the same maximum, with $\ell_{\rm eff}=-243.3$. The resulting profile rises from $\varepsilon'=3.62$ to 6.28 over $z_g=16.7$~m, with $\tan\delta=0.001$.

For the two-layer model, we use four starting points. These are the two maxima found in the 2.1--4.0~MHz fit, the fit obtained over 2.1--16~MHz and one additional point. They converge to four distinct maxima. The highest has $\ell_{\rm eff}=-221.9$ and consists of a layer with $\varepsilon'=3.46$ over a substrate with $\varepsilon'=6.99$, with an interface at 10.5~m and $\tan\delta=0.023$. Compared with the two-layer profile fitted jointly over 2.1--16~MHz, this interface is 1.0~m shallower and the deep permittivity is higher, 6.99 rather than 5.92.

The second maximum has $\ell_{\rm eff}=-235.5$. It has a top layer with $\varepsilon'=2.34$ over a substrate with $\varepsilon'=3.35$, with an interface at 11.7~m. Its loss tangent reaches the upper bound of the search, 0.050. The third maximum has an interface at 15.5~m, separating permittivities of 4.47 and 7.80, with $\tan\delta=0.032$. Its score is 17.8 below the highest two-layer maximum at 8.1--16.0~MHz. However, it gives the best prediction of the three highest two-layer maxima at 2.1--4.0~MHz, with $\ell_{\rm eff}=-406.3$. This is still 29.1 below the two-layer fit made directly at those frequencies.

At 2.1 + 4.0~MHz, the graded profile fitted at 8.1--16.0~MHz scores $-434.4$, while the highest two-layer solution scores $-424.4$. Both perform less well than profiles fitted directly at 2.1--4.0~MHz, which score $-395.4$ and $-377.2$ for the graded and two-layer descriptions, respectively (Table~\ref{tab:sep}). At 32.1~MHz, all the fits predict the measurements within 1.4 of the best description in Table~\ref{tab:holdout}. As in the forward test, the graded profiles are represented by 40 layers during fitting and evaluated using 320 layers.

Taken together, these tests show that the comparison is not driven by a particular choice of starting point, transmitter current or modest changes in the range assignments. The main result of the forward test also persists when the noise floors, correlation weights and receiver height are varied. The loss tangent affects the amplitude of the predicted interference ripple, and a two-layer profile fitted jointly over a wider frequency range can describe both bands more successfully than the particular two-layer profile fitted at 2.1--4.0~MHz. The reverse test further shows that fitting the higher frequencies alone does not reproduce the low-frequency data as well as fitting them directly. The evidence therefore concerns the predictive performance of the specific profiles selected by the fits, rather than establishing that the lunar regolith must have one particular functional form.

\section{Statistical methods}
\label{app:stats}

\subsection{Censored measurements and channel offsets}

Each SEP channel has an independent offset $o_c$, which accounts for the unknown calibration level of that channel. Both terms in Eq.~(\ref{eq:likelihood}) are concave functions of $o_c$. Consequently, the equation $\partial\ln\mathcal L_c/\partial o_c=0$ has a single root. We find it by Newton iteration, starting from the mean residual of the detected samples. When no samples are censored, the mean residual is itself the exact solution. When samples are censored, it is only an initial estimate. In the SEP data, up to 22\% of the samples in a channel are censored. Solving for the exact offset increases $\ell_{\rm eff}$ by 1.2--2.0 at 2.1--4.0~MHz and by up to 0.3 at 8.1--16.0~MHz. The resulting changes in the differences between descriptions are at most 0.8. The complete fitting procedure is also tested on synthetic data with a known graded profile in Appendix~\ref{app:pipeline}.

The fits maximize $\ell_{\rm eff}$ subject to physical bounds on the model parameters. For the half-space, we use $1.5\le\varepsilon'\le9$. For the two-layer model, the bounds are $1.5\le\varepsilon'_1\le6$, $2\le\varepsilon'_2\le9$, and a layer thickness of 1--60~m. For the graded and transition models, we use $1.5\le\varepsilon'_0,\varepsilon'_{\rm top}\le5$, $3\le\varepsilon'_\infty,\varepsilon'_{\rm deep}\le9$, $0.5\le z_g\le100$~m, $1\le z_t\le40$~m and $0.05\le \zeta\le30$~m. The loss tangent is restricted to $0.001\le\tan\delta\le0.05$ for all descriptions.

We use the trust-region reflective method, started from several points for each model. At every iteration, the channel offsets are optimized separately for the current model parameters. The minimization can be expressed in terms of a residual vector. Up to an additive constant, $-2\ell_{\rm eff}$ is the squared norm of this vector. Its components for the detected and censored samples are, respectively,
\[
\Big(\frac{N_{{\rm eff},c}}{N_c}\Big)^{1/2}\,\frac{D_{ci}-D^{\rm mod}_{ci}-o_c}{\sigma_d}
\]
and
\[
\Big(\frac{N_{{\rm eff},c}}{N_c}\Big)^{1/2}\,[-2\ln\Phi(\cdot)]^{1/2},
\]
where $\Phi(\cdot)$ is the censored term of Eq.~(\ref{eq:likelihood}). The effective sample size $N_{{\rm eff},c}$ accounts for the correlation between samples within a channel, as described below.

The optimization can have several local maxima. At 2.1--4.0~MHz, the half-space has optima at $\varepsilon'=1.68$, 2.20 and 3.00, whose values of $\ell_{\rm eff}$ differ by up to 180. The two-layer model also has a secondary optimum with a 34~m layer, which lies 68 below its best fit. The starting-point tests and the resulting maxima are discussed in Appendix~\ref{app:robust}.

\subsection{Effective sample size}

The effective sample size accounts for correlations between successive measurements along the traverse. We estimate the integrated autocorrelation time in Eq.~(\ref{eq:tauint}) from the residuals of the detected samples in each channel. The samples are ordered by range, and the channel offset is removed before calculating the autocorrelation.

We use the initial monotone sequence estimator of \citet{Geyer1992}. The sample autocorrelations are grouped in pairs,
\[
\Gamma_m=\gamma_{c,2m}+\gamma_{c,2m+1},
\]
with $\gamma_{c,0}=1$. The pairs are included while $\Gamma_m$ remains positive and does not increase. The resulting integrated autocorrelation time is
\[
\tau_c=-1+2\sum_m\Gamma_m.
\]
For a first-order autoregressive sequence with lag-one autocorrelation $\gamma_1$, this estimator gives $\tau_c=(1+\gamma_1)/(1-\gamma_1)$.

An alternative measure of correlation is the lag at which the autocorrelation first falls below $1/e$. For an autoregressive sequence of this kind, that lag is $-1/\ln\gamma_1$. When $\gamma_1$ is close to one, it is about half of $\tau_c$ and would therefore count about twice as many independent samples in the strongly correlated channels.

The effective sample size is calculated as $N_{{\rm eff},c}=N_c/\tau_c$. We use the same value of $N_{{\rm eff},c}$ for every description being compared. If the autocorrelation time were recalculated from the residuals of each description, $\tau_c$ would change in steps as the model changes. This would introduce artificial jumps in $\ell_{\rm eff}$ between descriptions with nearly identical predictions.

We therefore estimate $N_{{\rm eff},c}$ once for each set of measurements, taking the median over six reference descriptions. These are the fitted half-space and graded profiles, the two-layer description at the lower of its two maxima, the Grimm-like profile, and the graded and Grimm-like profiles evaluated with the impedance boundary. The resulting values are then held fixed during the model comparisons. At 2.1--4.0~MHz, we iterate the weights and the fits until the weights change by less than 2\%.

We also test the sensitivity to the reference descriptions used to calculate the median. Including the higher two-layer maximum instead changes $N_{{\rm eff},c}$ by at most 8\%, with the largest change occurring in the broadside channels at 2.1~MHz. The differences in $\ell_{\rm eff}$ between descriptions change by at most 1.1, except that the half-space falls 7.0 less far behind at 2.1--4.0~MHz. Other choices of correlation weights are examined in Appendix~\ref{app:robust}.

\subsection{Correlated residuals and the OU likelihood}

We also compare the models using an explicit covariance description of the residuals. For each channel and model, we represent the residuals of the detected samples at ranges $r_i$ as a zero-mean Gaussian process with covariance
\[
\sigma_{\rm OU}^2e^{-|r_i-r_{i'}|/r_{\rm OU}}
\]
between samples $i$ and $i'$. This is an Ornstein--Uhlenbeck (OU) process. Its Markov property allows the likelihood to be evaluated sequentially along the traverse, including when the sample spacing is unequal.

The first residual is Gaussian with variance $\sigma_{\rm OU}^2$. Each subsequent residual, conditioned on the preceding one, is Gaussian with mean equal to the preceding residual multiplied by $e^{-(r_i-r_{i-1})/r_{\rm OU}}$. Its conditional variance is $\sigma_{\rm OU}^2[1-e^{-2(r_i-r_{i-1})/r_{\rm OU}}]$. These conditional distributions give the exact likelihood $\ln\mathcal L_{\rm OU}$ for the adopted covariance model. Its evaluation requires time proportional to the number of samples.

The channel offset is obtained as a generalized-least-squares mean in closed form. We maximize the likelihood numerically over $\sigma_{\rm OU}$, constrained to be at least 0.5~dB, and $r_{\rm OU}$, using three starting values of the correlation length, 20, 60 and 200~m. Censored samples are excluded from this covariance comparison.

Unlike the effective-sample-size score, the OU likelihood accounts explicitly for correlations between residuals at different ranges. A description that leaves additional structure in the residuals is penalized both through the residuals themselves and through the determinant of the covariance matrix. The channels are treated as independent in this model, so it does not account for discrepancies shared between channels whose coupling varies with frequency. The common-block bootstrap described below retains such shared fluctuations in the resampling.

\subsection{Bootstrap schemes}

We use bootstrap resampling to test how stable the model comparisons are against fluctuations in the measured fields. For each channel, a bootstrap replicate is constructed by drawing blocks of $\max(3,\lceil\tau_c\rceil)$ consecutive bins at random with replacement. The blocks are joined until the replicate contains $N_c$ bins, and the result is truncated to that length. Censored bins are resampled together with their neighbors, preserving their position relative to the surrounding measurements.

The schemes differ in how these blocks are selected. Moving blocks \citep{Kunsch1989} are restricted to the observed traverse. Consequently, bins within one block length of either end are sampled less often than those in the interior. Circular blocks \citep{Politis1992} wrap from the far end of the traverse to its beginning, so every bin has the same probability of being selected.

For common blocks, we draw stretches of range whose length is set by the largest autocorrelation length among the measurements in the set. The same stretches are used for every channel in that set. This preserves fluctuations shared by different channels at the same range, such as those caused by variations in the ground. We also consider a variant in which the common stretches are half as long. The stationary bootstrap \citep{PolitisRomano1994} draws circular blocks with geometrically distributed lengths and the same mean block length. The stratified scheme draws circular blocks separately within three range intervals, 0--300~m, 300--800~m and beyond 800~m.

For each of 500 replicates of the 2.1--4.0~MHz measurements, we refit the half-space, two-layer and graded models with the weights held fixed, and then rescore the fixed descriptions. Repeating the full electromagnetic calculation for every replicate would be computationally expensive. We therefore linearize each model's prediction about its best-fitting parameters, $\bm\Psi_{\rm fit}$,
\[
D^{\rm mod}(\bm\Psi)\simeq D^{\rm mod}(\bm\Psi_{\rm fit})+\frac{\partial D^{\rm mod}}{\partial\bm\Psi}\,(\bm\Psi-\bm\Psi_{\rm fit}),
\]
where the derivatives are calculated using central differences of the full calculation.

For the predictions at the held-out frequencies and for the comparison between sharp and gradual rises, the model parameters remain fixed at their fitted values. Only the measurements are resampled. We check the accuracy of the linearization using three additional sets of replicates that are refitted with the full calculation. These comprise eight moving-block replicates, 16 circular-block replicates and 29 replicates drawn from a separate set of 80 common-block replicates. The 29 common-block replicates include all 15 cases in which the linearized $\varepsilon'_0$ reaches its bound, four other replicates selected at random and all ten additional cases in which the linearized graded and two-layer fits differ by less than 10 in $\ell_{\rm eff}$. The comparison between the full and linearized refits is given in Appendix~\ref{app:robust}.

We quote bootstrap fractions with 95\% Wilson intervals. These intervals describe the uncertainty arising from the finite number of bootstrap replicates. They do not represent posterior probabilities for the competing physical models.

\subsection{Validation with synthetic data}
\label{app:pipeline}

We test the statistical procedure of Sec.~\ref{sec:sep} using synthetic data for which the input regolith profile is known. The data are generated with the same forward model used in the analysis, using the transmitter in its actual configuration and a receiver 2.5~m above the surface. The input profile is graded, with $(\varepsilon'_0,\varepsilon'_\infty,z_g,\tan\delta)=(2.8,6.0,8~{\rm m},0.010)$. We generate the vertical field and horizontal power at 2.1 and 4.0~MHz along a traverse at $60^\circ$ to the transmitter axis. This orientation avoids symmetry conditions that would cause any of the measured components to vanish. The synthetic measurements are placed in 10~m bins from 50 to 1500~m.

We add several sources of measurement uncertainty to each channel. These include a random calibration offset, noise correlated along the traverse, 1~dB of white noise and a noise floor set at the 15th percentile of the channel level. The correlated noise is a first-order autoregressive sequence with a correlation of 0.8 between neighboring bins and an rms of 1.5~dB. The combination of the noise components gives a total noise of 1.8~dB rms. The adopted noise floor censors about 30\% of the bins. We then apply the fitting procedure described in Sec.~\ref{sec:sepstats} and the preceding subsections.

For the 96--104 detected bins in each channel, the effective sample-size estimator gives $N_{{\rm eff},c}=10$--28. For the input noise, the autocorrelation at lag $m$ is $0.69\times0.8^m$, giving $\tau_c=6.5$ and $N_{{\rm eff},c}\approx15$. The spread in the estimated values is consistent with estimating the autocorrelation time from about 100 correlated samples.

The fitted graded profile is $(2.87,\,6.00,\,8.1~{\rm m},\,0.007)$. Its residuals have an rms of 1.6--1.9~dB, consistent with the input noise of 1.8~dB. The graded model is preferred over the two-layer description by 4.7 in $\ell_{\rm eff}$ and over the half-space by 82. The corresponding leads under the covariance model are 23 and 203. As in the Apollo~17 comparison, the two-layer description is the closest competitor because it replaces the gradual permittivity increase with a single interface. Even when the true profile is known to be graded, the two frequencies and the chosen range coverage distinguish it from a single interface by only a moderate margin. For the Apollo~17 data, the stronger separation comes from the higher frequencies that are withheld from the fit.

A single realization demonstrates that the procedure can recover a known input profile, but it cannot establish the bias and scatter of the estimator. We therefore repeat the test for 300 realizations with different noise, calibration offsets and noise floors. To make these calculations practical, we linearize the model predictions as in the bootstrap analysis. The graded model is linearized about the parameters used to generate the data, while the half-space and two-layer models are linearized about their full fits to the first realization.

Across the 300 realizations, the median recovered graded parameters are 2.77, 6.00, 7.98~m and 0.0096. The half-widths of their 16th--84th percentile ranges are 0.20, 0.012, 0.45~m and 0.0034, respectively. The median biases are $-0.035$, $-0.002$, $-0.02$~m and $-0.0004$, all small compared with the corresponding scatter.

The graded model leads the two-layer model in every realization. The 16th--84th percentile range of the lead in $\ell_{\rm eff}$ is 3.0--6.6, with a median of 4.5. Under the covariance model, the corresponding range is 14.5--24.7. The graded model also leads the half-space by 84--137. At the spread of the refitted parameters, the full calculation differs from the linearized predictions by at most 0.22~dB rms and changes $\ell_{\rm eff}$ by at most 0.33. For the first realization, the linearized and full fits give leads of 4.65 and 4.70, respectively, over the two-layer model.

The ensemble of 300 realizations therefore tests the bias and scatter of the estimator, while the full nonlinear fit is checked only for the first realization. This distinction is important. The test verifies the statistical procedure and its ability to recover a known input profile. It does not independently validate the electromagnetic forward model, because the synthetic data are generated using the same model that is subsequently fitted.

\subsection{Fisher matrices and the linear bias}
\label{app:fisherbias}

For data $D_m$ with independent Gaussian errors $\sigma_m$ about a model $D^{\rm mod}_m(\bm\Xi)$, the curvature of the log-likelihood at its maximum defines the Fisher matrix,
\[
F_{ij}=\sum_m\frac{1}{\sigma_m^2}\,\frac{\partial D^{\rm mod}_m}{\partial\Xi_i}\,\frac{\partial D^{\rm mod}_m}{\partial\Xi_j}.
\]
Here, $\bm\Xi$ is the vector of all fitted parameters. Equations~(\ref{eq:fishercal}) and (\ref{eq:fisher}) are applications of this expression to the calibrator passes and the radiometer spectrum, respectively. For the calibrator, $D^{\rm mod}_m$ is the received power in dB. For the radiometer, it is $S_V(\nu_m)$.

The derivatives with respect to the regolith parameters are calculated using central differences of the full model. The marginal uncertainty on $\Xi_i$ is $[(F^{-1})_{ii}]^{1/2}$. A Gaussian prior of width $\sigma^{\rm pr}_i$ adds $1/(\sigma^{\rm pr}_i)^2$ to the corresponding diagonal element $F_{ii}$. A parameter known exactly is removed from the Fisher matrix.

The same formalism gives the first-order bias in fitted parameters when the model is evaluated with an incorrect regolith. Suppose the data are generated with $(\bm\Theta_0,\bm\Psi_0)$ and analyzed with the regolith fixed at $\bm\Psi_0+\Delta\bm\Psi$. To first order, the model is
\[
D^{\rm mod}_0+\frac{\partial D^{\rm mod}}{\partial\bm\Theta}\,\Delta\bm\Theta+\frac{\partial D^{\rm mod}}{\partial\bm\Psi}\,\Delta\bm\Psi,
\]
where $D^{\rm mod}_0$ is the model at $(\bm\Theta_0,\bm\Psi_0)$ and every row is divided by its $\sigma_m$. The best fit minimizes the squared norm of the difference between this model and the data. The resulting shift in the fitted parameters is
\[
\Delta\bm\Theta=-F_{\Theta\Theta}^{-1}F_{\Theta\Psi}\,\Delta\bm\Psi,
\]
which is Eq.~(\ref{eq:fisherbias}). The same expression applies to other modeling errors. Replacing $(\partial D^{\rm mod}/\partial\bm\Psi)\Delta\bm\Psi$ by the difference between the assumed model and the data gives the corresponding bias from the foreground representation or the thermal profile in Sec.~\ref{sec:timeforecast}.

\subsection{Antenna-fixed correction of the calibrator passes}
\label{app:calharm}

The calibrator analysis in Sec.~\ref{sec:calibrator} predicts the received power in dB at 5, 10, 20 and 40~MHz for every sample of every pass. Each pass is sampled at 241 directions between rising and setting at an elevation of 2$^\circ$. For the fiducial pass, this corresponds to approximately one sample every 5~s.

The fitted parameters comprise the four regolith parameters $\bm\Psi$ and one constant gain in dB for each pass and frequency. Thus, a single pass has four gains, while 30 passes have 120. We calculate the derivatives with respect to $\bm\Psi$ using central differences with steps of $10^{-3}$ of each parameter. The derivative with respect to a gain is 1 for samples belonging to its pass and frequency and 0 for all other samples.

To allow for an unmodeled angular dependence in the antenna response, we add an antenna-fixed correction to the predicted power in dB,
\begin{equation}
  \Delta P(\nu,\theta_c,\phi_c)=\sum_{\ell m}\Big[w_{\ell m}+w'_{\ell m}\ln\frac{\nu}{20~{\rm MHz}}\Big]\,Y_{\ell m}(\theta_c,\phi_c),
  \label{eq:calharm}
\end{equation}
where $(\theta_c,\phi_c)$ is the direction of the calibrator in the antenna frame. The sum runs over the degrees $\ell=0,\dots,\ell_{\max}$ and orders $m=-\ell,\dots,\ell$. The coefficients $w_{\ell m}$ and $w'_{\ell m}$ are expressed in dB.

The $Y_{\ell m}$ are real spherical harmonics. They are defined by $Y_{\ell0}=Y_\ell^0$, $Y_{\ell m}=\sqrt2\,{\rm Re}\,Y_\ell^m$ and $Y_{\ell,-m}=\sqrt2\,{\rm Im}\,Y_\ell^m$ for $m>0$, using complex spherical harmonics $Y_\ell^m$ normalized to unity over the sphere. The real harmonics are orthonormal over the sphere and smooth everywhere, including at the zenith. A harmonic of degree $\ell$ varies on angular scales of about $180^\circ/\ell$. Thus, an expansion to $\ell_{\max}=2$ or 3 describes structure broader than about 60--90$^\circ$.

The calibrator passes sample only the sky above 2$^\circ$ elevation. Over this restricted region, the harmonics are no longer orthogonal, but the fit requires only that they remain independent functions. The expansion contains $(\ell_{\max}+1)^2$ harmonics, each with two coefficients. It therefore has 18 coefficients for $\ell_{\max}=2$ and 32 for $\ell_{\max}=3$. These coefficients are common to all passes. The derivatives of the predicted power with respect to them are $Y_{\ell m}$ and $Y_{\ell m}\ln(\nu/20~{\rm MHz})$, evaluated at the direction and frequency of each sample.

The coefficients $w'_{\ell m}$ describe the frequency dependence of the correction. Since the correction is linear in $\ln\nu$ in dB, it multiplies the received power in each direction by a power of $\nu/(20~{\rm MHz})$. The exponent is $(\ln10/10)\sum_{\ell m}w'_{\ell m}Y_{\ell m}$ and depends on direction. This is the lowest-order smooth frequency dependence beyond a correction that is constant in frequency. It uses one additional coefficient per harmonic and keeps the model linear in its coefficients. Since the four observing frequencies are separated by octaves, the correction changes by the same amount in each direction between successive frequencies. At the reference frequency of 20~MHz, used also in the gain and foreground models of Sec.~\ref{sec:bias}, the correction is $\sum_{\ell m}w_{\ell m}Y_{\ell m}$.

With the antenna-fixed correction included, the Fisher matrix of Eq.~(\ref{eq:fishercal}) contains the four regolith parameters, the gains and the 18 or 32 harmonic coefficients. The $\ell=0$ terms are constant over the sky and can therefore be exchanged with the gains at each frequency. In addition, a single pass samples the antenna sky along only one track, leaving some combinations of the harmonic coefficients undetermined. The resulting Fisher matrix is singular. We use the Moore--Penrose pseudo-inverse to invert it. This excludes combinations that are not constrained by the samples, and we take the uncertainties of the regolith parameters from the diagonal of the pseudo-inverse.

For the lander test in Sec.~\ref{sec:calibrator}, the data are the passes calculated for the pseudo-dipole on the lander model (Sec.~\ref{sec:lander} and Appendix~\ref{app:lander}). The model used in the fit is the stand-alone 6~m dipole at the same height of 2.7~m above the same fiducial graded profile. We first fit the difference between the two sets of predictions using linear least squares. The rms differences and residuals quoted in Sec.~\ref{sec:calibrator} are those remaining after fitting the gains alone or the gains together with the harmonic correction.

To calculate the resulting bias in the regolith parameters, we also include the derivatives of the predicted passes with respect to $\bm\Psi$ in the fit. The fitted changes in $\bm\Psi$ give the parameter biases. This is the expression in Appendix~\ref{app:fisherbias}, with the difference between the two sets of passes replacing $(\partial D^{\rm mod}/\partial\bm\Psi)\Delta\bm\Psi$. We compare these biases with the uncertainties obtained from the Fisher matrix for the same model, assuming 0.1~dB uncertainty per sample.

We also consider a variant in which the harmonic coefficients are independent at each frequency. In this case, each harmonic has one coefficient per frequency instead of the pair $w_{\ell m}$ and $w'_{\ell m}$. For $\ell_{\max}=3$ at 5, 10 and 20~MHz, this gives 48 coefficients. Unlike the frequency-dependent correction in Eq.~(\ref{eq:calharm}), this variant makes no assumption about how the angular correction changes with frequency.

\section{Sky and signal models}
\label{app:signal}

This appendix describes the sky model and the Dark Ages signal used in the sensitivity analyses of Secs.~\ref{sec:thermal} and \ref{sec:bias}. The expressions are standard \citep{Furlanetto2006,PritchardLoeb2012}. We specify the inputs and numerical values used here so that the signal template can be reproduced. We also examine how the Fisher-matrix estimate depends on the number of foreground terms, which provides an additional test of the foreground model.

\paragraph{Sky-averaged spectrum.}
The Galactic background spectrum in Eq.~(\ref{eq:novaco}) turns over near 5~MHz because of free-free absorption by ionized Galactic gas. It gives temperatures of $3.7\times10^5$, $7.1\times10^4$ and $9.2\times10^3$~K at 10, 20 and 45~MHz, respectively. We represent this spectrum using a polynomial in $\log(\nu/20~{\rm MHz})$ for $\log T_{\rm sky}$. Over 5--45~MHz, the fractional differences from Eq.~(\ref{eq:novaco}) are $1.2\times10^{-4}$ with five terms, $5.8\times10^{-6}$ with six terms and $2.7\times10^{-7}$ with seven terms. For comparison, the Dark Ages signal at its trough is only $3.2\times10^{-7}$ of the sky brightness.

\paragraph{Angular sky patterns.}
In Sec.~\ref{sec:anisotropic}, we retain the sky-averaged spectrum of Eq.~(\ref{eq:novaco}) while varying the angular distribution of the foreground. We write the sky temperature as
$T_{\rm sky}(\nu,\hat{\bm n})=T_{\rm fg}(\nu)\,\mathcal M(\nu,\hat{\bm n})$,
where $T_{\rm fg}$ is the five-term representation of the sky-averaged spectrum and $\mathcal M$ is a sky map normalized by its mean over the sphere.

Our fiducial angular pattern combines two sky models. ULSA \citep{Cong2021} extrapolates the observed sky to low frequencies using a spectral index that varies with direction. It also includes free-free absorption by the warm ionized medium of the Galaxy and by discrete HII regions. Its direction-dependent spectral-index maps are available at 1, 3 and 10~MHz. The LWA1 Low Frequency Sky Model (LFSM) \citep{Dowell2017}, in the form distributed with the LWA Software Library \citep{Dowell2012LSL}, is based on surveys between 10 and 408~MHz. Below 21~MHz, however, its reconstruction contains pixels with negative brightness.

We therefore use ULSA at 1, 3 and 10~MHz and the LFSM at 22--60~MHz. To avoid differences in angular resolution between the two models, we smooth ULSA to the $5.1^\circ$ resolution of the LFSM. This also ensures that compact sources do not change across the transition between the maps. For each of the 49152 equal-area pixels, we interpolate $\log\mathcal M$ as a function of $\log\nu$ using a shape-preserving cubic through these frequency anchors.

The resulting pattern ranges from 0.024 to 8.4. Toward the Galactic center, $\mathcal M$ increases from 0.44 at 5~MHz to 8.3 at 22~MHz as the free-free absorption weakens. Toward the north Galactic pole, it decreases from 1.08 at 5~MHz to a minimum of 0.63. The LFSM also contains pixels near its low-frequency end in which the brightness increases with frequency. In 43 pixels, the brightness at 28~MHz exceeds that at 22~MHz by up to a factor of 2.1, while the median pixel decreases to 0.60 of its value. These features are retained in both the simulated data and the corresponding analysis.

\begin{table*}[t]
  \caption{Results for seven angular sky patterns, each normalized to the sky-averaged spectrum of Eq.~(\ref{eq:novaco}), for stand-alone crossed dipoles at 2.5~m above the Chang'E-4 reference profile. The second column gives the range of the beam-weighted sky $B$ at 20~MHz over a sidereal month. The next five columns give the rms residuals, in K, of the five-term fit to the month-averaged spectrum from 5 to 45~MHz. The two columns under ``Analysis with $B=1$'' use an analysis that ignores the beam. The first applies this analysis to data generated with the exact beam, while the second uses data generated with the beam fixed at its 20~MHz shape. In the latter case, the residual arises only from the frequency dependence of the sky pattern. The three ``Beam only'' columns give residuals obtained using the exact impedance and ground fraction but the beam of the indicated approximate description. The last column gives the statistical error on the Dark Ages amplitude for the fiducial graded profile, with the regolith known or fitted. The data and analysis use the same sky pattern. The signal amplitude is at most 0.040~K.\label{tab:skymodels}}
  \begin{ruledtabular}
  \begin{tabular}{lccccccc}
  & & \multicolumn{2}{c}{Analysis with $B=1$} & \multicolumn{3}{c}{Beam only} & $\sigma_A$ (mK) \\
  & $B$ at & data, & data, beam & & & & regolith \\
  Sky pattern & 20 MHz & exact beam & fixed at 20 MHz & IBC & Two-layer & Half-space & known/fitted \\
  \hline
  ULSA + LFSM (fiducial) & 0.68--1.61 & 469 & 481 & 296 & 306 & 308 & 281/345 \\
  ULSA + GSM2008 & 0.67--1.68 & 723 & 772 & 310 & 323 & 323 & 286/353 \\
  ULSA + GSM2016 & 0.72--1.53 & 302 & 135 & 301 & 318 & 317 & 276/342 \\
  ULSA + LFmap & 0.68--1.62 & 303 & 183 & 292 & 305 & 308 & 272/338 \\
  lowsky (simulated sky) & 0.79--1.26 & 522 & 39 & 510 & 564 & 526 & 262/324 \\
  ULSA, constant index & 0.73--1.52 & 566 & 10 & 507 & 515 & 530 & 275/341 \\
  LFSM at 22 MHz, fixed & 0.68--1.63 & 713 & 0 & 656 & 700 & 694 & 274/340 \\
  \end{tabular}
  \end{ruledtabular}
\end{table*}

We test the fiducial angular pattern against six alternatives (Sec.~\ref{sec:anisotropic}). Three replace the LFSM with other reconstructions of the radio surveys while retaining the same ULSA maps at low frequencies. These are the Global Sky Model of \citet{deOliveiraCosta2008} (GSM2008) at 22--60~MHz, its successor GSM2016 \citep{Zheng2017} at 30--60~MHz, and LFmap \citep{Polisensky2007} at 22--60~MHz. None provides a suitable reconstruction at lower frequencies.

GSM2016 contains negative pixels below 23~MHz. Up to 25~MHz, some of its pixels have brightnesses of only 0.0002--0.07 times the mean, compared with 0.29--0.40 in the LFSM and GSM2008. GSM2008 and LFmap also show the inner Galaxy at 10~MHz without the absorption seen in ULSA. Their brightnesses toward the Galactic center are 8.5 and 8.6 times the sky mean, respectively, compared with 1.1 in ULSA. The 10~MHz survey used in GSM2008 \citep{Caswell1976} does not cover the inner Galaxy, while LFmap extrapolates below its lowest input map at 22~MHz. Above 20~MHz, all three reconstructions correlate with the LFSM pattern over the sphere, with correlation coefficients of 0.96--0.99.

The fourth alternative is lowsky \citep{lowsky2026}, which calculates the radio sky from a three-dimensional model of the Galaxy by integrating free-free absorption along each line of sight. It covers 1--50~MHz without joining separate sky maps, and therefore varies smoothly through 10--22~MHz, where the fiducial pattern interpolates between ULSA and the LFSM. This is a realization of a Galactic model, not a map of the observed sky. Its correlation with the fiducial pattern increases from 0.49 at 5~MHz to 0.89 at 45~MHz.

The fifth alternative uses ULSA with a spectral index that is constant across the sky, using its maps at 1--10, 32 and 100~MHz. The sixth holds the LFSM pattern at 22~MHz fixed at all frequencies. We smooth GSM2008, LFmap and lowsky to the same resolution as ULSA. For GSM2016, we smooth the $5^\circ$ product by a further $1^\circ$, giving all patterns an angular resolution of about $5.1^\circ$. Table~\ref{tab:skymodels} summarizes the results for these patterns, including the range of the beam-weighted sky at 20~MHz.

We also test alternatives to the sky-averaged spectrum itself. In these tests, both the simulated data and the analysis use the spectrum of \citet{Cane1979}, a power law with index $-2.55$, or Eq.~(\ref{eq:novaco}) with its coefficient 3.28 changed to 5.0 or 2.0, giving a stronger or weaker turnover, respectively. The rms residual of the month-averaged spectrum without the beam is then 348--627~K, while the beam-only residuals are 245--384~K. For comparison, the corresponding values for the fiducial spectrum are 469~K and 296--308~K.

Finally, we test errors in the angular pattern by perturbing the fiducial pattern used to generate the data while retaining the unperturbed pattern in the analysis. The perturbed pattern is obtained by multiplying the fiducial pattern by the exponential of a random field with rms 0.15, together with a second random field of rms 0.05 multiplied by $\ln(\nu/20~{\rm MHz})$. We also consider larger perturbations, with rms values of 0.25 and 0.10, respectively. Both random fields are smooth on $10^\circ$ scales. Across 20 realizations, the median residual is 460 or 646~K with the exact beam and 497--777~K with an approximate beam.

\paragraph{Beam-weighted sky.}
The Moon's spin axis lies within $1.6^\circ$ of the ecliptic pole \citep{Archinal2018}. Consequently, the zenith at a given lunar site has an ecliptic latitude equal to the site's selenographic latitude. Its ecliptic longitude, $\lambda_z$, increases by $360^\circ$ during one sidereal month of 27.3 days.

We calculate the beam of Eq.~(\ref{eq:pattern}) at each frequency on a Gauss--Legendre grid of $96\times192$ nodes covering the upper hemisphere in the antenna frame, with the dipoles oriented along east and north. For each value of $\lambda_z$, we transform the direction of every grid node first into ecliptic coordinates and then into equatorial coordinates, using the J2000 obliquity of $23.44^\circ$. We then assign the node the sky temperature of the map pixel containing that direction. The angular integrals in Eq.~(\ref{eq:TA}) are evaluated as weighted sums over the grid nodes.

The numerator and denominator of Eq.~(\ref{eq:Bsky}) use the same angular grid and weights. As a result, a uniform sky gives $B=1$ exactly. We verify this calculation with a second implementation that sums directly over the sky-map pixels above the horizon for each $\lambda_z$. In this calculation, the beam is evaluated at the pixel centers. The reflection coefficients are interpolated in $\theta$ using cubic splines at 1201 points and are calculated from the characteristic-matrix product of Eq.~(\ref{eq:rstack}), rather than the recursion of Eq.~(\ref{eq:parratt}). The resulting sum of the beam reproduces the power radiated into the sky to $10^{-5}$ over 5--45~MHz (Table~\ref{tab:verification}).

The two implementations agree in $B$ to $1.0\times10^{-3}$ at individual values of $\lambda_z$. This difference arises from how the integration nodes sample the $0.9^\circ$ pixels. In the month average, the agreement improves to $3.5\times10^{-5}$. The month-averaged residuals in Sec.~\ref{sec:anisotropic} agree to 0.2\% when an approximate beam is used and to 0.3\% when the beam is omitted.

We also test the grid against a sky pattern for which the beam-weighted result can be calculated analytically. For the reference site on the lunar equator and $\mathcal M=1+0.5\,\hat{\bm n}\cdot\hat{\bm e}_0$, with $\hat{\bm e}_0$ the direction in the ecliptic plane at ecliptic longitude zero, the exact result is $B=1+0.5\,\langle\cos\theta\rangle\cos\lambda_z$, where $\langle\cos\theta\rangle$ is the beam average of $\cos\theta$. The numerical grid reproduces this result to $4\times10^{-5}$.

Further tests confirm convergence with angular and temporal resolution. Doubling the number of nodes in each angular direction changes $B$ by at most $1.1\times10^{-3}$ at individual values of $\lambda_z$. Reducing the step in $\lambda_z$ from $5^\circ$ to $2.5^\circ$ changes the month average by $2\times10^{-5}$. Increasing the moment-method current resolution from 80 to 160 segments changes $B$ by $8\times10^{-5}$. Rotating the dipoles by $45^\circ$ in azimuth changes the beam-only residuals by $-1.1\%$ to $+0.9\%$ and the residual without the beam from 469 to 538~K.

\paragraph{Dark Ages signal.}
We calculate the sky-averaged 21-cm brightness temperature, $T_{21}$, using the standard expressions of \citet{Furlanetto2006} and \citet{PritchardLoeb2012}. The signal is proportional to the difference between the hydrogen spin temperature and the CMB temperature, multiplied by the optical depth of the 21-cm line in the optically thin limit and reduced by the cosmological expansion.

Before the first stars form, the spin temperature is determined by the CMB and by collisions of hydrogen atoms with hydrogen atoms, electrons and protons. We use the collision-rate fit of \citet{Kuhlen2006} to the hydrogen rates of \citet{Zygelman2005}, the electron rates of \citet{Furlanetto2007a}, and a proton rate 3.2 times the hydrogen rate \citep{Furlanetto2007b}.

The ionized fraction is calculated using the effective three-level atom with a fudge factor of 1.14 \citep{Peebles1968,Seager1999} and the recombination coefficient of \citet{Pequignot1991}. The gas temperature evolves through adiabatic cooling and Compton heating by the CMB. We integrate the thermal and ionization histories from Saha equilibrium at redshift 1600, using the Planck 2018 cosmological parameters \citep{Planck2020}. The resulting residual ionized fraction is $2.7\times10^{-4}$ at redshift 100, and the gas temperature is 6.9~K at redshift 17.

The calculated signal reaches $-39.6$~mK at 16.1~MHz, corresponding to redshift 87. Its half-depth points are at 7.3 and 29.9~MHz. These values agree with published predictions for the Dark Ages signal \citep{PritchardLoeb2012,MondalBarkana2023}.

\paragraph{Dependence on the foreground model.}
We examine how the number of foreground terms affects the joint Fisher-matrix estimate in Sec.~\ref{sec:isotropic}. The calculation uses the stand-alone dipole at 2.5~m above the fiducial graded profile, an isotropic sky and 1000~h of integration. The data are generated using the same foreground model that is fitted in each case. We consider three to six foreground terms.

With the regolith known, the statistical errors on the Dark Ages amplitude are 31, 84, 267 and 319~mK for three, four, five and six terms, respectively. When the regolith is also fitted, the corresponding errors are 47, 89, 331 and 358~mK. The four foreground forms reproduce Eq.~(\ref{eq:novaco}) to $2.1\times10^{-2}$, $1.8\times10^{-3}$, $1.2\times10^{-4}$ and $5.8\times10^{-6}$, respectively. None reaches the required accuracy set by the signal, which is $3.2\times10^{-7}$ of the sky brightness at the trough. Seven terms are needed to reach this level.

\paragraph{Time-resolved analysis.}
For the calculations in Sec.~\ref{sec:timeforecast}, we evaluate the antenna coupling, ground fraction and absorption weights using the moment-method current at 81 frequencies between 5 and 45~MHz, with 0.5~MHz channel spacing. The beam-weighted sky $B$ is evaluated on a $1^\circ$ grid in $\lambda_z$ and averaged over each bin. We represent the regolith using 80 layers whose thickness increases with depth, starting from 5~mm at the surface. This surface resolution is finer than in the other calculations in the paper and is chosen to resolve the thermal weights.

For the month-averaged analysis, the resulting signal uncertainties are 280 and 359~mK in P0 and P1, respectively. These compare with 281 and 345~mK in Sec.~\ref{sec:anisotropic}. Derivatives with respect to the regolith parameters are calculated using central differences with steps of $10^{-4}$ of each parameter.

We also examine a finer description of the angular sky. The sphere is divided into the 12 base pixels of HEALPix \citep{Gorski2005}, in Galactic coordinates. Each of 11 regions has an independent amplitude and spectral index. The twelfth has no separate parameters, because a change common to all 12 regions is already represented by the foreground spectrum.

The nonlinear fits start from the parameters used to generate the data. We then iterate Gauss--Newton steps with Levenberg--Marquardt damping, including the P4 priors as additional residuals. Iteration stops when the change in $A$ falls below $10^{-7}$~K or when 60 iterations have been completed.

The map errors are simulated as Gaussian random fields expanded in spherical harmonics up to degree 18 and smoothed on $10^\circ$ scales, as in the month-averaged test. We also introduce errors in the spectral curvature by adding a field with rms 0.05 multiplied by $\ln^2(\nu/\nu_0)$ to the exponent. To test errors in free-free absorption, we multiply the pattern by $\exp[-\Delta_{\rm ff}(\nu/10~{\rm MHz})^{-2.1}]$. Here, $\Delta_{\rm ff}$ is a random field with rms 0.1 that decreases away from the Galactic plane as $\exp(-|{\rm latitude}|/10^\circ)$.

With the finer angular model, the nonlinear fits give biases in $A$ of $-244$~mK for three nights and $-254$~mK for twelve nights, compared with $-216$ and $-236$~mK with the eight sky parameters of P3 (Sec.~\ref{sec:timeforecast}).

\paragraph{Carbon recombination lines.}
Cold interstellar gas absorbs the Galactic background through carbon recombination lines. Between 5 and 45~MHz, there are 569 C$\alpha$ lines, separated by 65~kHz at 16~MHz. \citet{Yousuf2026CRRL} model these lines for LuSEE-Night, obtaining optical depths near $10^{-4}$ and $10^{-3}$ for their conservative and optimistic sky models, respectively.

We assess their possible effect by adding a forest of Lorentzian lines to the isotropic sky of Sec.~\ref{sec:isotropic}. The line forest extends 0.5~MHz beyond each edge of the observing band so that the outermost channels also include contributions from the wings of lines outside the band. We consider peak optical depths of $10^{-4}$ and $10^{-3}$ and widths of 10, 30 and 100~km\,s$^{-1}$. Both quantities are held constant across the band as a bracketing assumption. We then average the absorption over the frequency channels and fit the foreground and $A$ using the procedure described in Sec.~\ref{sec:isotropic}.

This is a screening test rather than a forecast of the carbon-line foreground. A realistic forecast requires line depths and widths that vary with frequency, together with the response of the spectrometer \citep{Yousuf2026CRRL}. In our test, the mean absorption in the channel at 16~MHz is 0.16--16~K. With five foreground terms, the residuals are 9--660~mK and the resulting biases in $A$ are $-5$ to $-590$~mK. With seven foreground terms, the biases increase to $-21$ to $-2400$~mK.

The mean absorption alone would have little effect. If it were distributed uniformly across frequency, it would bias $A$ by less than 0.5~mK. The bias instead arises from variations in the number of lines falling within each frequency channel. These variations produce fluctuations in the absorption of 7--10\% rms about its mean. A smooth foreground model cannot absorb this channel-to-channel structure.

\section{More on orbiting and surface experiments}
\label{app:secondary}

The reflectivities calculated in Sec.~\ref{sec:orbit} describe coherent specular reflection from a laterally uniform, stratified regolith. Decimeter-scale roughness reduces this contribution by at most a few percent at 5--50~MHz (Sec.~\ref{sec:limitations}). The radar value of 7\% \citep{Evans1969} is used only to set the permittivity of the uniform surface against which these results are compared.

For the Chang'E-4 reference profile, the disc-averaged reflectivity ranges from 0.11 to 0.22 across 5--50~MHz. At an orbital altitude of 300~km, its solid-angle average is 0.15 at 10~MHz and 0.17 at 30~MHz. The corresponding two-layer reduction oscillates more strongly, between 0.09 and 0.22. Because the emission originates meters below the surface, the disc-averaged brightness temperature changes little between noon and night. At 300~km, the Moon subtends 3.0~sr, so the change in antenna temperature quoted in Sec.~\ref{sec:orbit} is $3.0/4\pi$ times the change in reflectivity multiplied by $T_{\rm sky}$. Sensitivity studies for DSL, PRATUSH, CosmoCube and DARE \citep{Chen2021DSL,PRATUSH2023,Artuc2025,Burns2017} can use Eq.~(\ref{eq:TB}) in place of a constant reflectance and brightness temperature.

The same reflectivity also affects measurements of the global sky made from the ground during lunar occultations. The Murchison Widefield Array analysis at 72--230~MHz \citep{McKinley2018} fitted a constant lunar temperature of $180\pm12$~K together with the Galactic brightness, and modeled the reflected Galactic emission using an albedo of 0.07 \citep{Evans1969}. At 150~MHz, this corresponds to a reflected contribution of 25.3~K. For our far-side and Taurus-Littrow profiles, the disc-averaged reflectivity over 72--230~MHz is 0.09--0.16, giving a reflected contribution of 32--57~K at 150~MHz. This additional emission has nearly the same spectral shape as the Galactic background and therefore projects mainly onto the Galactic component in the fit. The inferred Galactic brightness at 150~MHz, $195\pm14$~K, would consequently increase by 7--31~K, comparable to or larger than its quoted uncertainty.

For two parallel elements lying on the surface and separated by a horizontal baseline $\bm b$, the coupling through the regolith is given by Eq.~(\ref{eq:Zr}) with $\tilde I(\kappa_x)\,\tilde I(-\kappa_x)$ replaced by $\tilde I_1(\kappa_x)\,\tilde I_2(-\kappa_x)\,e^{i\bm\kappa\cdot\bm b}$, where $\bm\kappa=(\kappa\cos\beta,\kappa\sin\beta)$. We compute only the beam of a single element. Its ground fraction is a different quantity. Unlike the beam, it is determined by the reactive near field of the wire and therefore depends on the wire radius, height and insulation. It must be calculated for the element in its actual configuration.

For a dipole lying on the surface (Sec.~\ref{sec:surfacearrays}), the zenith directivity relative to the power radiated into the sky is $4\pi\,(dP/d\Omega)_{\theta=0}/P_{\rm sky}$. Over the two homogeneous grounds used in the FARSIDE studies \citep{Burns2021FARSIDE,FARSIDEpol2025}, this quantity changes by less than 0.03\% across 5--45~MHz for a dipole 2~cm above the ground. Over the Chang'E-4 reference profile, the polarization asymmetry, defined as the response to Stokes~Q relative to Stokes~I for a uniform sky, lies between $-0.07$ and $+0.11$. Across the 72 tested far-side profiles, the largest zenith directivity exceeds the smallest by a factor of 1.1--1.7 for profiles with smooth permittivity changes or weak interfaces, and by a factor of 2.2--10.6 for profiles with sharp or strong interfaces (Table~\ref{tab:ensemble}).

These results characterize the single-element response and the regolith contribution to pairwise coupling. They do not describe the performance of a complete surface array. Mutual impedance and array calibration require a separate treatment and are not included in this work.

\bibliography{lunar}

\end{document}